\documentclass[prd,preprint,showpacs,superscriptaddress,floatfix]{revtex4}
\usepackage{epsfig}
\usepackage{graphicx}
\usepackage{dcolumn}
\usepackage{amsmath}
\usepackage{xspace}

\newcommand{\pbarp}{{\bar p}p}
\newcommand{\roots}{{\sqrt s}}
\newcommand{\pml} {\ \,\pm} 
\newcommand{\Et}{E_T}
\newcommand{\Pt}{p_T}
\newcommand{\Ht}{H_T}

\newcommand{\gt}{>}
\newcommand{\lt}{<}

\newcommand{\mett}{{\not\!\!E}_{T}}
\newcommand{\linespace}[1]{\protect\renewcommand{\baselinestretch}{#1}
  \footnotesize\normalsize}
\newcommand{\boxiso}{E^{iso}_{3x3}}
\newcommand{\coriso}{E^{iso}_{cone}}

\newcommand{\MeV}{\ensuremath{\mathrm{\ Me\kern -0.1em V}}\xspace}
\newcommand{\MeVc}{\ensuremath{\mathrm{\ Me\kern -0.1em V\kern -0.1em 
/\mathit{c}}}\xspace}
\newcommand{\MeVcsq}{\ensuremath{\mathrm{\ Me\kern -0.1em V\kern -0.1em 
/\mathit{c}^2}}\xspace}
\newcommand{\GeV}{\ensuremath{\mathrm{\ Ge\kern -0.1em V}}\xspace}
\newcommand{\GeVc}{\ensuremath{\mathrm{\ Ge\kern -0.1em V\kern -0.1em 
/\mathit{c}}}\xspace}
\newcommand{\GeVcsq}{\ensuremath{\mathrm{\ Ge\kern -0.1em V\kern -0.1em 
/\mathit{c}^2}}\xspace}
\newcommand{\TeV}{\ensuremath{\mathrm{\ Te\kern -0.1em V}}\xspace}
\newcommand{\nsGeV}{\ensuremath{\mathrm{Ge\kern -0.1em V}}\xspace}
\newcommand{\nsGeVc}{\ensuremath{\mathrm{Ge\kern -0.1em V\kern -0.1em 
/\mathit{c}}}\xspace}

\begin{document}
\linespace{1.0}
\begin{flushright}
FERMILAB-PUB-01/298-E \\
EFI-2001-45 \\
\end{flushright}

%\preprint{FERMILAB-PUB-01/298-E; EFI-2001-45}

\title[Search for New Physics in Photon-Lepton]
{Search for New Physics in Photon-Lepton Events 
in $p{\bar p}$ Collisions at $\roots= 1.8\TeV$}
%\begin{center}
%\vskip 0.5in

%\bf\large
%Search for New Physics in Photon-Lepton Events 
%in $p{\bar p}$ Collisions at $\roots= 1.8\TeV$
%\end{center}
\affiliation{Institute of Physics, Academia Sinica, Taipei, Taiwan 11529, Republic of China}
\affiliation{Argonne National Laboratory, Argonne, Illinois 60439}
\affiliation{Istituto Nazionale di Fisica Nucleare, University of Bologna, I-40127 Bologna, Italy}
\affiliation{Brandeis University, Waltham, Massachusetts 02254}
\affiliation{University of California at Davis, Davis, California 95616}
\affiliation{University of California at Los Angeles, Los Angeles, California 90024}
\affiliation{Instituto de Fisica de Cantabria, CSIC-University of Cantabria, 39005 Santander, Spain}
\affiliation{Enrico Fermi Institute, University of Chicago, Chicago, Illinois 60637}
\affiliation{Joint Institute for Nuclear Research, RU-141980 Dubna, Russia}
\affiliation{Duke University, Durham, North Carolina  27708}
\affiliation{Fermi National Accelerator Laboratory, Batavia, Illinois 60510}
\affiliation{University of Florida, Gainesville, Florida 32611}
\affiliation{Laboratori Nazionali di Frascati, Istituto Nazionale di Fisica Nucleare, I-00044 Frascati, Italy}
\affiliation{University of Geneva, CH-1211 Geneva 4, Switzerland}
\affiliation{Glasgow University, Glasgow G12 8QQ, United Kingdom}
\affiliation{Harvard University, Cambridge, Massachusetts 02138}
\affiliation{Hiroshima University, Higashi-Hiroshima 724, Japan}
\affiliation{University of Illinois, Urbana, Illinois 61801}
\affiliation{The Johns Hopkins University, Baltimore, Maryland 21218}
\affiliation{Institut f\"{u}r Experimentelle Kernphysik, Universit\"{a}t Karlsruhe, 76128 Karlsruhe, Germany}
\affiliation{Center for High Energy Physics: Kyungpook National University, Taegu 702-701; Seoul National University, Seoul 151-742; and SungKyunKwan University, Suwon 440-746; Korea}
\affiliation{High Energy Accelerator Research Organization (KEK), Tsukuba, Ibaraki 305, Japan}
\affiliation{Ernest Orlando Lawrence Berkeley National Laboratory, Berkeley, California 94720}
\affiliation{Massachusetts Institute of Technology, Cambridge, Massachusetts 02139}
\affiliation{University of Michigan, Ann Arbor, Michigan 48109}
\affiliation{Michigan State University, East Lansing, Michigan 48824}
\affiliation{University of New Mexico, Albuquerque, New Mexico 87131}
\affiliation{The Ohio State University, Columbus, Ohio 43210}
\affiliation{Osaka City University, Osaka 588, Japan}
\affiliation{University of Oxford, Oxford OX1 3RH, United Kingdom}
\affiliation{Universita di Padova, Istituto Nazionale di Fisica Nucleare, Sezione di Padova, I-35131 Padova, Italy}
\affiliation{University of Pennsylvania, Philadelphia, Pennsylvania 19104}
\affiliation{Istituto Nazionale di Fisica Nucleare, University and Scuola Normale Superiore of Pisa, I-56100 Pisa, Italy}
\affiliation{University of Pittsburgh, Pittsburgh, Pennsylvania 15260}
\affiliation{Purdue University, West Lafayette, Indiana 47907}
\affiliation{University of Rochester, Rochester, New York 14627}
\affiliation{Rockefeller University, New York, New York 10021}
\affiliation{Rutgers University, Piscataway, New Jersey 08855}
\affiliation{Texas A\&M University, College Station, Texas 77843}
\affiliation{Texas Tech University, Lubbock, Texas 79409}
\affiliation{Institute of Particle Physics, University of Toronto, Toronto M5S 1A7, Canada}
\affiliation{Istituto Nazionale di Fisica Nucleare, University of Trieste/Udine, Italy}
\affiliation{University of Tsukuba, Tsukuba, Ibaraki 305, Japan}
\affiliation{Tufts University, Medford, Massachusetts 02155}
\affiliation{Waseda University, Tokyo 169, Japan}
\affiliation{University of Wisconsin, Madison, Wisconsin 53706}
\affiliation{Yale University, New Haven, Connecticut 06520}

\author{D.~Acosta}
\affiliation{University of Florida, Gainesville, Florida 32611} 
\author{T.~Affolder} 
\affiliation{Ernest Orlando Lawrence Berkeley National Laboratory, Berkeley, California 94720} 
\author{H.~Akimoto} 
\affiliation{Waseda University, Tokyo 169, Japan}
\author{A.~Akopian} 
\affiliation{Rockefeller University, New York, New York 10021} 
\author{M.~G.~Albrow} 
\affiliation{Fermi National Accelerator Laboratory, Batavia, Illinois 60510} 
\author{P.~Amaral} 
\affiliation{Enrico Fermi Institute, University of Chicago, Chicago, Illinois 60637}
\author{D.~Amidei} 
\affiliation{University of Michigan, Ann Arbor, Michigan 48109} 
\author{K.~Anikeev} 
\affiliation{Massachusetts Institute of Technology, Cambridge, Massachusetts 02139} 
\author{J.~Antos} 
\affiliation{Institute of Physics, Academia Sinica, Taipei, Taiwan 11529, Republic of China}
\author{G.~Apollinari} 
\affiliation{Fermi National Accelerator Laboratory, Batavia, Illinois 60510} 
% ...

%\input{cdf1authorlist_aps}
%\author{Jeffrey Berryhill}
%\affiliation{The University of Chicago}
\collaboration{The CDF Collaboration}
\noaffiliation

\date{October 4, 2001}

\begin{abstract}
%\linespace{1.0}
\linespace{2.0}
We present the results of a search 
in $p{\bar p}$ collisions at $\roots= 1.8\TeV$ 
for anomalous production
of events containing a photon with large transverse energy 
and a lepton ($e$ or $\mu$) with large transverse energy,
using 86 pb$^{-1}$ of data collected with
the Collider Detector at Fermilab during the 1994-95 collider 
run at the Fermilab Tevatron.  
The presence of large missing transverse energy ($\mett$), 
additional photons, or additional leptons in these events is also analyzed. 
The results are consistent with standard model expectations, 
with the possible exception of photon-lepton events with
large $\mett$, for which the observed total is 16 events and
the expected mean total is $7.6\pm0.7$ events.
\linespace{1.0}
\end{abstract}

% PACS, the Physics and Astronomy Classification Scheme.

\maketitle
%\input{title_pp}
%\tableofcontents
%\listoftables
%\listoffigures
%\linespace{1.0}
\linespace{2.0}

\section{Introduction}
\label{introduction}
An important test of the standard model of particle physics~\cite{SM}
(and the extent of its validity) is to measure and understand the
properties of the highest-energy particle collisions.  The chief
predictions of the standard model for these collisions are the numbers
and varieties of fundamental particles, i.e., the fermions and gauge
bosons of the standard model, that are produced.  The observation of
an anomalous production rate of any combination of such particles is
therefore a clear indication of a new physical process.  This paper
describes an analysis of the production of a set of combinations
involving at least one photon and at least one lepton ($e$ or $\mu$),
using 86~pb$^{-1}$ of data from proton-antiproton collisions collected
with the Collider Detector at Fermilab (CDF)~\cite{cdfblurb} during
the 1994-95 run of the Fermilab Tevatron.

Production of these particular combinations of particles is of
interest for several reasons.  Events with photons and leptons are
potentially related to the puzzling ``$ee\gamma\gamma\mett$'' event
recorded by CDF~\cite{eegg}.  A supersymmetric model~\cite{susy}
designed to explain the $ee\gamma\gamma\mett$ event predicts the
production of photons from the radiative decay of the
$\tilde{\chi}^{0}_{2}$ neutralino, and leptons through the decay of
charginos, indicating $\ell\gamma\mett$ events as a signal for the
production of a chargino-neutralino pair.  Other hypothetical, massive
particles could subsequently decay to one or more standard model
electroweak gauge bosons, one of which could be a photon and the other
of which could be a leptonically decaying $W$ or $Z^0$ boson.  In
addition, photon-lepton studies complement similarly motivated
inclusive searches for new physics in diphoton~\cite{gg},
photon-jet~\cite{gj}, and photon-$b$-quark events~\cite{gb}.

The scope and strategy of this analysis are meant to reflect the
motivating principles.  Categories of photon-lepton events are defined
{\it a priori} in a way that characterizes the different possibilities
for new physics.  For each category, the inclusive event total is
compared with standard model expectations, and a few simple kinematic
distributions are presented for further examination.  The decay
products of massive particles are typically isolated from other
particles, and possess large transverse momentum and low rapidity.
This search is therefore limited to those events with at
least one isolated, central ($|\eta| < 1.0$) photon with $\Et > 25
\GeV$, and at least one isolated, central electron or muon with $\Et >
25 \GeV$.  Studying this class of events has the added advantage of
highly efficient detection and data acquisition.  These photon-lepton
candidates are further partitioned by angular separation.  Events
where exactly one photon and one lepton are detected nearly opposite
in azimuth ($\Delta\varphi_{\ell\gamma} > 150^{\circ}$) are
characteristic of a two-particle final state (two-body photon-lepton
events), and the remaining photon-lepton events are characteristic of
three or more particles in the final state (multi-body photon-lepton
events).  The inclusive event totals and kinematic properties of each
of these two categories are studied.  The multi-body photon-lepton
events are then further studied for the presence of additional
particles: photons, leptons, or the missing transverse energy
associated with weakly interacting neutral particles.

Section 2 describes the CDF detector.  Section 3 specifies the methods
for identifying photons and leptons, and the selection of
photon-lepton candidates.  Section 4 estimates the standard model
sources of photon-lepton candidates in the various search categories.
Section 5 compares the standard model expectations with the CDF data.
Section 6 presents the conclusions of the analysis.

\section{The CDF Detector}

\linespace{1.0}
\begin{figure}[!ht]
\begin{center}
\includegraphics*[width=1.0\textwidth]{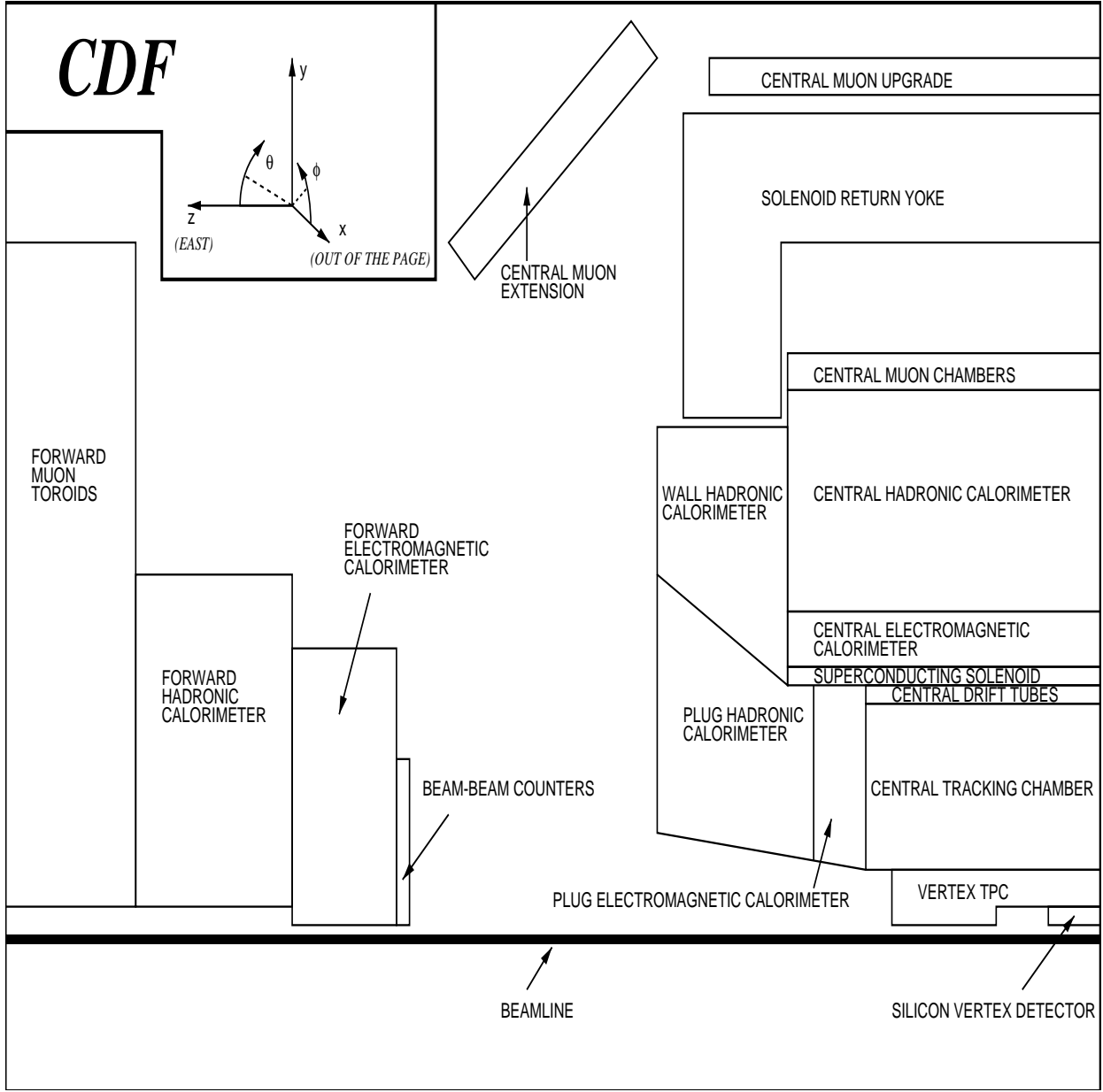}
\caption{A schematic drawing of one quadrant of the CDF detector.}
\label{detector}
\end{center}
\end{figure}
\linespace{2.0}

The CDF detector is a cylindrically symmetric, forward-backward
symmetric particle detector designed to study $\pbarp$ collisions at
the Fermilab Tevatron.  A schematic drawing of the major detector
components is shown in Figure~\ref{detector}.  A superconducting
solenoid of length 4.8~m and radius 1.5~m generates a magnetic field
of 1.4~T and contains tracking chambers used to detect charged
particles and measure their momenta.  Sampling calorimeters, used to
measure the electromagnetic and hadronic energy deposited by
electrons, photons, and jets of hadrons, surround the solenoid.
Outside the calorimeters are drift chambers used for muon detection.
In this section the subsystems relevant to this analysis are briefly
described; a more detailed description can be found
elsewhere~\cite{cdfblurb}.

A set of vertex time projection chambers (VTX)~\cite{VTX} provides
measurements in the $r$-$z$ plane up to a radius of 22~cm and detects
particle tracks in the region $|\eta| < 3.25$.  VTX tracks are used to
find the $z$ position of the $\pbarp$ interaction ($z_{event}$) and to
constrain the origin of track helices.  The 3.5-m-long central
tracking chamber (CTC) is a wire drift chamber which provides up to 84
measurements between the radii of 31.0~cm and 132.5~cm, efficient for
track detection in the region $|\eta| < 1.0$.  The CTC measures the
momenta of charged particles with momentum resolution ${\sigma_p}/{p}
< \sqrt{(0.0011 p)^2+(0.0066)^2}$, where $p$ is measured in\GeVc.

The calorimeter, segmented into towers projecting to the nominal
interaction point, is divided into three separate $\eta$ regions: a
central barrel which surrounds the solenoid coil ($|\eta|<1.1$),
`end-plugs' ($1.1<|\eta|<2.4$), and forward/backward modules
($2.4<|\eta|<4.2$).  The central barrel has an electromagnetic
calorimeter (CEM) which absorbs and measures the total energy of
electrons and photons and also a portion of the energies of
penetrating hadrons and muons.  The CEM is a sampling calorimeter
consisting of polystyrene scintillator sandwiched between lead
absorber sheets, and is segmented into 480 towers spanning
$15^{\circ}$ in $\varphi$ and $0.1$ in $\eta$.  The CEM is also
instrumented with proportional chambers (CES) embedded near shower
maximum at approximately 6 radiation lengths.  Wires and cathode
strips in the CES measure electromagnetic shower profiles in the
$\varphi$ and $z$ views, respectively.  Beyond the outer radius of the
CEM is a hadronic calorimeter (CHA) which absorbs and measures the
energy of hadrons and also a portion of the energy of penetrating
muons.  The CHA is a sampling calorimeter consisting of acrylic
scintillator sandwiched between iron absorber sheets, and is segmented
similarly to the CEM.  An endwall hadronic calorimeter (WHA) covers
the gap between the central barrel calorimeter and the end-plug
calorimeters, with construction similar to the CHA.  The end-plug
calorimeters, one on each side of the central barrel, have an
electromagnetic calorimeter (PEM) consisting of proportional chambers
sandwiched between lead absorber sheets, and a hadronic calorimeter
(PHA) consisting of proportional chambers sandwiched between iron
absorber sheets.  The PEM and PHA are both segmented into towers
spanning $5^{\circ}$ in $\varphi$ and $0.09$ in $\eta$.  The
forward/backward modules also have electromagnetic (FEM) and hadronic
(FHA) calorimeters, and are constructed similarly to the PEM and PHA.

\begin{figure}[!ht]
\begin{center}
\includegraphics*[height=0.75\textheight]{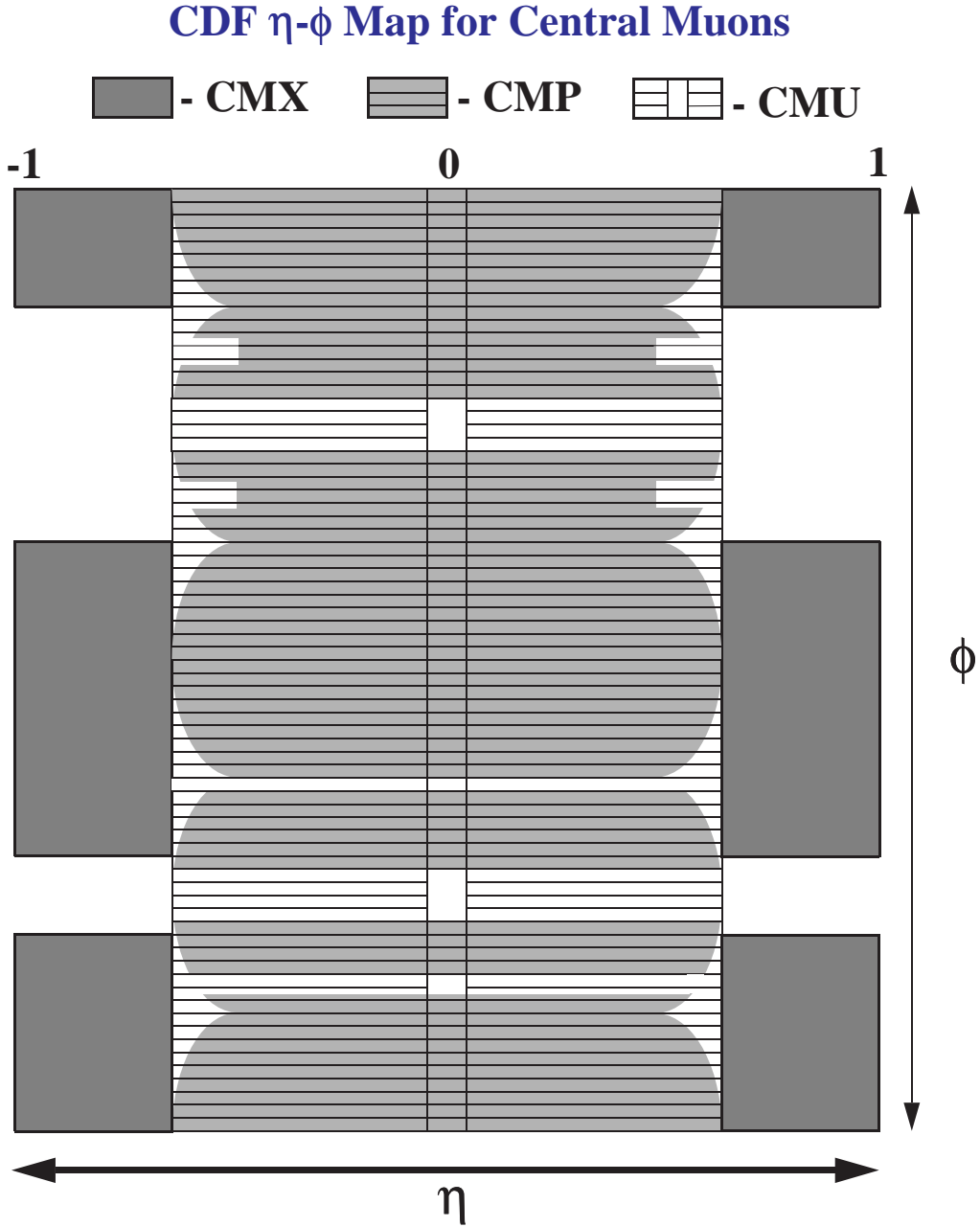}
\caption{The coverage in $\eta$-$\varphi$ space of the CDF 
central muon systems for the 1994-95 run~\cite{Drell-Yan}.}
\label{muon coverage 1b}
\end{center}
\end{figure}

Muons are detected with three systems of muon chambers situated
outside the calorimeters in the region $|\eta|<1.1$.  The central muon
detector (CMU) system consists of four layers of drift chambers
directly outside the central hadronic calorimeter, covering 84\% of
the solid angle for $|\eta| < 0.6$.  Outside of the CMU system is
0.6~m of steel shielding, followed by the central muon upgrade (CMP)
system.  The CMP system consists of four layers of drift chambers
covering 63\% of the solid angle for $|\eta| < 0.6$.  About 53\% of
the solid angle for $|\eta| < 0.6$ is covered by both the CMU and the
CMP.  The central muon extension (CMX) system consists of eight layers
of drift tubes sandwiched between scintillation counters.  The CMX
detector covers 71\% of the solid angle for $0.6<|\eta|<1.0$.
Figure~\ref{muon coverage 1b} shows the coverage in
$\eta$-$\varphi$~space for the three muon detection systems.  In each
muon system the drift chambers reconstruct the position of charged
particles using the time-to-distance relationship in the transverse
($r$-$\varphi$) plane, and charge division in the longitudinal
($r$-$z$) plane.  Three-dimensional muon track segments (``muon
stubs'') consist of position measurements in at least three layers of
chambers, in both the $r$-$\varphi$ and $r$-$z$ planes.

\linespace{1.0}
\linespace{2.0}
    
A three-level multipurpose trigger is used to select $p\bar p$ collisions
for analysis.  Each level is a logical \texttt{OR} of a number of triggers
designed to select events with electrons, muons, photons, or jets.  The 
function of each trigger level is briefly described here; 
the particular trigger combinations employed in this analysis 
are specified in Section~\ref{selection}. 

The first trigger stage, ``Level 1'', uses fast outputs from the three
central muon detectors for muon triggers, and fast outputs from all
the calorimeters for electron and jet triggers.  The second trigger
stage, ``Level 2'', combines tracking data and clusters of energy in
the calorimeters to form muon, electron, photon, and jet candidates.
A list of calorimeter clusters is provided by a nearest-neighbor
hardware cluster finder.  For each cluster, the $\Et$, average
$\varphi$, and average $\eta$ are determined.  Jet candidates are
selected from this list of clusters, and clusters that predominantly
consist of electromagnetic calorimeter energy are identified as
electron or photon candidates.  A list of $r$-$\varphi$ tracks is
provided by the central fast tracker (CFT)~\cite{CFT}, a hardware
track processor, which uses fast timing information from the CTC as
input.  A list of muon stubs is obtained from the central muon
detectors, and they are matched to CFT tracks to form muon candidates.
CFT tracks can also be matched to electromagnetic energy clusters to
form electron candidates.  A decision by the Level 2 hardware to
accept the event initiates full readout of the CDF detector data.  The
last trigger stage, ``Level 3'', performs full event reconstruction
using software executed by commercial processors.  Electron, muon,
photon, and jet candidates are selected using algorithms similar to
those employed in the final offline analysis, and a final trigger
decision selects events to be recorded for later analysis.

\section{Selection of Photon-Lepton Candidates}
\label{selection}
   
Photon-lepton candidates are obtained from three different samples of
events selected by the Level 3 trigger: inclusive photon events and
inclusive muon events, from which photon-muon candidates are selected;
and inclusive electron events, from which photon-electron candidates
are selected.  The methods for lepton identification~\cite{cdf lepton
id} and photon identification~\cite{eegg,gb} are very similar to
those of previous analyses.  The offline identification requirements
of photons and the selection of photon-muon candidates from the
inclusive photon sample are described in Section~\ref{photon id}; the
offline identification requirements of muons and the selection of
photon-muon candidates from a muon trigger sample are described in
Section~\ref{muon id}.  The offline identification requirements of
electrons and the selection of photon-electron candidates are
described in Section~\ref{electron id}.  The identification
requirements of missing transverse energy, additional photons, or
additional leptons in the photon-lepton sample are described in
Section~\ref{additional id}.  A description of the subsamples of
photon-lepton candidates to be analyzed is given in
Section~\ref{photon lepton id}.

All CDF data samples described in this paper satisfy the following
requirements: $|z_{event}|$ is less than 60~cm, so that the collision is
well-contained by the CDF detector; and there is no measurable energy in the
calorimeters recorded out of time (more than 20 ns early or more than
35 ns late, as measured by TDC's within the CHA) with the $p{\bar p}$
collision time, in order to suppress cosmic ray events and backgrounds
related to the Main Ring accelerator.
   
\subsection{Photon Identification}
\label{photon id}
Photon selection criteria are listed in Table~\ref{photoncuts} 
and are described below.  
For the energies considered here,
the response of the CEM to photons is nearly identical 
to that of electrons; 
the reconstruction and identification of 
electrons and photons are therefore very similar, the chief
difference being the high momentum track left by the 
former and the absence of any tracks left by the latter. 
% CEM energy clustering
Photon or electron candidates in the CEM are chosen from clusters of 
energy in adjacent CEM towers.  A cluster starts from seed towers 
exceeding 3\GeV in energy, and spans three towers in $\eta$ 
by one tower in $\varphi$, 
with no sharing of towers between different clusters.  The total 
photon or electron energy is the sum of the energies of the towers
in a cluster, where the energy scales of the 
CEM towers are calibrated by electrons from $Z^0$ decays. 
The energy resolution of a CEM electron or photon is given by~\cite{W mass}
\begin{eqnarray}
\bigglb( \frac{\delta E}{E} \biggrb)^{2} & = & 
\bigglb( \frac{(13.5\pm0.7)\% \GeV^{1/2}}{\sqrt{\Et}} \biggrb)^{2} + (1.5\pm0.3\%)^{2}.
\end{eqnarray}
The resolution for $\Et \gt 25\GeV$ is better than 3\%.

For photons or electrons, 
the CES shower position is determined by the energy-weighted
centroid of the highest energy clusters of those strips and wires in the CES
corresponding to the seed tower of the CEM energy cluster.
For electrons, the shower position is determined by the clusters of strips and 
wires in the CES closest to the position of the electron track, when the track 
is extrapolated to the CES radius. 
Similarly, the photon direction is determined 
by the line connecting the primary event vertex 
to the CES shower position, and 
the electron direction is determined by the electron track.

\linespace{1.25}
\begin{table}[!ht]
\begin{center}
\begin{tabular}{lcl}
\hline \hline
\multicolumn{3}{c}{Photon candidates} \\ 
\hline
CEM fiducial photon &  & \\  
Photon $\Et$ & $\gt$ & 25\GeV \\ 
Tracks with  $\Pt \gt  1\GeVc$ & $=$  & 0\\
Tracks with  $\Pt \leq 1\GeVc$ & $\leq$  & 1\\
$E_{HAD}/E_{EM}$ & $\lt$ & 
    $0.055 + 0.00045\GeV^{-1}\times E^{\gamma}$ \\ 
$\chi^2_{avg}= (\chi^2_{strip}+ \chi^2_{wire})/2$ &
   $\lt$ & 20\\ 
$E^{CES}_{2nd}$ & $\lt$ & 
$2.39\GeV + 0.01 \times E^{\gamma}$ \\ 
$\Et$ in a cone of 0.4, $\coriso$ & $\lt$ & 2\GeV \\ 
$\Pt$ of tracks in a cone of 0.4 & $\lt$ & 5\GeVc \\ 
\hline \hline
\end{tabular}
\caption{The selection criteria used to identify photon candidates.}
\label{photoncuts}
\end{center}
\end{table}
\linespace{2.0}

% CEM fiducial volume
To ensure that events are well measured, the shower positions
of electron or photon candidates 
are required to fall within the fiducial volume of the CEM.
To be in the fiducial region,
the shower position is required to lie
within 21~cm of the tower center 
in the $r$-$\varphi$ view
so that the shower is fully contained in the active region.
The region $|\eta|<0.05$, where the two halves of the detector
meet, is excluded. 
The region $0.77<\eta<1.0, 75^{\circ}<\varphi<90^{\circ}$
is uninstrumented because it is the penetration for the
cryogenic connections to the solenoidal magnet. 
In addition, the region $1.0<|\eta|<1.1$ is excluded 
because of the smaller depth of the electromagnetic
calorimeter in that region. 
The fiducial CEM coverage per photon or electron is 81\%
of the solid angle in the region defined by $|\eta|<1.0$.  

\linespace{1.25}

\begin{table}[!ht]
\begin{center}
\begin{tabular}{lcl}\hline \hline
\multicolumn{3}{c}{Inclusive photon trigger} \\ \hline
\multicolumn{3}{l}{CEM photon}\\ 
$\Et$ & $\gt$ & 23\GeV \\ 
\multicolumn{3}{l}{Fiducial CES cluster} \\ 
\multicolumn{3}{l}{$\boxiso \lt 4\GeV$ 
OR $\Et \gt 50\GeV$}\\ \hline
\multicolumn{3}{c}{Inclusive muon trigger} \\ \hline 
\multicolumn{3}{l}{CMNP, CMUP, or CMX muon}\\ 
$\Pt$ & $\gt$ & 18\GeVc \\ 
CHA energy & $\lt$ & 6\GeV\\ 
\multicolumn{3}{l}{Track-stub matching:}\\ 
\multicolumn{1}{l}{\ \ \ \ \ $|\Delta x_{stub}|$} & $\lt$ & 
  5 cm (CMNP, CMUP)\\ 
\multicolumn{1}{l}{\ \ \ \ \ $|\Delta x_{stub}|$} & $\lt$ & 10 cm (CMX)\\ 
\hline
\multicolumn{3}{c}{Inclusive electron trigger} \\ \hline
\multicolumn{3}{l}{CEM electron}\\ 
$\Et$ & $\gt$ & 18\GeV \\ 
$\Pt$ & $\gt$ & 13\GeVc \\
$E_{HAD}/E_{EM}$ & $\lt$ & 0.125 \\ 
$\chi^2_{strip}$ & $\lt$ & 10 \\
$L_{shr}$ & $\lt$ & 0.2   \\ 
\multicolumn{3}{l}{Track-CES matching:}\\
\multicolumn{1}{l}{\ \ \ \ \ $|\Delta x_{CES}|$} & $\lt$ & 3 cm \\  
\multicolumn{1}{l}{\ \ \ \ \ $|\Delta z_{CES}|$} & $\lt$ & 5 cm \\ 
\hline \hline
\end{tabular}
\caption{Level 3 trigger criteria for the 
inclusive photon, inclusive muon, and inclusive electron samples.}
\label{inclusive lepton cuts}
\end{center}
\end{table}
\linespace{2.0}

% photon CEM shower properties
Photon candidates are required to have tracking and CEM shower
characteristics consistent with that of a single, neutral,
electromagnetically interacting particle.  No CTC tracks with
$\Pt~\gt~1\GeVc$ may point at the CEM towers in the photon cluster; at
most one track with $\Pt < 1\GeVc$ is allowed to point at these same
towers.  The ratio, $E_{HAD}/E_{EM}$, of the total energy $E_{HAD}$ of
the CHA towers located behind the CEM towers in the photon cluster to
the total energy $E_{EM}$ of those CEM towers, is required to be less
than $0.055+0.00045 \GeV^{-1}\times E^{\gamma}$, where $E^{\gamma}$ is
the energy of the photon candidate.  A $\chi^2$ statistic is used to
compare the energy deposited in the CES wires ($\chi^2_{wire}$) and
cathode strips ($\chi^2_{strip}$) to that expected from test beam
data.  The average of the two measurements, $\chi^2_{avg}$, is
required to be less than 20.  The CES cluster of second highest energy
in the CEM seed tower, $E^{CES}_{2nd}$, is required to be less than
$2.39+0.01\times E^{\gamma}$ in units of\GeV.  The last two
requirements suppress CEM clusters arising from hadrons, since hadron
decay typically results in two or more closely spaced photons.
 
% photon isolation 
Calorimeter and tracking data in a cone of $\eta$-$\varphi$ space,
defined by a radius of $R \equiv \sqrt{\Delta\eta^2 + \Delta\varphi^2} \lt
0.4$ surrounding the photon cluster, are used to discriminate photons
produced in isolation from those originating in jets of hadrons.  The
total transverse energy deposited in the calorimeters in a cone of
$R=0.4$ around the photon shower position is summed, and the photon
$\Et$ is subtracted.  If there are multiple $p{\bar p}$ interactions
in the event, the mean transverse energy in a cone of $R=0.4$ per
additional interaction (0.23\GeV/interaction) is also subtracted.
The mean transverse energy leakage of the photon shower
into CEM towers outside the photon cluster, as a function of photon
shower position, is also subtracted.  The remaining energy in the cone
is the photon isolation energy, $\coriso$, which is required to be
less than 2\GeV.  As an additional indicator of photon isolation, the
sum of the momenta of CTC tracks incident upon a cone of $R=0.4$
around the photon shower position must be less than 5\GeVc.

% inclusive photon sample
An inclusive photon sample is selected with the CDF trigger
requirements described below and summarized in Table~\ref{inclusive
lepton cuts}.  At Level 1, events are required to have at least one
CEM trigger tower~\cite{trigger tower} with $\Et$ exceeding 8\GeV.  At
Level 2, a low-threshold, isolated photon trigger selects events with
CEM clusters exceeding 23\GeV in $\Et$ (computed assuming $z_{event} =
0.0$).  In addition, a CES energy cluster is required to accompany the
CEM cluster, and the additional transverse energy deposited in an
array of calorimeter towers spanning three towers in $\eta$ by three
towers in $\varphi$ surrounding the CEM cluster, $\boxiso$, is
required to be less than approximately 4\GeV.  Alternatively at Level
2, a high-threshold photon trigger selects events with CEM clusters
exceeding 50\GeV in $\Et$.  At Level 3, the full offline CEM
clustering is performed and events passing the low-threshold isolated
photon trigger are required to have fiducial CEM clusters with
$\Et \gt 23\GeV$; events passing the high-threshold photon trigger are
required to have fiducial CEM clusters with $\Et \gt 50\GeV$.  Events
selected by these photon triggers are then required to have at least
one photon candidate, satisfying all offline photon selection
requirements, with $25\GeV \lt \Et \lt 55\GeV$ for events passing the
low-threshold trigger, or with $\Et \geq 55\GeV$ for events passing
the high-threshold trigger.  This results in an inclusive photon
sample of 314,420 events.  The trigger efficiency for the
low-threshold trigger increases from 43\% to 89\% as photon $\Et$
increases from 25\GeV to 31\GeV, and remains constant at 89\% from
31\GeV to 55\GeV.  The trigger efficiency for the high-threshold
trigger is greater than 99\%.  The detection efficiency of the offline
photon selection criteria is $86.0\pm0.7\%$~\cite{jwb thesis}.

Photon-muon candidate events are selected from the inclusive photon sample by
requiring at least one muon in addition to the photon in the event.
The muon can have any of the 
central muon stub types described in Section~\ref{muon id}, 
the muon track must have $\Pt \gt 25\GeVc$, 
and all of the offline muon selection requirements must be satisfied, as
described in
Section~\ref{muon id} and summarized in Table~\ref{leptoncuts}.
This results in a photon-muon sample of 28 events.  

\linespace{1.25}
\begin{table}[!ht]
\begin{center}
\begin{tabular}{lcl}\hline \hline
\multicolumn{3}{c}{Electron candidates} \\ \hline
\multicolumn{3}{l}{CEM fiducial electron}\\
Electron $\Et$ & $\gt$ & 25\GeV\\
$\Pt \times c$ & $\gt$ & $5/9 \times \Et$ \\
Track-CES matching:  & & \\
\multicolumn{1}{l}{\ \ \ \ \ $|\Delta x_{CES}|$} & $\lt$ & 1.5 cm \\ 
\multicolumn{1}{l}{\ \ \ \ \ $|\Delta z_{CES}|$} & $\lt$ & 3 cm \\
\multicolumn{3}{l}{Track-vertex matching:} \\
\multicolumn{1}{l}{\ \ \ \ \ $|\Delta z_{event}|$} & $\lt$ & 5 cm \\
$E_{HAD}/E_{EM}$ & $\lt$ & 0.05 \\ 
$\chi^2_{strip}$ & $\lt$ & 10 \\
$|L_{shr}|$ & $\lt$ & 0.2\\ 
\multicolumn{3}{l}{Photon conversion removal}\\ 
Isolation $\Et$ & $\lt$ & $0.1 \times \Et$ \\ \hline
\multicolumn{3}{c}{Muon candidates} \\ \hline
\multicolumn{3}{l}{CMNP, CMUP, CMX, CMP, or CMU muon} \\
Track $\Pt$ & $\gt$ & 25\GeVc \\ 
\multicolumn{3}{l}{Track-stub matching:}\\ 
\multicolumn{1}{l}{\ \ \ \ \ $|\Delta x_{stub}|$} & $\lt$ & 5 cm (CMP, CMX) \\
\multicolumn{1}{l}{\ \ \ \ \ $|\Delta x_{stub}|$} & $\lt$ & 2 cm (all other)\\ 
\multicolumn{3}{l}{Track-vertex matching:}\\
\multicolumn{1}{l}{\ \ \ \ \ $|d_0|$} & $\lt$ & 0.3 cm \\
\multicolumn{1}{l}{\ \ \ \ \ $|\Delta z_{event}|$} & $\lt$ & 5 cm \\
CEM energy & $\lt$ & 2\GeV \\ 
CHA energy & $\lt$ & 6\GeV \\ 
CEM+CHA energy & $\gt$ & 0.1\GeV \\ 
Isolation $\Et$ & $\lt$ & $0.1c \times \Pt$ \\
\hline \hline
\end{tabular}
\caption{The selection criteria used to identify 
electron and muon candidates.} 
\label{leptoncuts}
\end{center}
\end{table}
\linespace{2.0}

\subsection{Muon Identification}
\label{muon id}

%muon id
% muon matching cuts
Muons are identified by extrapolating CTC tracks
through the calorimeters, and the extrapolation
must match to a stub in either the CMU, CMP, or CMX.  
There are five different types of track-stub matches: tracks which intersect
only the CMU and match a CMU stub (CMNP muons), tracks which intersect both the
CMU and CMP and match stubs in both (CMUP muons), tracks which intersect
both the CMU and CMP and match a stub in the CMU only (CMU muons),
tracks which intersect the CMP and match a stub in the CMP only (CMP muons),
and tracks which intersect the CMX and match a stub in the CMX (CMX muons). 
%The RMS spread of track-stub matching distance
%$\Delta x_{stub}$ (chamber radius $\times \Delta \varphi$) 
%is typically 0.5~cm.  
For offline identification, CMP and CMX muons are required to have a
matching distance ($\Delta x_{stub}$) less than 5~cm, and all other
muon types are required to have a matching distance less than 2~cm.
% muon track cuts
CTC tracks that are matched to muon stubs are required to be
well-measured and to be consistent with originating from the primary
event vertex.  The muon track is required to have a minimum of six
layers of CTC wire measurements, at least three of which must be axial
wire measurements and at least two of which must be stereo wire
measurements.  The distance of closest approach of the CTC track to
the primary event vertex must be less than 3~mm in the $r$-$\varphi$
view ($d_0$), and less than 5~cm in the $z$ direction ($\Delta
z_{event}$).  Muon tracks which match with $z_{event}$ are refit with
the additional constraint of originating from the primary event vertex
(``beam-constrained''), which improves muon momentum resolution by a
factor of approximately two.  The curvature resolution for
beam-constrained muons satisfying all offline selection requirements
is given by
\begin{eqnarray}
\delta(1/\Pt) & = & (0.091\pm0.004)\times 10^{-2}(\nsGeVc)^{-1},
\end{eqnarray}
corresponding to a $\Pt$ resolution of 2--8\%
for muons with $\Pt$ ranging from 25--100\GeVc~\cite{W mass}.

% muon cal and iso cuts
High energy muons are typically isolated, 
minimum-ionizing particles which have limited calorimeter activity.
A muon traversing the CEM deposits an average energy of 0.3\GeV; 
muon candidates are therefore required to deposit less than 2\GeV total
in the CEM tower(s) the muon track intersects.  
Similarly, muons traversing the CHA
deposit an average energy of 2\GeV, and so muon candidates are required
to deposit less than 6\GeV total in the intersecting CHA tower(s).  
An additional requirement that the sum of all energies in the 
intersecting CEM and CHA towers exceeds 0.1\GeV is imposed in order to 
suppress hadrons or cosmic rays which may have passed through cracks
in the central calorimeters.
Finally, in order to further suppress hadrons and muons arising from
the decay of hadrons, 
the total transverse energy deposited in the calorimeters,
in a cone of $R=0.4$ around the muon track direction,
must be less than $0.1$ times the muon track transverse momentum in\GeVc.   
The detection efficiency of the offline
muon selection criteria is $93.0\pm0.3\%$~\cite{mk thesis}.      

% inclusive muon sample
Photon-muon candidates are obtained from CDF muon triggers as follows.
At Level 1, a muon stub is required in either the CMU or CMX.  Muon
$\Pt$ is inferred from the angle of incidence of the muon stub due to
deflection by the magnetic field of the solenoid; CMU stub $\Pt$ must
exceed 6\GeVc, and CMX stub $\Pt$ must exceed 10\GeVc.  In addition,
a minimum energy of 300\MeV is required in the CHA tower associated
with the muon stub.  At Level 2, a CFT track with $\Pt \gt 12\GeVc$
is required to point within $5^{\circ}$ of a CMUP, CMNP, or CMX muon
stub triggered at Level 1.  Level 2 inclusive muon triggers are
prescaled due to bandwidth limitations; more restrictive (but not
prescaled) triggers at Level 2 must be employed to increase the
selection efficiency for photon-muon candidates.  To this end, a Level
2 trigger with no prescaling selects events which pass the Level 2
muon trigger requirements and which also have a calorimeter energy
cluster with Level 2 cluster $\Et \gt 15\GeV$.  At Level 3, as
summarized in Table~\ref{inclusive lepton cuts}, a fully reconstructed
CMUP, CMNP, or CMX muon is required, with maximum track-stub matching
distances of 5~cm, 5~cm, and 10~cm, respectively.  The muon track
$\Pt$ must exceed 18\GeVc, and the energy deposited in a CHA tower by
the muon must be less than 6\GeV.  Photon-muon candidates are selected
from 313,963 events passing the Level 3 muon triggers by requiring at
least one CMUP, CMNP, or CMX muon candidate satisfying all offline
muon selection requirements, as described in Table~\ref{leptoncuts},
and at least one photon candidate satisfying all offline photon
selection requirements, as described in Table~\ref{photoncuts}.  This
results in a photon-muon sample of 20 events.  When combined with the
28 photon-muon events from the photon triggers in
Section~\ref{photon id}, a sample of 29 unique photon-muon events is
obtained.  Of those 29 events, 9 events satisfied only the photon
trigger requirements, 1 event satisfied only the muon trigger
requirements, and 19 events satisfied both the photon and muon trigger
requirements.

The efficiency for CMUP photon-muon or CMNP photon-muon candidates is
$84\pm3\%$; the efficiency for CMX photon-muon candidates is
$68\pm5\%$~\cite{jwb thesis}.  When photon-muon candidates from the
muon triggers are combined with those from the photon triggers in
Section~\ref{photon id}, the combined trigger efficiency varies with
photon $\Et$ and muon stub type, with an average efficiency exceeding
90\%.

\subsection{Electron Identification}
\label{electron id}

%electron id 
%electron matching cuts and track cuts  
Electrons are identified in the CEM by matching
high momentum CTC tracks to high energy CEM clusters, as
summarized in Table~\ref{leptoncuts}.
The track of highest $\Pt$ which intersects one of the towers in a
CEM cluster is defined to be the electron track.  An electron candidate
is required to have a track with $\Pt$ (in\GeVc) $\gt~5/9$ 
of the CEM cluster $\Et$ (in\GeV).   
The track position, as extrapolated to the CES radius, is required
to fall within 1.5~cm of the CES shower position of the cluster in 
the $r$-$\varphi$ view ($\Delta x_{CES}$), 
and within 3~cm of the CES shower position 
in the $z$ direction ($\Delta z_{CES}$).        
The distance of closest approach of the
CTC track to the primary event vertex must be
less than 5~cm in the $z$ direction ($\Delta z_{event}$).        

% electron cal and iso cuts
The CEM shower characteristics of electron candidates must be
consistent with that of a single charged particle.  The ratio,
$E_{HAD}/E_{EM}$, of the total energy of the CHA towers located behind
the CEM towers in the electron cluster to that of the electron itself
is required to be less than 0.05.  A statistic comparing the energy
deposited in the CES cathode strips to that expected from test beam
data, $\chi^2_{strip}$, is required to be less than 10.  A comparison
of the lateral shower profile in the CEM cluster with test beam data
is parameterized by a dimensionless quantity, $L_{shr}$, which is
required to have a magnitude less than 0.2~\cite{r papers}.  Electrons
from photon conversions are removed using an algorithm based on
tracking information~\cite{cdf lepton id}.  Finally, as an additional
isolation requirement, the total transverse energy deposited in the
calorimeters, in a cone of $R=0.4$ around the electron track, must be
less than 10\% of the electron $\Et$.  The detection efficiency of the
offline electron selection criteria is $81.0\pm0.2\%$~\cite{mk thesis}.

% inclusive electron sample
Photon-electron candidates are obtained from a CDF electron trigger as
follows.  At Level 1, events are required to have at least one CEM
trigger tower~\cite{trigger tower} with $\Et$ exceeding 8\GeV.  At
Level 2, two CEM clusters with $\Et \gt 16\GeV$ are required, and also
the ratio $E_{HAD}/E_{EM}$ of each cluster is required to be less than
0.125.  The Level 3 electron trigger, summarized in
Table~\ref{inclusive lepton cuts}, requires a CEM cluster with
$\Et \gt 18\GeV$ matched to a CTC track with $\Pt \gt 13\GeVc$.  In
addition, a set of electron identification criteria less selective
than offline identification criteria is imposed: $E_{HAD}/E_{EM}$ is
required to be less than 0.125, the CES cathode strip $\chi^2$ is
required to be less than 10, the magnitude of $L_{shr}$ is required to
be less than 0.2, and the electron track must match the CES position
by 3~cm in $\Delta x_{CES}$ and by 5~cm in $\Delta z_{CES}$.

Photon-electron candidates are selected from 
474,912 events passing the Level 3 electron trigger by requiring 
at least one electron candidate satisfying 
all offline electron selection requirements, 
as described in Table~\ref{leptoncuts}, 
and at least one photon candidate satisfying all 
offline photon selection requirements, 
as described in Table~\ref{photoncuts}.  
This results in a photon-electron sample of 48 events.
The efficiency of the CDF electron trigger requirements 
for photon-electron candidates is $98.5\pm1.5\%$~\cite{jwb thesis}.  

\subsection{Selection of Additional Objects}
\label{additional id}

In addition to inclusive photon-lepton production, this analysis
investigates the associated production of other photons, other
leptons, and large missing transverse energy.  Identification of
additional photon candidates is the same as that described in
Section~\ref{photon id} and summarized in Table~\ref{photoncuts}.  The
identification of additional leptons is less selective, because the
presence of the primary photon and lepton provides good trigger
efficiency and reduces the sources of misidentified particles.

The selection of additional electron candidates 
is identical to that of previous CDF analyses~\cite{r papers} and 
is summarized in Table~\ref{loose lepton cuts}.
%2nd central electron
Additional electron candidates in the CEM (``LCEM electrons'') are
identified with criteria similar to, but looser than that of the
primary electron candidates in Section~\ref{electron id}: electron
$\Et$ must be 20\GeV or greater; electron track $\Pt$ (in\GeVc) must
exceed half of the electron $\Et$ (in\GeV); the ratio $E_{HAD}/E_{EM}$
for the electron must be less than 0.1; and the total transverse
energy deposited in the calorimeters, in a cone of $R=0.4$ around the
electron direction, must be less than 10\% of the electron $\Et$.  The
detection efficiency of these electron selection criteria is
$88.9\pm0.4\%$ for candidates with $\Et \gt 20\GeV$.

\linespace{1.25}

\begin{table}[!ht]
\begin{center}
\begin{tabular}{lcl} \hline \hline
\multicolumn{3}{c}{LCEM electron}\\ \hline
$\Et$ & $\gt$ & 20\GeV \\
$\Pt \times c$ & $\gt$ & $1/2 \times \Et$ \\
$E_{HAD}/E_{EM}$ & $\lt$ & 0.1 \\ 
Isolation $\Et$ & $\lt$ & $0.1 \times \Et$ \\ \hline
\multicolumn{3}{c}{PEM electron}\\ \hline
$\Et$ & $\gt$ & 15\GeV \\
$E_{HAD}/E_{EM}$ & $\lt$ & 0.1 \\ 
$\chi^2_{3\times3}$ & $\lt$ & 3.0 \\
%VTX occupancy & $\gt$ & 0.8 \\ 
Isolation $\Et$ & $\lt$ & $0.1 \times \Et$ \\ \hline
\multicolumn{3}{c}{FEM electron}\\ \hline
$\Et$ & $\gt$ & 10\GeV \\
$E_{HAD}/E_{EM}$ & $\lt$ & 0.1 \\ 
%VTX occupancy & $\gt$ & 0.8 \\ 
Isolation $\Et$ & $\lt$ & $0.1 \times \Et$ \\ \hline
\multicolumn{3}{c}{CMI muon}\\ \hline
$\Pt$ & $\gt$ & 20\GeVc \\ 
$|\eta_{\mu}|$ & $\lt$ & 1.2 \\
\multicolumn{3}{l}{Track-vertex matching:}\\
\multicolumn{1}{l}{\ \ \ \ \ $|d_0|$} & $\lt$ & 0.3 cm \\
\multicolumn{1}{l}{\ \ \ \ \ $|\Delta z_{event}|$} & $\lt$ & 5 cm \\
CEM energy & $\lt$ & 2\GeV \\ 
CHA energy & $\lt$ & 6\GeV \\ 
CEM+CHA energy & $\gt$ & 0.1\GeV \\ 
$\Pt$ of tracks in a cone of 0.4 & $\lt$ & $0.1 \times \Pt$  \\ 
Isolation $\Et$ & $\lt$ & $0.1c \times \Pt$ \\
\hline \hline
\end{tabular}
\caption{The selection criteria used to 
identify additional lepton candidates.} 
\label{loose lepton cuts}
\end{center}
\end{table}
\linespace{2.0}

%plug and forward electron id
Additional electron identification is extended 
to the endplug and forward regions of the calorimeter.
Electron candidates originate with
clusters of energy in the PEM or FEM with cluster $\Et$
in excess of 15\GeV and 10\GeV, respectively.   
For PEM electrons, 
a statistic comparing the energy deposited in 
a $3\times3$ array of PEM towers surrounding the PEM cluster
to that expected from test beam data, $\chi^2_{3\times3}$, is required to
be less than 3.  
The ratio $E_{HAD}/E_{EM}$ of the total energy of the PHA (FHA) towers located 
behind the PEM (FEM) towers in the 
electron cluster to that of the electron itself, is 
required to be less than 0.1.
As an isolation requirement, the total transverse energy deposited in the 
calorimeters, in a cone of $R=0.4$ around the cluster direction,
must be less than 10\% of the cluster $\Et$.   
The detection efficiency of these selection criteria is 
$87.4\pm0.7\%$ for PEM electrons with $\Et \gt 15\GeV$ 
and $75.4\pm2.6\%$ for FEM electrons with $\Et \gt 10\GeV$.

%CMUO or CMIO id
Additional muon candidates include the following: any muon satisfying
the criteria in Table~\ref{leptoncuts}, with the muon $\Pt$
requirement lowered to 20\GeVc; or an isolated CTC track consistent
with that of a minimum ionizing particle (CMI muons), the criteria for
which are summarized in Table~\ref{loose lepton cuts}.  CTC tracks in
the central region of the detector ($|\eta_{\mu}| \lt 1.2$) which do
not extrapolate to any of the central muon chambers are required to
have beam-constrained $\Pt \gt 20\GeVc$, and are required to satisfy
all of the muon selection requirements in Section~\ref{muon id}, with
the following exceptions: the muon stub matching requirement is no
longer employed; and the isolation requirements are supplemented by
the requirement that the sum of the momenta of CTC tracks, incident
upon a cone of $R=0.4$ around the muon track, be less than $0.1$ of
the muon track $\Pt$.  The detection efficiency of these selection
criteria is $91.3\pm1.3\%$ for CMI muons with $\Pt \gt 20\GeVc$.

% \mett
% raw \mett
The missing transverse energy of an event, $\mett$, is calculated as
follows.  For each tower of each calorimeter, a vector $\roarrow{E}^{i}_{T}$
is defined whose magnitude equals the calorimeter transverse energy,
as determined by the line directed from the primary event vertex 
to the calorimeter tower center, 
and whose direction is that of the same line projected into the 
plane transverse to the beam direction.
The opposite of the vector sum over all calorimeter towers,  
\begin{eqnarray}
\roarrow{\not\!\!E}_{T}(raw) & = & -\sum_i \roarrow{E}^{i}_{T},
\end{eqnarray}
is a first approximation of $\mett$.  
In this paper, the measurement of $\mett$ is improved 
by the identification of jets, 
muons, electrons, and photons, as described below. 
   
%jet id and jet corrections
Jets of hadrons are identified via clusters of energy measured by the
calorimeters.  A jet reconstruction algorithm~\cite{jet id} 
finds clusters of energy deposited in cones of fixed radius $R = 0.4$.  
The jet energy and jet direction are measured 
using the total energy and the energy-weighted centroid, 
respectively, of the calorimeter towers contained in the cone.  
The jet energy is then corrected for non-linearity in the response of the 
calorimeters, the leakage of energy between
calorimeter towers, the energy deposited outside of the jet cone,
the energy from the underlying $p\bar p$ collision debris,
and the energy from any additional $p\bar p$ interactions.
These corrections result in mean increases of 70\% (35\%) to the 
raw jet $\Et$, 
for jets with raw $\Et$ of 10\GeV (100\GeV)~\cite{cdf lepton id}.

% jet correction to \mett
An estimate of $\mett$ which takes into account the corrected jet energies,
$\mett(j)$, is obtained from $\mett(raw)$
by adding for each jet the raw jet momentum vector,  
$\roarrow{E}^{j}_{T}(raw)$,
and subtracting the corrected jet momentum vector,       
$\roarrow{E}^{j}_{T}(cor)$:
\begin{eqnarray}
\roarrow{\not\!\!E}_{T}(j) & = & 
\roarrow{\not\!\!E}_{T}(raw) - \sum_{j} \bigglb( 1 - 
    \frac{E^{j}_{T}(raw)}
         {E^{j}_{T}(cor)}
    \biggrb) \roarrow{E}^{j}_{T}(cor).
\label{eq:metj}
\end{eqnarray}
The jets included in this sum are required to have 
$E^{j}_{T}(raw) \gt 8\GeV$
and $|\eta_{j}| \lt 2.4$.

% muon correction
Muons penetrate the calorimeters, so their energy is not accounted for in
$\mett(raw)$ and must be included separately.     
Muons with any combination of stubs in the central muon chambers
are included in the $\mett$ calculation, provided that 
the beam-constrained muon track $\Pt$ exceeds 10\GeVc, 
less than 6\GeV of energy is deposited in intersecting CHA towers, 
less than 2\GeV of energy is deposited in intersecting CEM towers, 
and $\Delta x_{stub}$ satisfies the requirements in Table~\ref{leptoncuts}.
High momentum tracks without matching muon chamber stubs are also included, 
provided that all of the CMI muon criteria in 
Table~\ref{loose lepton cuts} are satisfied, except for the following 
differences:
the track need not extrapolate to regions uninstrumented by muon chambers;
the isolation requirements in Table~\ref{loose lepton cuts} are rescinded;
and in their place is added the requirement that 
the total transverse energy deposited in the 
calorimeters, in a cone of $R=0.4$ around the track direction,
must be less than 5\GeV.
An estimate of $\mett$ which takes into account the muons described above,
$\mett(j\mu)$, is obtained from $\mett(j)$
by subtracting for each muon the muon momentum vector, 
$\roarrow{p}^{\mu}_{T}$,
and adding the transverse energy vector, 
$\roarrow{E}^{\mu}_{T}$, 
of the total energy deposited in intersecting CHA and CEM towers:             
\begin{eqnarray}
\roarrow{\not\!\!E}_{T}(j\mu) & = & 
\roarrow{\not\!\!E}_{T}(j) - \sum_{\mu} \bigglb( 1 - 
    \frac{E^{\mu}_{T}}{cp^{\mu}_{T}}
    \biggrb) c\roarrow{p}^{\mu}_{T}.
\end{eqnarray}

% electron and photon correction
The response of the calorimeters to high energy electrons and photons 
differs from that of jets of hadrons, so their energy is not properly
accounted for by $\mett(j\mu)$.  
The following types of electrons and photons are included in this
correction:
any CEM photon satisfying the criteria in Table~\ref{photoncuts};
and any CEM, PEM, or FEM electron satisfying criteria identical to that 
listed in Table~\ref{loose lepton cuts}, except that the isolation 
requirements are rescinded.    
The final estimate of $\mett$ which takes into account the electron and
photon candidates described above,
$\mett(j\mu e \gamma)$, is obtained from $\mett(j\mu)$
by subtracting for each electron or photon its transverse energy vector, 
$\roarrow{E}^{e,\gamma}_{T}$,
and adding the transverse energy vector of the 
jet energy cluster corresponding to it, 
$\roarrow{E}^{j_{e,\gamma}}_{T}(cor)$: 
\begin{eqnarray}
\roarrow{\not\!\!E}_{T} & \equiv & \roarrow{\not\!\!E}_{T}(j\mu e \gamma) 
\nonumber \\*
 & = & 
\roarrow{\not\!\!E}_{T}(j\mu) - \sum_{e,\gamma} 
         \bigglb(\roarrow{E}^{e,\gamma}_{T} - 
         \roarrow{E}^{j_{e,\gamma}}_{T}(cor)\biggrb).
\end{eqnarray}

\subsection{Photon-Lepton Samples}
\label{photon lepton id}

\begin{figure}[!ht]
\begin{center}
\includegraphics*[height=0.65\textheight]{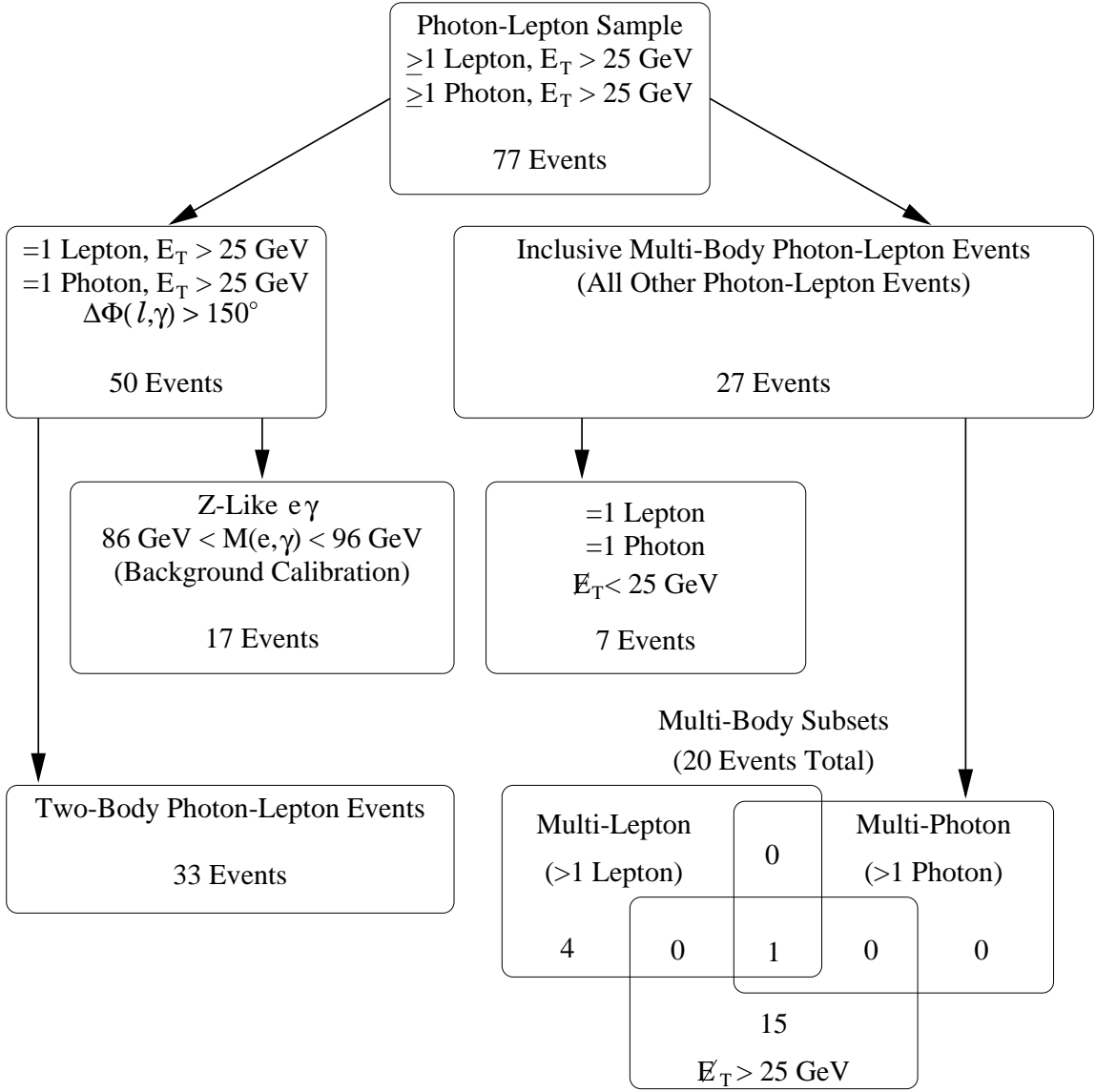}
\end{center}
\caption{The subsets of 
inclusive photon-lepton events analyzed in this paper.  The multi-body
photon-lepton subcategories of $\ell\gamma\mett$, multi-lepton, and
multi-photon events are not mutually exclusive.}
\label{lgamma path}
\end{figure}

The selection of 29 photon-muon events and 48 photon-electron events
results in the ``inclusive photon-lepton sample'' of 77 events total.
The purpose of this paper is to sort and analyze the inclusive and
exclusive combinations of particles produced for events in this sample,
the method for which is summarized in Figure~\ref{lgamma path}.

\linespace{1.0}
\linespace{2.0}

% two-body vs. multi-body
The first step in understanding the sample composition is through the
angular separation between the lepton and the photon. A two-particle
final state is indicated by the identification of a single lepton and
a single photon that are nearly opposite in azimuth.  Since a
two-particle photon-lepton final state would violate the conservation
of lepton number, such events arise from the standard model in one of
two ways: either the lepton or photon has been misidentified, or is
associated with a jet of hadrons; or a second lepton which restores
conservation of lepton number has evaded identification.  The former
is characterized by a photon and a lepton opposite in azimuth, while
the latter is suppressed in this geometry, so such a sample isolates
the majority of events with misidentified photons or leptons.  To this
end, the inclusive photon-lepton sample is analyzed as two subsamples:
a ``two-body photon-lepton sample'' typical of a two-particle final
state; and a ``multi-body photon-lepton sample'' typical of three or
more particles in the final state.  The selection requirements of the
two-body photon-lepton sample are as follows: exactly one photon and
exactly one lepton satisfying the criteria summarized in
Tables~\ref{photoncuts} and~\ref{leptoncuts}; no additional leptons
satisfying the criteria in Table~\ref{loose lepton cuts}; and the
nearest distance in azimuth between the photon and lepton, $\Delta
\varphi_{\ell\gamma}$, must exceed $150^{\circ}$.  The region $\Delta
\varphi_{\ell\gamma} \gt 150^{\circ}$ was chosen by requiring it to
include 95\% of $Z^{0}$ boson events decaying to two CEM electrons,
which are a source of misidentified photons.  Excluded from the
two-body photon-lepton sample are those two-body photon-electron
events for which the photon-electron invariant mass, $M_{e\gamma}$, is
within 5\GeVcsq of $M_Z$.  This ``$Z^{0}$-like'' control sample is used
to estimate the photon misidentification rate from electrons, as
described in Section~\ref{zee}.  The multi-body sample is composed of
the remaining inclusive photon-lepton events.

The multi-body sample is then further analyzed for the presence of 
large $\mett$, additional leptons, or additional photons.  Multi-body
events with $\mett \gt 25\GeV$, the ``multi-body $\ell\gamma\mett$ sample'',
and multi-body events with one or more additional photons or leptons 
satisfying the criteria described in Section~\ref{additional id},
the ``multi-photon and multi-lepton sample'',  are studied concurrently
with the two-body sample and the inclusive multi-body sample.  
The $\mett$ threshold of 25\GeV was chosen 
from previous analyses~\cite{eegg, r papers} as a significant indicator of a 
neutrino arising from leptonic decays of the $W$ boson.
Among these samples, the following properties are analyzed:
the total event rate; the distribution of lepton $\Et$, photon $\Et$,
and $\mett$; the distribution of the invariant mass of any relevant
combinations of particles; and the angular distributions of any
relevant combinations of particles.

\section{Standard Model Sources}
\label{background}
\subsection{$W\gamma$ and $Z^0\gamma$ Production}
\label{diboson}

The dominant source of photon-lepton events at the Tevatron is
electroweak diboson production, wherein an electroweak boson ($W$ or
$Z^0$) decays leptonically ($\ell \nu$ or $\ell\ell$) and a photon is
radiated from either the initial state quark, a charged electroweak
boson ($W$), or a charged final state lepton.  The number of
photon-lepton events from electroweak diboson production is estimated
from a Monte Carlo event generator program~\cite{diboson mc}.  The
event generator program outputs 4-vectors of particles emanating from a
diboson production event, and this output is used as input to a CDF
detector simulation program, which outputs simulated data in a format
identical to that of an actual CDF event.  Simulated photon-lepton
events can then be analyzed in a manner identical to that of CDF data.

\begin{figure}[!ht]
\begin{center}
\includegraphics*[width=0.4\textwidth]{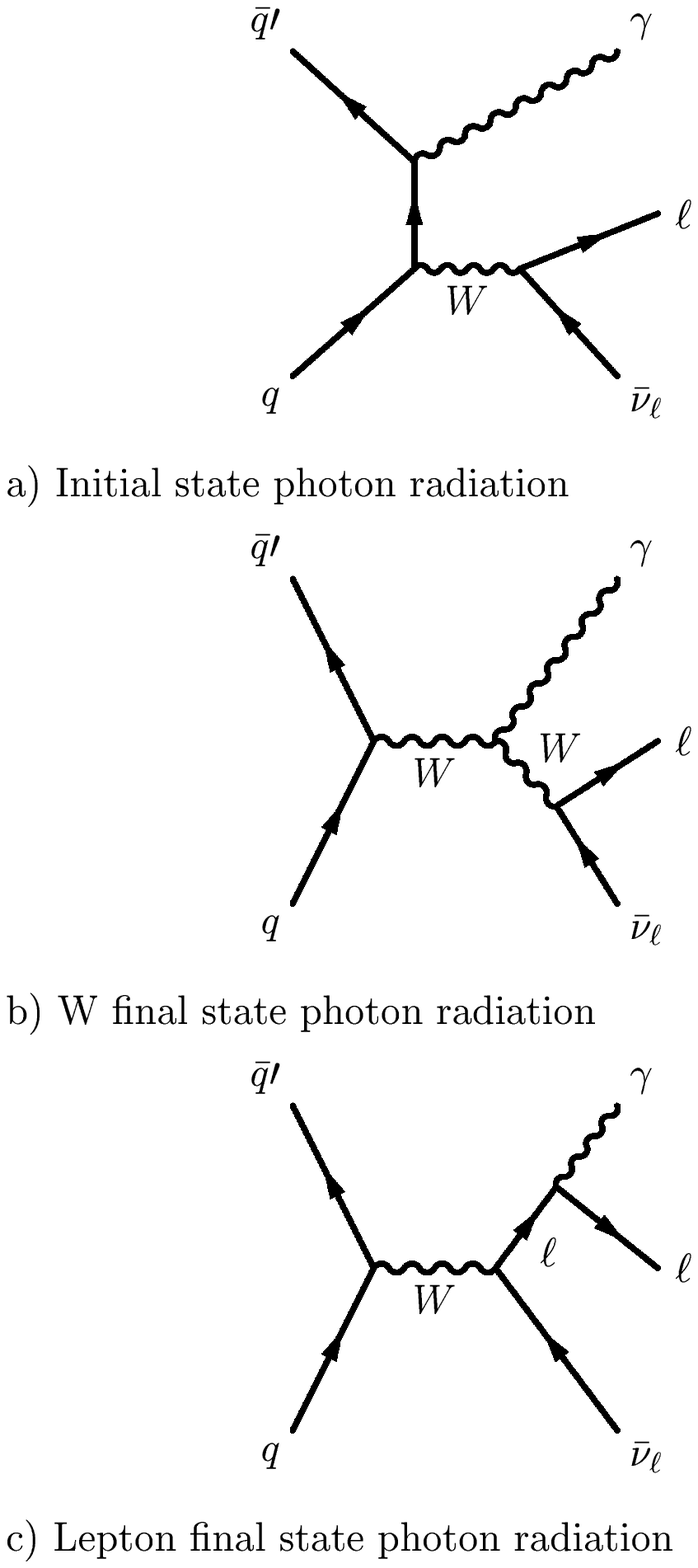}
\end{center}
\caption{The leading-order Feynman diagrams for photon radiation
in the process
$q\bar q\prime \rightarrow \ell \bar\nu_{\ell} \gamma$.}
\label{feynmf wg}
\end{figure}

\begin{figure}[!ht]
\begin{center}
\includegraphics*[width=0.4\textwidth]{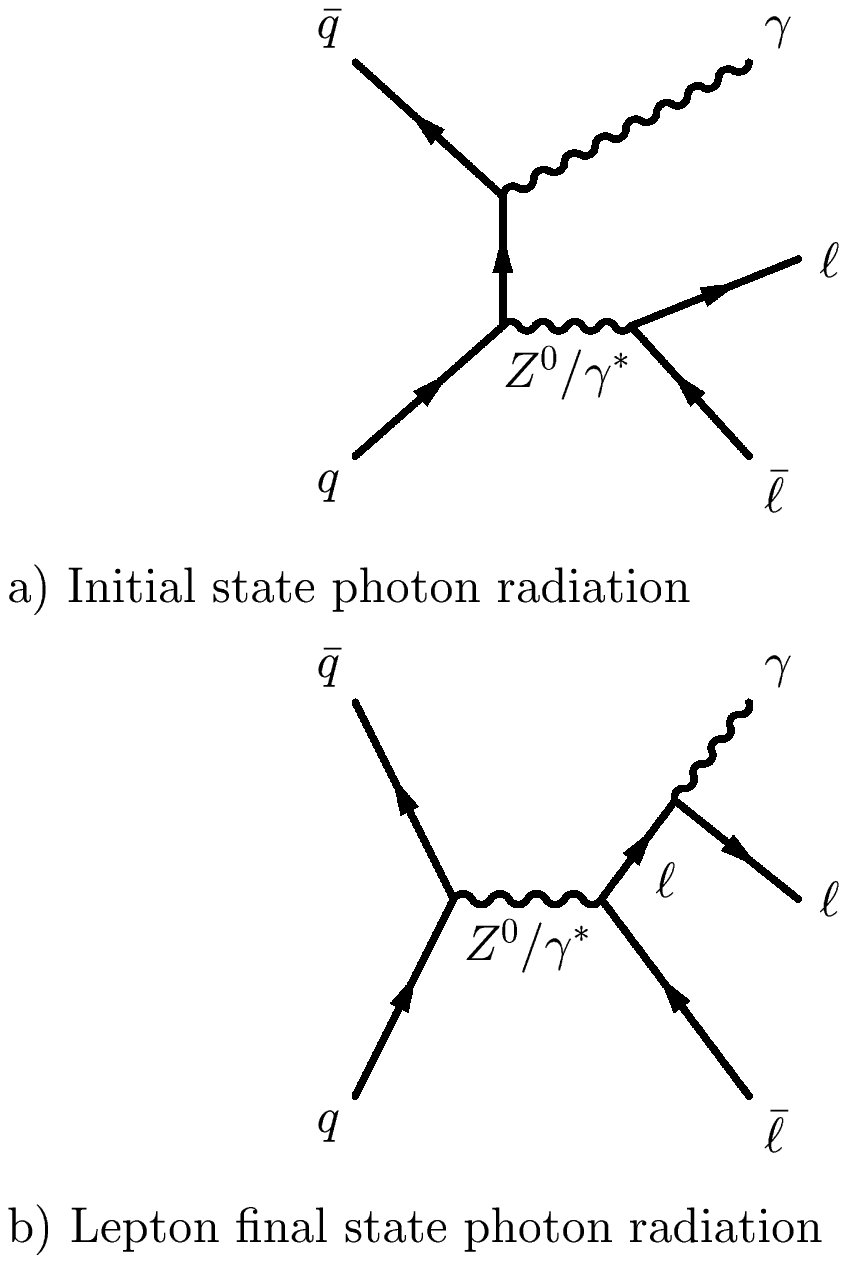}
\end{center}
\caption{The leading-order Feynman diagrams for photon radiation 
in the process
$q\bar q \rightarrow \ell \bar\ell \gamma$.}
\label{feynmf zg}
\end{figure}

The event generator program consists of a set of leading-order matrix
element calculations~\cite{diboson matrix} which was incorporated into
the general-purpose event generator program
\textsc{pythia}~\cite{PYTHIA}.  The matrix element calculation for
$W\gamma$ ($Z^{0}\gamma$) includes all tree-level diagrams with a
$q\bar q\prime$ ($q\bar q$) initial state and a $\ell \nu_{\ell}
\gamma$ ($\ell \ell \gamma$) final state, where $\ell$ is an $e$,
$\mu$, or $\tau$, and the mediating electroweak boson is a real or
virtual $W$ ($Z^0$ or $\gamma^*$).  Figure~\ref{feynmf wg} shows the
leading-order Feynman diagrams for $q\bar q\prime \rightarrow \ell
\nu_{\ell} \gamma$.  Figure~\ref{feynmf zg} shows the leading-order
Feynman diagrams for $q\bar q \rightarrow \ell \bar\ell \gamma$.

\linespace{1.0}
\linespace{2.0}

\linespace{1.0}
\linespace{2.0}

The region of phase space where the final state lepton and photon are
collinear is carefully sampled, taking into account the lepton mass
for each lepton flavor.  This allows reliable calculations to be made
for all photon-lepton separation angles and for photon $\Et$ well
below ($< 1 \GeV$) those considered in this analysis.
\textsc{Pythia} generates, fragments, and hadronizes the partons 
described by the matrix elements.  Event rates in $p\bar{p}$ collisions 
are obtained from the parton-level matrix elements through convolution with 
the leading-order proton structure function CTEQ5L~\cite{CTEQ5}.
The \textsc{tauola}~\cite{TAUOLA} program is used to
compute the decays of any $\tau$ leptons generated.  Each generated
event is assigned a weight proportional to the probability of its
occurrence as determined by the event rate calculation.

Generated events are used as input to a program which simulates the
CDF detector response to the final state particles.  The simulation
includes the following features relevant to this analysis: the
$z_{event}$ distribution of $p{\bar p}$ collisions observed in CDF
data, the geometric acceptance of all CDF detector subsystems, charged
tracks measured by the CTC, the tower-by-tower response of the
calorimeters to final state particles, the CES response to
electromagnetic showers, and the response of the central muon chambers
to penetrating charged particles.  The program is not used to simulate
the CDF trigger, the $z_{event}$ distribution beyond $|z_{event}| =
60$~cm, nor the energy-out-of-time distribution; the event selection
efficiencies for these must be applied as separate corrections to the
simulated event rates.  There also exist 6-8\% differences between the
lepton (and photon) detection efficiencies found in CDF data and the
efficiencies similarly computed in simulated data, as described in
Section~\ref{selection}.  Simulated event rates containing particles
of type $X$ are therefore adjusted by a ratio $C_X$ of detection
efficiencies in CDF data to that of simulated events,
\begin{eqnarray}
C_X & = & \epsilon^{data}_{X ID}/\epsilon^{sim}_{X ID},
\end{eqnarray} 
where $\epsilon^{data}_{X ID}$ is the detection efficiency of $X$ in
CDF data and $\epsilon^{sim}_{X ID}$ is the corresponding efficiency
in simulated data.  The detection efficiencies are obtained from
samples of $Z$ boson candidates decaying to pairs of leptons,
specifically those events which have one lepton candidate satisfying
the selection criteria of Tables~\ref{leptoncuts}~and~\ref{loose
lepton cuts}, a second lepton candidate satisfying the fiducial and
kinematic selection criteria from those Tables, and a dilepton mass
$M_{\ell\ell}$ within 10\GeVcsq of $M_Z$.  The efficiency is extracted
from that fraction of events where the second lepton satisfies all
selection criteria.  Photon identification efficiency is similarly
measured with electron pair data.  Particle identification
efficiencies in simulated data are obtained with the same procedure
using a sample of $Z$ boson events created by the \textsc{pythia}
event generator and a detector simulation.  The systematic uncertainty
of $C_X$ is estimated to be half of the difference between $C_X$ and
unity.  Table~\ref{cid table} lists the corrections for the various
types of leptons and photons analyzed.

Simulated events with PEM electrons are an exception to this
procedure, since the PEM shower shape quantity $\chi^2_{3\times3}$ is
not included in the detector simulation.  The PEM electron detection
efficiency for all the requirements in Table~\ref{loose lepton cuts},
except the $\chi^2_{3\times3}$ requirement, is measured and corrected
for in the same way as other leptons; the correction is listed in
Table~\ref{cid table}.  The efficiency of the $\chi^2_{3\times3}$
requirement for PEM electrons which satisfy all other requirements,
$\epsilon_{PEM \ \chi^2}$, is then measured separately using CDF data
to be $95.0\pm0.5\%$~\cite{jwb thesis}.  This is an additional
correction to the identification efficiency for simulated events with
PEM electrons.

\linespace{1.25}
\begin{table}[!ht]
\begin{center}
\begin{tabular}{lrrr}\hline\hline 
Particle & 
\multicolumn{1}{c}{$\epsilon^{data}_{X ID}$} & 
\multicolumn{1}{c}{$\epsilon^{sim}_{X ID}$} & 
\multicolumn{1}{c}{$C_{X ID}$} \\ 
\hline
CEM photon        & 0.86 & 0.93 & $0.93\pm0.04$ \\
CEM electron      & 0.81 & 0.88 & $0.92\pm0.04$ \\
2nd CEM electron  & 0.89 & 0.97 & $0.91\pm0.05$ \\
PEM electron      & 0.92 & 0.99 & $0.94\pm0.03$ \\
FEM electron      & 0.75 & 0.98 & $0.77\pm0.12$ \\
central muon      & 0.93 & 0.99 & $0.94\pm0.03$ \\
CMI muon          & 0.91 & 0.99 & $0.92\pm0.04$ \\
\hline\hline
\end{tabular}
\caption{Corrections to the simulated particle identification efficiencies 
obtained from CDF data~\cite{jwb thesis}.  Included are the
efficiencies measured directly from CDF data ($\epsilon^{data}_{X
ID}$), the efficiencies measured from simulated data
($\epsilon^{sim}_{X ID}$), and the corrections to simulated rates
($C_{X ID}$).}
\label{cid table}
\end{center}
\end{table}
\linespace{2.0}

The complete set of correction factors to the detection efficiencies
of simulated events, $C_{sim}$, is given by 
\begin{eqnarray}
C_{sim} & = &
\epsilon_{z60} \times \epsilon_{EOT} \times 
\prod_{X} C_{XID}^{N_X} \times 
\epsilon_{PEM \ \chi^2}^{N_{PEM}},
%\nonumber \\*
\label{eq:c sim}
\end{eqnarray}
where
\begin{itemize} 
\item{$\epsilon_{z60}$ is the efficiency for the requirement 
$|z_{event}| < 60$~cm, measured to be $0.95\pm0.02$.}
\item{$\epsilon_{EOT}$ is the efficiency for the requirement 
$\Et$(out-of-time) = 0, measured to be $0.975\pm0.004$~\cite{eegg}.}
\item{$\prod_{X} C_{XID}^{N_X}$
is the product, over each type $X$ of lepton or photon identified in
this analysis, of corrections $C_{XID}$ to the simulated particle
identification efficiencies listed in Table~\ref{cid table}.  Each
factor has an exponent $N_X$ equal to the number of particles of type
$X$ identified by the detector simulation.}
\item{and $\epsilon_{PEM \ \chi^2}^{N_{PEM}}$ is 
an additional correction factor for PEM electrons, measured to be
$0.953\pm0.005$ and described above, with an exponent $N_{PEM}$ equal
to the number of PEM electrons identified by the detector simulation.}
\end{itemize}

\linespace{1.25}
\begin{table}[!ht]
\begin{center}
\begin{tabular}{lcr}\hline\hline
Item & Value & Relative Uncertainty \\ \hline

$K_{NLO}$                       & $1.30\pm0.10$                & 7.7\%  \\  
$\sigma_{LO}$                   & $105.0\pm5.3$ pb             & 5.0\%  \\
$\sum w_{pass}/\sum w_{tot}$    & $(2.57\pm0.12)\times10^{-4}$ & 4.7\%  \\
$\epsilon_{trig}$               & $0.985\pm0.015$              & 1.5\%  \\
$C_{sim}$                       & $0.792\pm0.052$              & 6.6\%  \\
$\int L dt$                     & $86.34\pm3.52$ pb$^{-1}$     & 4.1\%  \\
\hline
$\bar N_{e\gamma}$              & $2.36\pm0.31$                & 13.1\% \\
\hline \hline
\end{tabular}
\caption{The mean number of multi-body photon-electron
events, $\bar N_{e\gamma}$, expected from $W(\rightarrow e\nu)+\gamma$.  
The factors used in Equation~\ref{eq:mc rate} and their uncertainties
are also shown.}
\label{wzg sample table}
\end{center}
\end{table}
\linespace{2.0}
     
The mean contribution to photon-lepton candidates in CDF data,
$\bar N_{\ell\gamma}$, for a particular generated process is given by
\begin{eqnarray}
\bar N_{\ell\gamma} & = &
\sigma_{LO}
\times K_{NLO}
\times \epsilon_{trig}
\times C_{sim} 
\times \int L dt \nonumber \\*
& & \times ( \sum w_{pass})/(\sum w_{tot}),
\label{eq:mc rate}
\end{eqnarray}
where
\begin{itemize}
\item{$\sigma_{LO}$ is the leading order cross section computed 
by the event generator for a given process with a given set of
generator-level selection requirements and thresholds.  The
uncertainty due to generator statistics is negligible.  The
uncertainty due to PDF normalization is taken to be $\pm5\%$, as
recommended in~\cite{MRST}.}
\item{$K_{NLO}$ is the next-to-leading order (NLO) QCD K-factor for
$W\gamma$ ($Z^0\gamma$) production estimated from 
NLO calculations~\cite{K factor}.  The K-factors used are $1.30\pm0.10$
for $W\gamma$ production and $1.25\pm0.05$ for $Z^0\gamma$ production, 
where the uncertainties are estimated from the QCD renormalization 
scale dependence of the NLO cross section.}  
\item{$\epsilon_{trig}$ is the trigger efficiency for photon-lepton
events. For photon-electron events $\epsilon_{trig} = 98.5\pm1.5\%$;
for photon-muon events $\epsilon_{trig}$ varies with muon type and
photon $\Et$, with an average efficiency of 94\% for simulated
$W\gamma$ events satisfying all selection criteria.  The uncertainty
of the photon-muon trigger efficiency is $\pm6\%$~\cite{jwb thesis}.}
\item{$C_{sim}$ is the product of the correction factors to the detection
efficiencies computed by the CDF detector simulation, as described above.}
\item{$\int L dt$ is the integrated luminosity for the 1994-5 run employed
in this analysis, $86.34\pm3.52$ pb$^{-1}$~\cite{lumint}.}
\item{$\sum w_{pass}$ is the sum of the weights of the simulated 
events satisfying all selection criteria.  The uncertainty is given
by $\sqrt{ \sum w_{pass}^2 }$, which is typically a few percent.}
\item{$\sum w_{tot}$ is the sum of the weights of all simulated events,
with uncertainty given by $\sqrt{ \sum w_{tot}^2 }$, which is typically 
negligible.}
\end{itemize}

Table~\ref{wzg sample table} shows a sample calculation for
multi-body photon-electron events originating from $W(\rightarrow
e\nu)+\gamma$ production.  The uncertainty in the mean rate has
roughly equal contributions from the NLO K-factor, simulation
systematics, luminosity, proton structure, and generator statistics.
Other simulated processes have similar uncertainties.

\linespace{1.25}
           
\begin{table*}[!ht]
\begin{center}
\begin{tabular}{lllllll}\hline \hline
 & \multicolumn{2}{c}{Two-Body Events} 
 & \multicolumn{2}{c}{Multi-Body Events} 
 & \multicolumn{2}{c}{Multi-Body Events} 
\\
Process 
    & \multicolumn{1}{c}{$e\gamma X$}   & \multicolumn{1}{c}{$\mu\gamma X$} 
    & \multicolumn{1}{c}{$e\gamma X$}   & \multicolumn{1}{c}{$\mu\gamma X$} 
    & \multicolumn{1}{c}{$e\gamma\mett X$}   & \multicolumn{1}{c}{$\mu\gamma\mett X$} 
\\
\hline
\multicolumn{7}{l} {$\gamma+W$ production}  \\
\hline 
$\gamma+W\rightarrow \ell\nu$      
    & $1.1\pml0.1$ & $1.4\pml 0.2$ 
    & $2.4\pml0.3$ & $2.5\pml0.3$ & $1.9\pml0.3$ & $1.9\pml0.3$ \\
$\gamma+W\rightarrow \tau\nu$      
    & $0.08\pm0.02$ & $0.09\pm0.02$
    & $0.08\pm0.02$ & $0.06\pm0.01$ & $0.04\pm0.01$ & $0.05\pm0.01$ \\ 
\hline
Subtotal 
    & $1.2\pml0.2$ & $1.5\pml0.2$ 
    & $2.4\pml0.3$ & $2.5\pml0.3$ & $1.9\pml0.3$ & $2.0\pml0.3$ \\ 
\hline 
\multicolumn{7}{l} {$\gamma+Z^{0}$ production}  \\
\hline 
$\gamma+Z^{0}\rightarrow \ell\ell$ 
    & $5.1\pml0.5$ & $6.5\pml0.8$ 
    & $4.9\pml0.5$ & $4.5\pml0.5$ & $0.3\pml0.1$ & $0.9\pml0.1$ \\
$\gamma+Z^{0}\rightarrow \tau\tau$ 
    & $0.3\pml0.1$ & $0.5\pml0.1$ 
    & $0.13\pm0.03$ & $0.10\pm0.02$ & $0.03\pm0.01$ & $0.05\pm0.01$ \\ 
\hline
Subtotal                           
    & $5.4\pml0.6$ & $7.1\pml0.8$
    & $5.0\pml0.5$ & $4.6\pml0.5$ & $0.3\pml0.1$ & $1.0\pml0.2$ \\ 
\hline
Total                              
    & $6.6\pml0.7$ & $8.6\pml1.0$
    & $7.5\pml0.8$ & $7.1\pml0.8$ & $2.3\pml0.3$ & $3.0\pml0.4$   \\  
\hline\hline
\end{tabular}
\caption{The estimated $W\gamma$ and $Z^{0}\gamma$ backgrounds for
two-body photon-lepton events, inclusive multi-body photon-lepton
events, and multi-body $\ell\gamma\mett$ events.  There exist
correlated uncertainties between the different photon-lepton sources.
The symbol $X$ denotes the allowed inclusion of any other
combination of particles, except where explicitly prohibited.}
\label{wzg table}
\end{center}
\end{table*}
\linespace{2.0}

Table~\ref{wzg table} shows the results of all simulated processes,
for inclusive two-body events, inclusive multi-body events, and
multi-body $\ell\gamma\mett$ events.  The slightly larger contribution
of two-body $\mu\gamma$ events relative to $e\gamma$ events is due to
the explicit exclusion of $e\gamma$ events whose invariant mass is
``$Z^0$-like'' ($86 \GeVcsq < M_{e\gamma} < 96 \GeVcsq$).  There are
no significant differences between the inclusive multi-body rates for
$e\gamma$ and $\mu\gamma$ production. In the case of $Z^0\gamma$
production, there is a larger number of multi-body $\mu\gamma\mett$
events (1.0) relative to $e\gamma\mett$ events (0.3).  The difference
is due to events where the second muon falls outside the solid angle
in which muons can be detected ($|\eta_{\mu}| \gt 1.2$), subsequently
inducing missing $\Et$ equal to the $\Pt$ of the second muon.  Leptons
from $\tau$ decays contribute to the total photon-lepton rate at a
level far below the leptonic branching ratio of a $\tau$ (about 3\%
accepted compared to a leptonic branching ratio of 18\%) because the
average lepton $\Et$ is much lower than that of leptons from the
direct decay of a $W$ or $Z^{0}$.

Table~\ref{wzg table2} shows the results
of all simulated processes for multi-body photon-lepton events with
additional leptons or photons, respectively.  More
$ee\gamma$ events than $\mu\mu\gamma$ events are expected due to the
larger detector acceptance for additional electrons, which are
identified in the central, plug, and forward calorimeters.

\linespace{1.5}
           
\begin{table*}[!ht]
\begin{center}
\begin{tabular}{lccccc}\hline \hline
  &  \multicolumn{5}{c}{Multi-Body Events} \\
Process & $ee\gamma$ & $\mu\mu\gamma$ & $e\mu\gamma$  & $e\gamma\gamma$ & $\mu\gamma\gamma$ \\
\hline
\multicolumn{6}{l} {$\gamma+W$ production}  \\
\hline 
$\gamma+W\rightarrow \ell\nu$ & 
       ---  & ---            & ---           & ---             & --- \\
$\gamma+W\rightarrow \tau\nu$ & 
       ---           & ---            & ---           & ---             & --- \\ 
\hline
Subtotal & 
       --- & ---            & ---           & ---             & --- \\ 
\hline 
\multicolumn{6}{l} {$\gamma+Z^{0}$ production}  \\
\hline 
$\gamma+Z^{0}\rightarrow \ell\ell$ & 
       $3.3\pm0.4$ & $2.2\pm0.3$  & ---           & $0.012\pm0.012$   & $0.004\pm0.004$ \\
$\gamma+Z^{0}\rightarrow \tau\tau$ & 
       ---           & ---            & $0.05\pm0.01$ & ---             & --- \\ 
\hline
Subtotal & 
       $3.3\pm0.4$ & $2.2\pm0.3$  & $0.05\pm0.01$ & $0.012\pm0.012$   & $0.004\pm0.004$ \\ 
\hline
Total & 
       $3.3\pm0.4$ & $2.2\pm0.3$  & $0.05\pm0.01$ & $0.012\pm0.012$   & $0.004\pm0.004$ \\  
\hline\hline
\end{tabular}
\caption{The estimated $W\gamma$ and $Z^{0}\gamma$ backgrounds for multi-body
photon-lepton samples with additional leptons and photons.}
\label{wzg table2}
\end{center}
\end{table*}
\linespace{2.0}

\subsection{Jets Misidentified as Photons}
\label{fakephoton}

A jet of hadrons initiated by a final state quark or gluon can contain
mesons which decay to photons, such as the $\pi^0$, $\eta$, or
$\omega$. If one or more of these photons constitute a sufficiently
large fraction of the jet momentum, then the hadron jet can be
misidentified by the CDF detector as a single prompt photon.  Such a
jet, when produced in association with a lepton candidate, contributes to the
detected photon-lepton candidates.

The contribution of lepton plus misidentified jet events is determined
by counting the number of jets in CDF lepton data, $N_{\ell jet}$, and
then multiplying that number by an estimate of the probability of a
jet being misidentified as a photon, $P^{jet}_{\gamma}$, to obtain the
number of photon-lepton candidates,
\begin{equation}
N_{\ell\gamma} = N_{\ell jet} \times P^{jet}_{\gamma}.  
\end{equation}

Lepton-jet candidates are selected from inclusive electron and muon
triggers as follows.  The Level 1 trigger and Level 3 trigger
requirements are identical to those enumerated in Sections~\ref{muon
id}~and~\ref{electron id} of Section 3.  The Level 2 trigger
requirements differ from those of the photon-lepton sample due to the
absence of the photon.  Electron-jet events must be accepted by a
Level 2 electron trigger, which requires a CEM energy cluster with
$\Et \gt 16 \GeV$; the ratio $E_{HAD}/E_{EM}$ for that cluster $\lt
0.125$; and a CFT track matching the CEM cluster with $\Pt \gt
12\GeVc$.  The efficiency of these electron trigger requirements has
been measured to be $\epsilon_e = 90.9\pm0.3\%$~\cite{r papers}.
Muon-jet events are selected from the Level 2 inclusive muon triggers,
which have the same efficiency as the muon triggers described in
Section~\ref{muon id}, except that they are prescaled due to bandwidth
limitations.  The prescaling results in a reduction of the trigger
efficiency by a factor of $0.43\pm0.02$ for CMX muons, $0.43\pm0.02$
for CMNP muons, and $1.0$ (no prescale) for CMUP muons.  Requiring a
Level 2 muon trigger precludes the use of CMP or CMU muons.

The requirements for lepton-jet candidates are as follows: one or more
lepton candidates satisfying the criteria in Table~\ref{leptoncuts};
and one or more jets with $|\eta_j|~<~1.0$, jet $\Et~>~25\GeV$, and a
separation distance of the jet from the lepton in $\eta-\varphi$
space, $\Delta R_{\ell j}$, greater than 0.5.  As a further step to
prevent electrons from $Z^{0}$ boson decays being counted as jets, jet
candidates must have electron-jet separation $\Delta R_{ej}~>0.5$ for
all central electrons satisfying the selection criteria for additional
electrons listed in Table~\ref{loose lepton cuts}.  Table~\ref{ljet
table} shows the raw total number of jets, summed over all lepton-jet
candidate events, for the various signal regions of this analysis.

Because the lepton trigger requirements of the lepton-jet sample are
less efficient than the trigger requirements of the photon-lepton
sample, the effective number of jets which potentially contribute to
the photon-lepton candidates must be augmented by a ratio of the
efficiencies of the different trigger paths.  For electron-jet 
events with exactly one electron,
this is simply a constant, $\epsilon_{e\gamma}/\epsilon_e =
1.08\pm0.02$; for muon-jet events with exactly one muon, 
the efficiency ratio, $R_{\mu_i \gamma}$, varies with muon stub 
type and jet $\Et$,
\begin{equation}
R_{\mu_i \gamma} = \frac{\epsilon_{\mu_i} + 
                (1-\epsilon_{\mu_i})\times\epsilon_{\gamma}(\Et)}
                {P_{\mu_i}\epsilon_{\mu_i}}
\label{eq:R mu}
\end{equation}
where $\epsilon_{\mu_i}$ is the trigger efficiency for muons of stub
type $i$, $P_{\mu_i}$ is the inclusive muon trigger prescale factor 
for muons of stub type $i$, and $\epsilon_{\gamma}(\Et)$ is the trigger
efficiency of the photon candidate a jet would produce in the event of
jet misidentification, as a function of photon $\Et$.  This
ratio is evaluated for each jet in each event, and the sum over all
jets in all events gives the total effective number of jets.  Because
CMU and CMP muons have been excluded from the lepton-jet sample, the
number of jets in muon-jet events must be additionally multiplied by
a factor of $1.14\pm0.03$ to compensate for the acceptance lost
relative to that of photon-lepton events.  This lost acceptance was
calculated from the $W\gamma$ and $Z\gamma$ simulation described in
Section~\ref{diboson}.

For lepton-jet events with multiple leptons, the presence of the
additional lepton increases the efficiency of the lepton trigger
requirements, and the efficiency ratio of such events relative to the
corresponding photon-lepton events must be accounted for separately.
For electron-jet events with an additional CEM electron, the trigger
efficiency for both electron-jet and photon-electron events is
nearly 100\%, so that the trigger efficiency ratio of such events is
assumed to be unity.  Electron-jet events with additional PEM or FEM
electrons have the same efficiency ratio as that of single
electron-jet events above.  For muon-jet events with an additional
CMNP, CMUP, or CMX muon, the trigger efficiency ratio depends upon the
muon trigger efficiencies of the two muon stub types:
\begin{eqnarray}
R_{\mu_i\mu_j\gamma} & = & \frac{\epsilon_{\mu_i\mu_j} + 
(1-\epsilon_{\mu_i\mu_j})\times\epsilon_{\gamma}(\Et)}  
{P_{\mu_i}\epsilon_{\mu_i} + (1-P_{\mu_i}\epsilon_{\mu_i})\times
P_{\mu_j}\epsilon_{\mu_j}},
\label{eq:R mumu}
\end{eqnarray}
where $\epsilon_{\mu_i}$ and $\epsilon_{\mu_j}$ are the muon trigger
efficiencies of the two different muon stub types, $P_{\mu_i}$ and
$P_{\mu_j}$ are the inclusive muon trigger prescales of the two
different muon stub types, and $\epsilon_{\mu_i\mu_j}$ is the
efficiency of the logical OR of the two muon triggers,
\begin{eqnarray}
\epsilon_{\mu_i\mu_j} & \equiv & \epsilon_{\mu_i} + 
(1-\epsilon_{\mu_i})\times\epsilon_{\mu_j}.  
\label{eq:epsilon mumu}
\end{eqnarray}
Muon-jet events with additional CMU, CMP, or CMI muons have the 
same efficiency ratio as that of single muon-jet events above. 
 
The total effective number of jets in lepton-jet candidate events
after all corrections have been applied is also given in
Table~\ref{ljet table}.  There are more electron-jet candidates than
muon-jet candidates because the angular coverage of the CEM is larger
than that of the central muon chambers, particularly at higher lepton
$|\eta|$.  Comparing Table~\ref{ljet table} with Tables~\ref{wzg
table} and~\ref{wzg table2}, it is concluded that in order to measure
photon-lepton processes with electroweak-sized cross sections and a
signal-to-background ratio greater than 1:1, $P^{jet}_{\gamma}$ must
be less than approximately $10^{-3}$.

Mesons which decay to photons are typically only a portion of a shower
of hadrons initiated by a high $\Et$ quark or gluon.  Other hadrons in
the shower will deposit energy in the calorimeter close to the
electromagnetic shower produced by these photons.  Prompt photons (or
electrons, which shower similarly) produced in the hard scattering of
partons do not exhibit additional nearby energy in the calorimeters;
the additional $\Et$ measured in a cone of $R = 0.4$ around
the electromagnetic shower position, $\coriso$, therefore serves as a
discriminant between prompt photons and misidentified jets.  This
discriminant is already employed in the photon selection
(Table~\ref{photoncuts}), by requiring $\coriso~<~2\GeV$.  If the
distribution of $\coriso$ is relatively flat for misidentified jets,
the distribution of $\coriso$ of the photon candidates which fail this
requirement can be extrapolated linearly to estimate the number of
misidentified jets which satisfy it.

The probability that a jet is misidentified as a photon is determined
from samples of jets and photons in events with a lepton trigger.  
Lepton candidates in lepton-triggered jet
events are selected with the same trigger requirements as the
lepton-jet sample described above.  Instead of applying the full
lepton selection criteria in Table~\ref{leptoncuts}, the minimal set
of Level 3 lepton trigger requirements, listed in Table~\ref{inclusive
lepton cuts}, is applied in this selection, so as to maximize the
sample size.  Along with exactly one such loose lepton candidate,
lepton-triggered jet events are required to have exactly one jet with
$|\eta_j|~<~1.0$, $\Et~>~25\GeV$, and $\Delta R_{\ell j} \gt 0.5$.
The lepton-triggered jet sample consists of 46091 electron-triggered
jet events and 12875 muon-triggered jet events.

Lepton candidates in lepton-triggered photon events are selected with
the same trigger requirements as the lepton-triggered jet events
described above, except that the prescaled Level 2 inclusive muon
trigger requirements are replaced by the muon-jet trigger 
described in Section~\ref{muon id}.  Lepton-triggered photon events
are required to have exactly one loose lepton candidate as above, and
are required to have exactly one photon candidate satisfying all of
the photon selection criteria in Table~\ref{photoncuts}, except for
the isolation requirements.  Specifically, the requirement that the
sum of the $\Pt$ of all tracks in a cone of $R=0.4$ around the photon
be less than 5\GeVc is rescinded, and the $\coriso$ requirement is
loosened from 2\GeV to 12\GeV.  The lepton-triggered photon sample
consists of 121 photon-electron and 38 photon-muon events.

Since the muon-triggered jet sample has a less efficient trigger
path than the muon-triggered photon sample, an unbiased comparison
of the two samples requires that the number of muon-triggered jet
events must be augmented on an event-by-event basis
by the ratio of trigger efficiencies of the two samples.  
The ratio for each event in this case is simply the 
inverse of the Level 2 muon trigger 
prescale factor for the stub type of the muon, $1/P_{\mu_i}$.  
The effective number of muon-triggered jet events increases 
from 12875 to 17745.

Photon candidates in the lepton-triggered photon sample consist of a
combination of prompt photons, electrons misidentified as photons, and
jets misidentified as photons, where only the jet component is
relevant to the evaluation of $P^{jet}_{\gamma}$.  The distribution of
$\coriso$ of the other two components is measured using a sample of
CEM electrons from $Z^{0}$ decays.  Dielectron events are selected
from events satisfying the same trigger criteria as that of the
photon-electron candidates described in Section~\ref{electron id}.
From these triggers, $Z^{0}$-like dielectron events are
selected which have exactly two CEM electrons passing the electron
criteria in Table~\ref{leptoncuts}, excepting the isolation
requirement (that the total $\Et$ deposited in the calorimeters, in a
cone of $R=0.4$ around the electron track, be less than 10\% of the
electron $\Et$), and which have dielectron invariant mass within 5\GeV
of $M_Z$.  The distribution of $\coriso$ normalized to unity,
$dN_Z/d\coriso$, for the 3300 electrons in this sample is shown in
Figure~6.  CEM electron showers---which have the same
calorimeter response as CEM showers from prompt photons---exhibit
$\coriso \lt 2\GeV$ 95\% of the time.

\linespace{1.0}
\begin{figure}[!ht]
\begin{center}
\includegraphics*[width=0.75\textwidth]{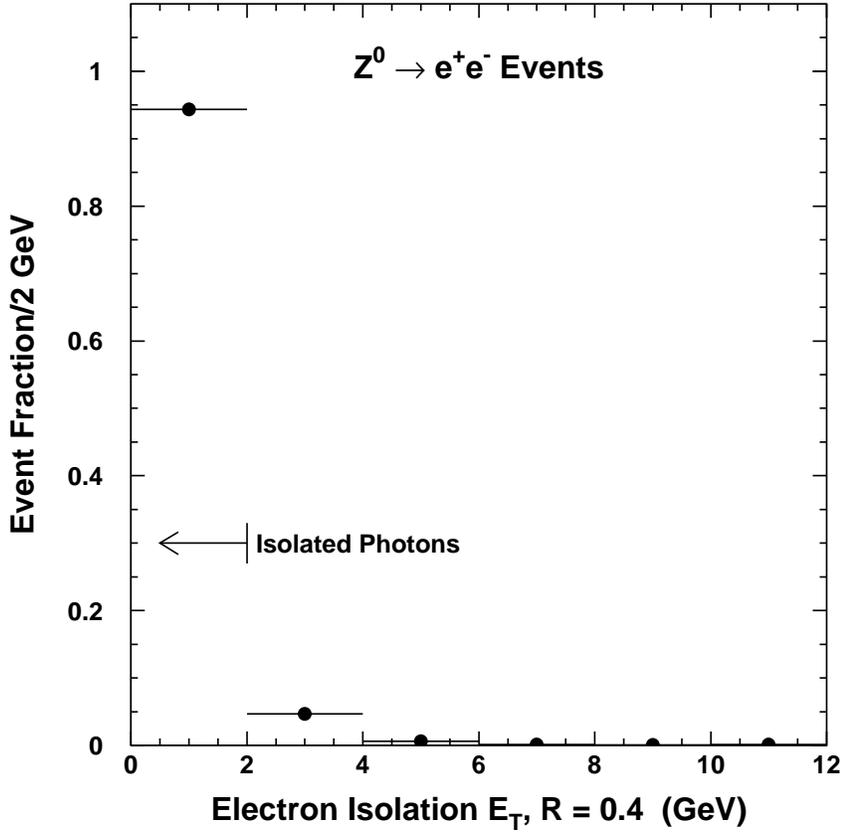}
\end{center}
\label{zeeiso}
\caption{The distribution of $\coriso$ for CEM electrons from $Z^{0}$ decays, 
normalized to unity.}
\end{figure}
\linespace{2.0}

Using the measured distribution $dN_Z/d\coriso$ for prompt photons or
electrons, and assuming a linear distribution in $\coriso$ for jets
misidentified as photons, the total number of photon candidates as a
function of $\coriso$, $dN/d\coriso$, is given by
\begin{eqnarray}
\frac{dN}{d\coriso} & = & A_1 \times \frac{dN_Z}{d\coriso} 
+ A_2 + A_3 \times \coriso, 
\label{eq:coriso}
\end{eqnarray}
where $A_1$, $A_2$, and $A_3$ are free parameters to be fit to the
data.  If the bin size is chosen to be equal to the $\coriso$
threshold for isolated photon candidates (2\GeV), then 
the number of prompt photon (or electron misidentified as photon)
candidates with $\coriso < 2\GeV$ is given by 
\begin{eqnarray}
A_1 \times \frac{dN_Z}{d\coriso}\Bigg\vert_{bin \ 1}  & = & A_1 \times 0.95, 
\label{eq:a1}
\end{eqnarray}
and the number of jets misidentified as photons with $\coriso < 2\GeV$
is given by 
\begin{eqnarray}
A_2 + A_3 \times \coriso \Bigg\vert_{bin \ 1}  & = & A_2 + A_3 \times 1 \GeV. 
\label{eq:a2pla3}
\end{eqnarray}
If in addition the normalization of the distribution is chosen to be
the ratio of the number of lepton-triggered photon events (121
photon-electron and 38 photon-muon) to that of the effective number of
lepton-triggered jet events (46091 electron-jet and 17745 muon-jet),
then $A_2 + A_3 \times 1 \GeV$ is identically the jet
misidentification rate $P^{jet}_{\gamma}$.

Employing these conventions, the distribution $dN/d\coriso$ for
lepton-triggered photon events is shown in Figure~7.  
The distribution (solid points) is peaked in the first bin
corresponding to isolated photon candidates, followed by a linearly
falling tail of non-isolated photon candidates.  The minimum $\chi^2$
fit of the data to the functional form of Equation~\ref{eq:coriso}
(solid line) is shown in Figure~7, along with the
linear portion of the fit obtained from $A_2$ and $A_3$ (dashed line).
The functional form chosen describes the data well ($\chi^2$/d.o.f. =
0.38), yielding an average jet misidentification rate
$P^{jet}_{\gamma}$ of $3.8\pm0.7\times10^{-4}$.  The best fit
parameters are shown in Table~\ref{fakerate table}.

Also shown in Figure~7 is an estimate of
$dN/d\coriso$ obtained from a simulation of $W$-jet production
(cross-hatched histogram), using the \textsc{pythia} event generator
and the detector simulation described in Section~\ref{diboson}.  The
leading-order Feynman diagrams for $W$-jet production employed by the
\textsc{pythia} event generator are shown in Figure~8.  
Simulated events are selected which satisfy the same requirements as
the lepton-triggered jet and lepton-triggered photon samples obtained
from the data, and photon candidates are required to arise solely from
hadron decay.  The simulated results for $dN/d\coriso$ exhibit a shape
consistent with a linear functional form, as well as a magnitude
consistent with the observed jet misidentification rate.
 
\linespace{1.0}
\begin{figure}[!ht]
\begin{center}
\includegraphics*[width=0.75\textwidth]{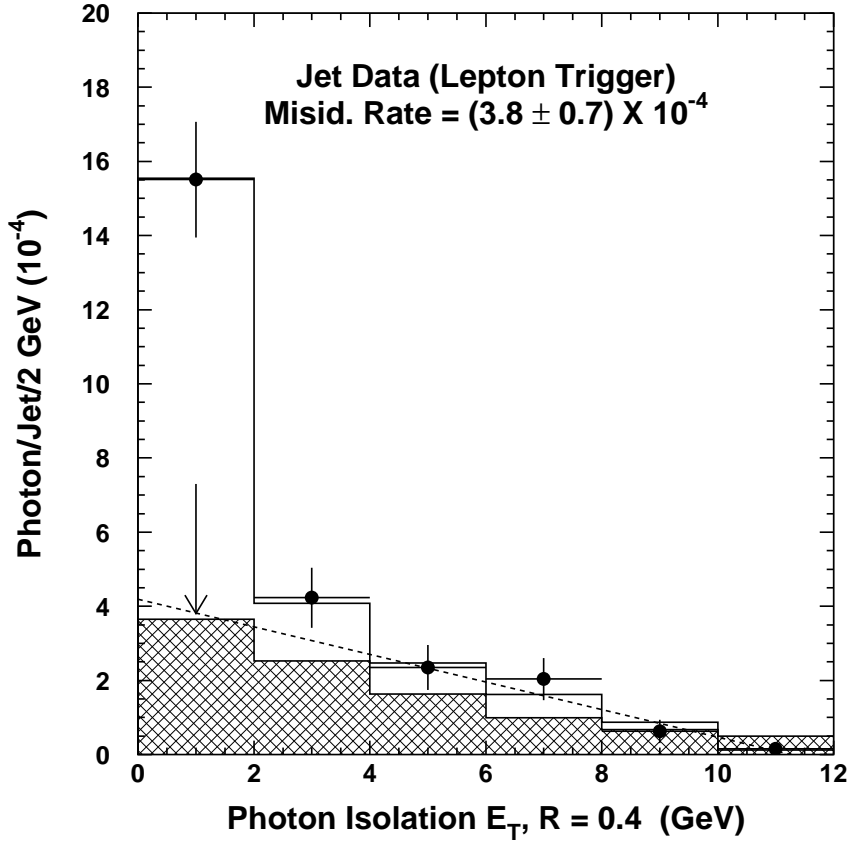}
\end{center}
\label{wjet_niso_eormu}
\caption{The number of photon candidates per jet, as a function
of $\coriso$, for CDF jet data obtained with a lepton trigger. 
Included are the results of CDF data (points), the fit of CDF data
to Equation~\ref{eq:coriso} (solid line), the linear portion of the
same fit (dotted line), an estimate of this distribution from
a simulation of $W$ plus jet events performed by \textsc{pythia}
(cross-hatched histogram), 
and an arrow indicating the value of $P^{jet}_{\gamma}$.}
\end{figure}
\linespace{2.0}

\linespace{1.25}
\begin{table}[!ht]
\begin{center}
\begin{tabular}{lrrr}\hline\hline
 & \multicolumn{3}{c}{Lepton-Jet Samples} \\   
 &  $ej$ & $\mu j$ & $\ell j$ \\   
\hline
Photons                      & 121            & 38          & 159           \\
Jets                         & 46091          & 17745       & 63836         \\
$A_1 (10^{-4})$ &$13\ \ \,\pm2\ \ \,$&$14\ \ \,\pm4\ \ \,$&$13\ \ \,\pm2\ \ \,$\\
$A_2 (10^{-4})$              & $4.7\pm0.9$    & $2.4\pm1.5$ & $4.2\pm0.7$   \\
$A_3 (10^{-4}/\nsGeV)$       & $-0.4\pm0.1$ & $-0.2\pm0.2$  & $-0.4\pm0.1$  \\
$P^{jet}_{\gamma} (10^{-4})$ & $4.3\pm1.0$    & $2.2\pm1.5$ & $3.8\pm0.7$   \\
$\chi^2$/d.o.f.              & $0.38$         & $0.44$      & $0.42$        \\
\hline\hline
\end{tabular}
\caption{The results of fitting $dN/d\coriso$ to photon candidates in CDF jet
data obtained with a lepton trigger.
Included are the number of photons and jets in each sample, the best fit 
parameters $A_i$, the $\chi^2$ per degree of freedom for the fit, and the
jet misidentification rate $P^{jet}_{\gamma}$.}
\label{fakerate table}
\end{center}
\end{table}
\linespace{2.0}

\linespace{1.0}
\begin{figure}[!ht]
\begin{center}
\includegraphics*[width=0.4\textwidth]{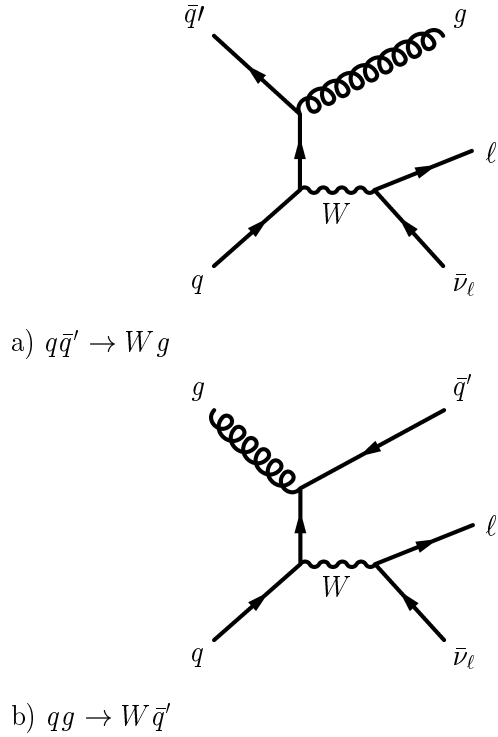}
\end{center}
\caption{The leading-order Feynman diagrams for $W$-jet production.}
\label{feynmf wjet}
\end{figure}
\linespace{2.0}

Figure~9 shows the distribution $dN/d\coriso$
computed for electron-triggered photon events and muon-triggered photon 
events separately.  The separate jet misidentification rates obtained from
these distributions, also shown in Table~\ref{fakerate table}, are
statistically consistent with each other.  

\linespace{1.0}
\begin{figure}[!ht]
\begin{center}
\includegraphics*[width=0.75\textwidth]{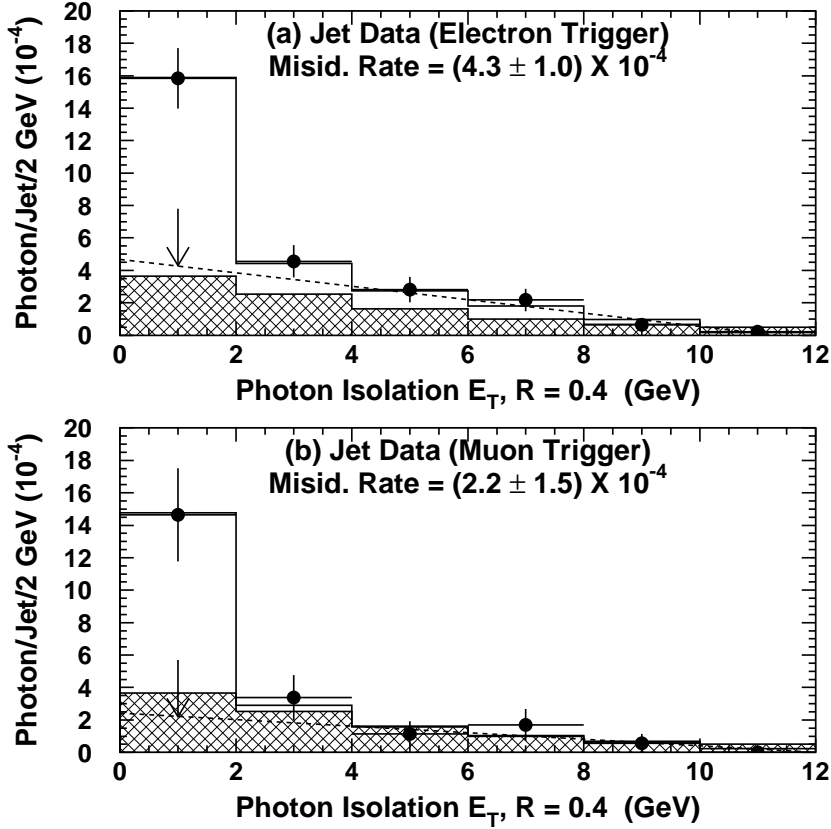}
\end{center}
\label{wjet_niso_eandmu}
\caption{The number of photon candidates per jet, as a function of $\coriso$,
for CDF jet data obtained with (a) an electron trigger or (b) a muon
trigger.  Included are the results of CDF data (points), the fit of
CDF data to Equation~\ref{eq:coriso} (solid line), the linear portion of the
same fit (dotted line), an estimate of this distribution from
a simulation of $W$ plus jet events performed by \textsc{pythia}
(cross-hatched histogram), 
and an arrow indicating the value of $P^{jet}_{\gamma}$.}
\end{figure}
\linespace{2.0}

Additional evidence for the linear behavior of $dN/d\coriso$ in
misidentified jets is obtained from a sample of lepton-triggered
events enriched with $\pi^0$'s.  Lepton candidates in these
lepton-triggered $\pi^0$ events are selected with the same trigger
requirements as the lepton-triggered photon events described above.
Lepton-triggered $\pi^0$ events are required to have exactly one loose
lepton candidate as above, and are required to have exactly one
$\pi^0$ candidate which satisfies requirements similar to photon
candidates in Table~\ref{photoncuts}, with the following differences:
the isolation requirements are not applied, as done for the
lepton-triggered photon sample; the requirements for additional CES
energy clusters are not applied; and the $\chi^2_{avg}$ is required to
be \textit{greater} than 20.  The lepton-triggered $\pi^0$ sample
consists of 38 electron-$\pi^0$ and 11 muon-$\pi^0$ events.

The distribution $dN/d\coriso$ for lepton-triggered $\pi^0$ events is
shown in Figure~10.  The distribution (solid
points) is consistent with that of a linearly decreasing tail.  Also
shown in Figure~10 is an estimate of $dN/d\coriso$
obtained from a simulation of $W$-jet production (cross-hatched
histogram) as described above, except with the lepton-triggered
$\pi^0$ selection applied.  As with lepton-triggered photons, the
simulated results for $dN/d\coriso$ exhibit a shape consistent with a
linear functional form, as well as a magnitude consistent with the
observed $\pi^0$ rate.
  
\linespace{1.0}
\begin{figure}[!ht]
\begin{center}
\includegraphics*[width=0.75\textwidth]{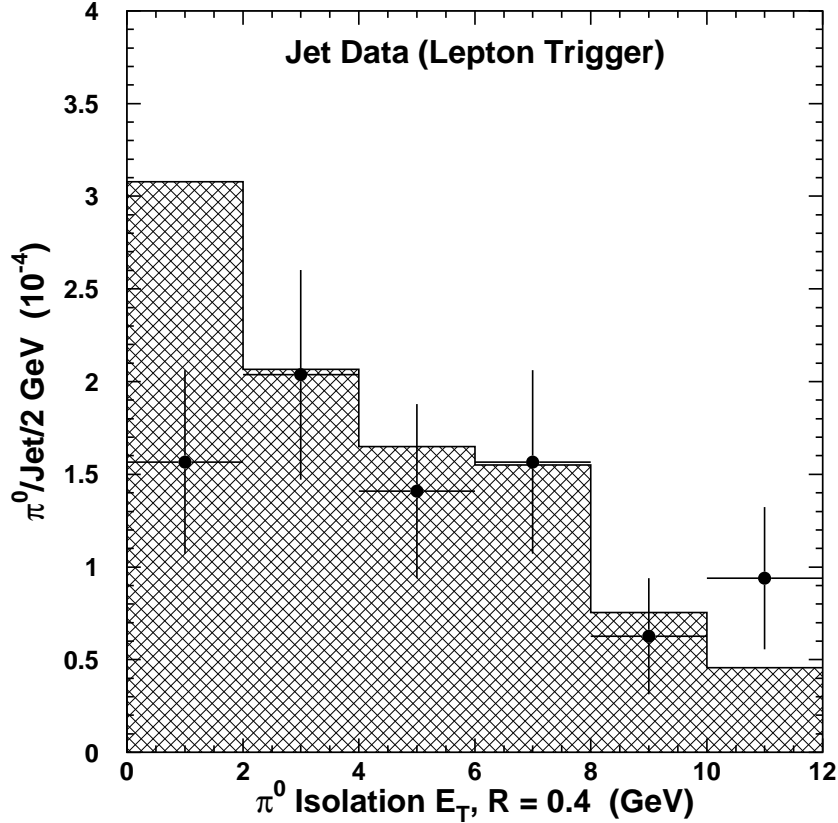}
\end{center}
\label{wjet_nchi_eormu}
\caption{The number of $\pi^{0}$ candidates per jet, as a function of
$\coriso$, for CDF jet data obtained with a lepton trigger.  Included
are the results of CDF data (points) and an estimate of this distribution 
from a simulation of $W$ plus jet events performed by
\textsc{pythia} (cross-hatched histogram).}
\end{figure}
\linespace{2.0}
  
Table~\ref{ljet table} shows the mean number of photon-lepton events
expected to originate from misidentified jets, for the various subsets
of photon-lepton events to be analyzed.  The uncertainties in these
estimates are dominated by the uncertainty in $P^{jet}_{\gamma}$,
which in turn is limited in precision by the number of exclusive
photon-lepton events.  The total number of two-body and
multi-body events expected is 1-2 events per category per lepton species, 
with roughly equal contributions in photon-electron and photon-muon
events. The number of multi-lepton events arising from misidentified jets
is an order of magnitude smaller. 
The number of $e\mu\gamma$, $e\gamma\gamma$,
and $\mu\gamma\gamma$ events arising from misidentified jets is
negligible, due to the small number of jets in $e\mu$, $e\gamma$, and
$\mu\gamma$ events, respectively.

\linespace{1.25}
\begin{table}[!ht]
\begin{center}
\begin{tabular}{lrrl}\hline\hline
  & 
\multicolumn{1}{c}{$N_{raw}$} & 
\multicolumn{1}{c}{$N_{\ell jet}$} & 
\multicolumn{1}{c}{$N_{\ell\gamma}$} \\ 
\hline
\multicolumn{4}{c}{Two-Body Events}\\ 
\hline
$e\gamma X$         & 4530 & 4909 & $1.9\pml0.3$ \\
$\mu\gamma X$       & 1983 & 3844 & $1.5\pml0.3$ \\
\hline
\multicolumn{4}{c}{Multi-Body Events}\\ 
\hline
$e\gamma X$         & 4235 & 4565 & $1.7\pml0.3$ \\
$\mu\gamma X$       & 2024 & 3855 & $1.5\pml0.3$ \\
$e\gamma\mett X$    & 2584 & 2798 & $1.1\pml0.2$ \\
$\mu\gamma\mett X$  & 1369 & 2633 & $1.0\pml0.2$ \\
$ee\gamma X$        &  479 &  496 & $0.19\pm0.03$ \\
$\mu\mu\gamma X$    &  226 &  346 & $0.13\pm0.02$ \\
$e\mu\gamma X$      &   16 &   19 &   \multicolumn{1}{c}{---} \\
$e\gamma\gamma X$   &    3 &    3 &   \multicolumn{1}{c}{---} \\
$\mu\gamma\gamma X$ &    3 &    4 &   \multicolumn{1}{c}{---} \\
\hline\hline
\end{tabular}
\caption
{The contributions $N_{\ell\gamma}$ to the various categories of
photon-lepton candidates from jets misidentified as photons, using the
measured jet misidentification rate $3.8\pm0.7\times10^{-4}$.
Included are the raw number $N_{raw}$ of jets in inclusive lepton data
and the effective number of jets $N_{\ell jet}$ which potentially
contribute to each category.}
\label{ljet table}
\end{center}
\end{table}
\linespace{2.0}
\subsection{Electrons Misidentified as Photons}
\label{zee}
The dominant source of misidentifed particles in photon-electron
events is $Z^{0}\rightarrow e^+e^-$ production, wherein one of the
electrons undergoes hard photon bremsstrahlung in the detector
material, or the CTC fails to detect one of the electron tracks, and
that electron is subsequently misidentified as a prompt photon.  There
are approximately 1000 central electron pairs in the CDF data, so a
electron misidentification rate as low as $1\%$ will give rise to 20
photon-electron events, which would be unacceptably high for finding
sources of new physics comparable to $W/Z^{0}+\gamma$ production (see
Tables~\ref{wzg table} and \ref{wzg table2}).  It is therefore
necessary to either obtain independently the electron
misidentification rate to sufficient accuracy that a background
subtraction can be performed, or to assume that those photon-electron
events in the CDF data which are sufficiently similar in their
kinematics to $Z^{0}$ production are not a significant source of new
physics, and that such events may be used to estimate misidentified
photon-electron events elsewhere.  The latter method is employed in
what follows.

\begin{figure}[!ht]
\begin{center}
\includegraphics*[width=0.75\textwidth]{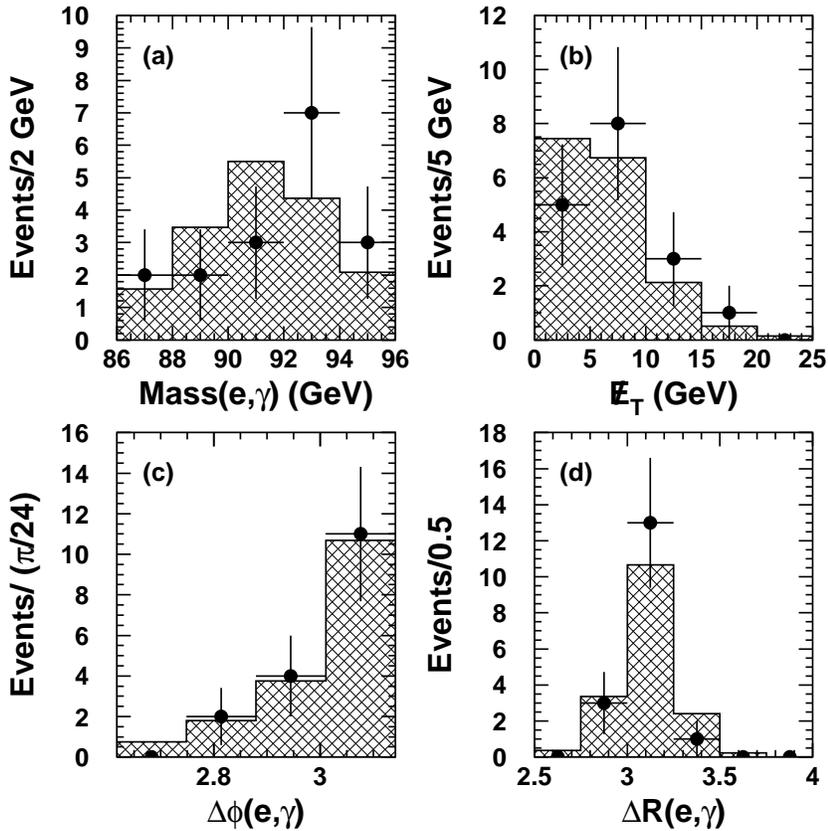}
\end{center}
\caption{The distributions for 
(a) $M_{e\gamma}$, (b) $\mett$,
(c) $\Delta\varphi_{e\gamma}$, and (d) $\Delta R_{e\gamma}$ 
in $Z^0$-like events.  
The points are the $Z^0$-like photon-electron sample; 
the cross-hatched histogram is electron-electron events from CDF data 
with the same kinematic requirements, normalized to the control sample.}
\label{egbb}
\end{figure}

A control sample of $Z^{0}$-like events is selected from
photon-electron candidates with the following requirements:
exactly one photon and exactly one electron 
satisfying the criteria summarized in Tables~\ref{photoncuts} 
and~\ref{leptoncuts}; no additional leptons
satisfying the criteria in Table~\ref{loose lepton cuts}; the
nearest distance in azimuth between the photon and the electron, $\Delta
\varphi_{e\gamma}$, must exceed $150^{\circ}$; and the invariant
mass of the photon-electron pair, $M_{e\gamma}$, must be within 5\GeVcsq of
the $Z^{0}$ mass (91\GeVcsq).  There are 17 such events in the CDF data,
and their characteristics are shown in Figure~\ref{egbb}.  In order to
check the assumption that these are predominantly $Z^{0}\rightarrow
e^+e^-$ events, a sample of $Z^{0}\rightarrow e^+e^-$ events is
selected from the inclusive electron sample which have exactly two
electrons passing the electron criteria in Table~\ref{leptoncuts}, and
which have the same kinematic requirements as the photon-electron
control sample.  There are 1235 such events, and their distributions,
normalized to the photon-electron control sample, are also shown in
Figure~\ref{egbb}; the shapes of the distributions of the two samples
are statistically consistent with each other.

Some of the photon-electron events in the control sample will arise
from real photons from $W/Z^{0}+ \gamma$ production, or from jets
misidentified as photons.  In order to avoid double-counting these as a
source of background, the diboson Monte Carlo calculations described
in Section~\ref{diboson} and the jet misidentification calculations
described in Section~\ref{fakephoton} are used to estimate the number
of photon-electron events passing the control sample requirements, and
this is subtracted from the total number of control sample events to
give a corrected of misidentified photon-electron events.
Out of 17 events, $1.24\pm0.13$ events ($1.01\pm0.12$ from diboson
events, $0.23\pm0.04$ from misidentified jets) on average are expected
to have real photons, which are subtracted to give $15.8\pm4.3$
misidentified photon-electron events in the control sample.
  
The number of misidentified photon-electron events in the control
sample, $N^{ctrl}_{e\gamma}$, divided by the number of
electron-electron events with the same kinematics, $N^{ctrl}_{ee}$,
gives the misidentified photon-electron rate per central electron
pair. 
For any other particular subset of central electron pairs, the total
contribution to the corresponding photon-electron sample is the product
of the number of central electron pairs with this misidentification  
rate.
For multi-body photon-lepton
events, a sample of dielectron events is selected from events
satisfying the same trigger criteria as that of the photon-electron
candidates described in Section~\ref{electron id}.  From these
triggers a sample of multi-body dielectron events is selected which
has exactly two electrons satisfying the electron criteria in
Table~\ref{leptoncuts}, and which has the same angular separation
requirements ($\Delta \varphi_{ee}
\lt 150^{\circ}$) as the multi-body photon-lepton sample. There are
132 such events.  The estimated number of misidentified
photon-electron events in multi-body photon-electron events is
therefore
\begin{eqnarray}
N^{mult}_{e\gamma} & = & [(15.8\pm4.3)/1235]\times 132 \nonumber \\* 
& = & 1.7\pm0.5 \text{ events.}
\end{eqnarray}
Similar calculations are made for the other photon-lepton samples 
analyzed, and the results are summarized in Table~\ref{zee table}.
The number of multi-photon and multi-lepton events is negligible,
due to the low number of $ee\gamma$ and $eee$ events in the CDF data.

\linespace{1.0}
\linespace{2.0}
        
\linespace{1.25}
\begin{table}[!ht]
\begin{center}
\begin{tabular}{lrl}\hline \hline
 & \multicolumn{1}{c}{$N_{ee}$} & \multicolumn{1}{c}{$N_{e\gamma}$} \\ \hline
Two-Body $e\gamma X$        & 321         & $4.1\pml1.1$        \\ 
Multi-Body $e\gamma X$      & 132         & $1.7\pml0.5$        \\ 
Multi-Body $e\gamma\mett X$ &   8         & $0.10\pm0.04$        \\ 
\hline\hline
\end{tabular}
\caption{The expected mean number of photon-electron candidates $N_{e\gamma}$
from $Z^{0}$ electrons misidentified as photons, for the various categories 
analyzed.  The number of dielectron events $N_{ee}$ which 
potentially contribute to each category is also included.}
\label{zee table}
\end{center}
\end{table}
\linespace{2.0}
\subsection{Light Hadrons Misidentified as Muons}
\label{pthru}

A hadron jet can contain charged hadrons, which may occasionally
penetrate the calorimeters and be detected by the muon chambers,
(``hadron punchthrough''), or which may decay to a muon before
reaching the calorimeters (``hadron decay-in-flight'').  If one of
these hadrons constitutes a sufficiently large fraction of the jet
momentum, then the hadron jet can be misidentified by the CDF detector
as a single prompt muon.  Such a jet produced in association with a
photon candidate contributes to the detected photon-muon candidates.
The contribution of photon plus misidentified jet events is determined
by analyzing a sample of isolated, high-momentum tracks in CDF photon
data, determining the probability of each track being misidentified
as a muon, and computing the total contribution by summing this
probability over all tracks in the sample.

Starting with the inclusive photon events described in
Section~\ref{photon id}, a photon-track sample is selected by
requiring one or more photon candidates satisfying the criteria in
Table~\ref{photoncuts} and one or more CTC tracks with $\Pt \gt 25
\GeVc$ which extrapolate to the CMU, CMP, or CMX detectors.  The
selected CTC tracks must also satisfy the same track requirements as
those of muon tracks, as described in Section~\ref{muon id}: a minimum
of six layers of CTC wire measurements, at least three of which must
be axial wire measurements and at least 2 of which must be stereo wire
measurements; an impact parameter $d_0 \lt 0.3$~cm; the distance in
$z$ of the CTC track to the primary event vertex $|\Delta z_{event}|
\lt 5$~cm; and, as an isolation requirement, the sum of the momenta of
other CTC tracks incident upon a cone of $R=0.4$ around the candidate
track direction must be less than 10\% of the $\Pt$ of the candidate
track.  The photon-track sample consists of 394 events containing 398
track candidates.

Because the photon trigger requirements of the photon-track sample are
less efficient than the trigger requirements of the photon-muon
sample, the effective number of tracks which potentially contribute to
the photon-muon candidates must be augmented by a ratio of the
efficiencies of the different trigger paths, for each track in each
event of the sample.  The efficiency ratio $R_{\gamma t}$
varies with photon $\Et$ and the muon stub type $\mu_i$ that the track
$t$ would produce in the event of hadron punchthrough or
decay-in-flight:
\begin{equation}
R_{\gamma t} = \frac{\epsilon_{\mu_i} + 
                (1-\epsilon_{\mu_i})\times\epsilon_{\gamma}(\Et)}
                {\epsilon_{\gamma}(\Et)},
\label{eq:R gamma}
\end{equation}
where $\epsilon_{\mu_i}$ is the trigger efficiency for muons of stub
type $i$, and $\epsilon_{\gamma}(\Et)$ is the trigger efficiency of
photon candidates as a function of photon $\Et$.

The fraction of track candidates which give rise to hadron
punchthrough is computed from the number of hadronic interaction
lengths traversed through the calorimeter to a muon chamber, for
high-momentum pions and kaons.  The thickness of the CDF calorimeter,
typically 5 absorption lengths for pions and 4.4 lengthx for kaons,
corresponds to a hadron rejection factor of about 150 (80) for the CMU
(CMX).  The CMP is additionally shielded from hadrons by 60~cm of
steel, which effectively absorbs all incident hadrons; the
contribution of hadron punchthrough to CMP or CMUP muon candidates is
henceforth assumed to be negligible.  The contribution to hadron
punchthrough of hadrons which partially shower in the calorimeter is
reduced to a negligible level by the muon identification requirements
of low calorimeter activity and a small track-stub matching distance.
It is therefore sufficient to consider only the case where a hadron
traverses the entire length of the calorimeter without interacting,
and subsequently enters the CMU or CMX.

For each track in the photon-track sample, the
probability of the track becoming hadron punchthrough, $P^{t}_{PT
\mu}$, is given by
\begin{eqnarray}
P^{t}_{PT \mu} & = & 
F_\pi \times \exp{(-\lambda_\pi(E^t)/\sin\theta_t)} \nonumber \\*
& & + F_{K} \times \exp{(-\lambda_{K}(E^t)/\sin\theta_t)},
\end{eqnarray}
where $F_{\pi}$ and $F_{K}$ are the relative $\pi:K$ fractions; and
$\lambda_\pi(E^t)$ and $\lambda_{K}(E^t)$ are the calorimeter
thicknesses in units of the interaction lengths~\cite{kpi int length}
for the corresponding particle type, as a function of the total energy
$E^t$ of the track $t$ and the sign of its charge.  The interaction
length for kaons is longer than that of pions, so $P_{PT \mu}$ is a
maximum for $F_{K} = 1.0$ and a minimum for $F_{K}$ = 0.0.  For the
central value estimate, an experimentally measured value $F_{K} =
0.33$ is used~\cite{ktopi}, with upper and lower systematic bounds
defined by $F_{K} = 1.0$ and $F_{K} = 0.0$.  This systematic
uncertainty is the dominant uncertainty for the hadron punchthrough
estimates.

\linespace{1.25}
\begin{table}[!ht]
\begin{center}
\begin{tabular}{lccc}
\hline\hline
\multicolumn{1}{c}{Stub Type} & 
Two-Body &
Multi-Body &
Multi-Body \\
&
\multicolumn{1}{c}{$\mu\gamma X$} &
\multicolumn{1}{c}{$\mu\gamma X$} &
\multicolumn{1}{c}{$\mu\gamma\mett X$} \\
\hline
CMUP     &    --- &    --- &    --- \\ 
CMNP     & $0.37$ & $0.12$ & $0.07$ \\
CMX      & $0.15$ & $0.08$ & $0.03$ \\
CMP      &    --- &    --- &    --- \\ 
CMU      & $0.90$ & $0.25$ & $0.09$ \\ 
\hline
Total    & $1.42\pm0.74$ & $0.45\pm0.25$ & $0.18\pm0.11$ \\ \hline\hline
\end{tabular}
\caption{The contribution to the photon-muon candidates of punchthrough
hadrons misidentified as muons, indexed by muon stub type, 
for various categories analyzed.}
\label{pthru table}
\end{center}
\end{table}
\linespace{2.0}

For any particular subset of the photon-track sample, the total
contribution to the corresponding photon-muon sample is 
the sum over all candidate tracks of the hadron punchthrough probabilities,
weighted by the appropriate trigger efficiency ratio for each track:
\begin{eqnarray}
N_{PT \mu} & = & \sum_{t} R_{\gamma t}\times P^{t}_{PT \mu}.
\end{eqnarray}
For example, in the case of multi-body
$\mu\gamma$ events, a subset of the punchthrough candidates is
selected for which the track extrapolates to the CMU or CMX detectors,
and $\Delta\varphi$ between the photon and the track is less than $150^\circ$.
There are 89 such tracks, corresponding to a background of
$0.45\pm0.25$ events from hadron punchthrough in the inclusive
multi-body $\mu\gamma$ sample.  Of these 89 tracks, 32 belong to
events with $\mett > 25\GeV$, corresponding to $0.18\pm0.11$
punchthrough events in the multi-body $\mu\gamma\mett$ sample.  The
results indexed by muon stub type are shown in Table~\ref{pthru
table}.

Each of the photon-track events described above also potentially
contributes to photon-muon events in the form of hadron
decay-in-flight; hadrons which decay to muons prior to interacting
with the central calorimeters will satisfy the requirements of prompt
muons.  The inner radius of the central calorimeters is 1.73~m, and
the radius beyond this corresponding to one hadronic interaction length is
approximately 2~m; hadrons decaying prior to a radius of 2~m are
therefore likely to be misidentified as muons.

For each track in the photon-track sample, the hadron
decay-in-flight probability $P^{t}_{DIF \mu}$ is given by
\begin{eqnarray}
P^{t}_{DIF \mu}& = & 
F_\pi\times{\textrm BR}(\pi^\pm \rightarrow \mu\nu) \nonumber \\*
& & \times(1-\exp{(-(2.0/c\tau_{\pi})(m_{\pi}/c\Pt))})  \nonumber \\*
& &+F_K\times{\textrm BR}(K^\pm \rightarrow \mu\nu)   \nonumber \\*
& &\times(1-\exp{(-(2.0/c\tau_{K})(m_{K}/c\Pt))}),
\end{eqnarray}  
where $\Pt$ is the transverse momentum of the track $t$, in\GeVc;
$F_\pi$ is the fraction of tracks which are pions,
BR$(\pi^\pm\rightarrow\mu\nu)$ is the branching ratio of pions to
muons ($\sim 1.0$), $c\tau_{\pi}$ is the pion proper decay length in
meters (7.8 m), and $m_{\pi}$ is the pion mass (0.140\GeV); $F_K$
is the fraction of tracks which are kaons, BR$(K^\pm \rightarrow
\mu\nu)$ is the branching ratio of kaons to muons (0.635), $c\tau_{K}$
is the kaon proper decay length in meters (3.7 m), and $m_{K}$ is
the kaon mass (0.494\GeV).  For tracks with transverse momentum
of 25\GeVc, the decay-in-flight probability is 0.67\% for kaons
and 0.14\% for pions.

For any particular subset of the photon-track sample, the
contribution to the corresponding photon-muon candidates 
of decay-in-flight hadrons is
the sum over all tracks of the decay-in-flight probabilities,
augmented by the trigger efficiency ratio:
\begin{eqnarray}
N_{DIF \mu} & = & \sum_{t} R_{\gamma t}\times P^{t}_{DIF \mu}
\end{eqnarray}
Due to the shorter kaon lifetime, the upper and lower bounds are again
determined by the results assuming kaon fractions of 1.0 and 0.0,
respectively, with the central value determined by $F_K = 0.33$.  The
results indexed by muon stub type are shown in Table~\ref{dif table}.
The contributions relative to those sources of photon-muon events
considered previously are small.

\linespace{1.25}
\begin{table}[!ht]
\begin{center}
\begin{tabular}{lccc}
\hline\hline
\multicolumn{1}{c}{Stub Type} & 
Two-Body &
Multi-Body &
Multi-Body \\
&
\multicolumn{1}{c}{$\mu\gamma X$} &
\multicolumn{1}{c}{$\mu\gamma X$} &
\multicolumn{1}{c}{$\mu\gamma\mett X$} \\
\hline
CMUP     & $0.35$ & $0.10$ & $0.03$ \\ 
CMNP     & $0.15$ & $0.04$ & $0.02$ \\
CMX      & $0.21$ & $0.11$ & $0.03$ \\
CMP      & $0.08$ & $0.04$ & $0.01$ \\ 
CMU      &    --- &    --- &    --- \\ 
\hline
Total    & $0.80\pm0.89$ & $0.28\pm0.31$ & $0.10\pm0.11$ \\ \hline\hline
\end{tabular}
\caption{The contribution to the photon-muon candidates of decay-in-flight
hadrons misidentified as muons, indexed by muon 
stub type, for the various categories analyzed.}
\label{dif table}
\end{center}
\end{table}
\linespace{2.0}
\subsection{Heavy-Flavored Hadron Decay to Leptons}

A hadron consisting of one or more quarks with heavy flavor (charm or
bottom) has a much shorter lifetime than those hadrons considered in
Section~\ref{pthru}; at the Tevatron, heavy-flavored hadrons typically
travel a few millimeters before decaying and do not produce a
measurable track in the CTC.  Consequently, the decay in flight of
heavy-flavored hadrons to leptons is not accounted for in the
estimates of Section~\ref{pthru}, which infer the number of
decay-in-flight hadrons from CTC tracks.  The contribution to
photon-lepton candidates that arises from heavy-flavored hadrons
produced in association with a prompt photon is instead accounted for
through Monte Carlo event generation and detector simulation, as in
Section~\ref{diboson}.

\linespace{1.0}
\begin{figure}[!ht]
\begin{center}
\includegraphics*[width=0.4\textwidth]{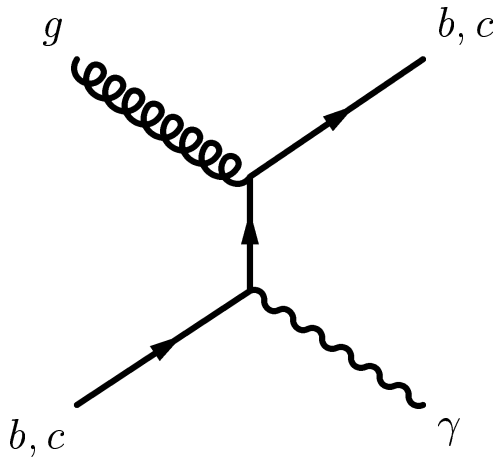}
\caption{The leading-order Feynman diagram for $\gamma + b,c$ production.}
\label{feynmf photonc}
\end{center}
\end{figure}
\linespace{2.0}

Figure~\ref{feynmf photonc} shows the leading-order Feynman diagram
for a heavy-flavored quark produced in association with a prompt
photon.  The leading-order matrix element for this process is
calculated with the \textsc{pythia}~\cite{PYTHIA} event generator
program, using the leading-order proton structure function
CTEQ5L~\cite{CTEQ5}.  \textsc{Pythia} also generates, fragments, and
hadronizes the partons produced in a simulated interaction.  The
\textsc{QQ} program, based on measurements of the CLEO
experiment~\cite{QQ}, is used to compute the decays of heavy-flavored
hadrons.  Previous measurements of photon-heavy-flavor events at the
Tevatron~\cite{cdf photonc} indicate agreement of CDF data with
next-to-leading order QCD predictions.  In order to obtain agreement
of the leading order simulation with next-to-leading order cross
section predictions, a next-to-leading order K-factor is applied to
the leading order cross section computed by \textsc{pythia}.  
In the previous measurements this K-factor was found to be $K_{NLO} =
1.9\pm0.2$.  Using this K-factor and the leading-order cross section
computed by \textsc{pythia} ($\sigma_{LO} = 7$~nb), the mean
contribution to photon-lepton candidates in CDF data for this process
is given by Equation~\ref{eq:mc rate} in Section~\ref{diboson}.

Table~\ref{hf table} shows, for the various signal regions of this
analysis, the number of simulated events which are photon-lepton
candidates, $N_{MC}$, out of 117 million events (equivalent to 8.4
fb$^{-1}$) generated; and the mean contribution expected in
86.34~pb$^{-1}$ of CDF data, $N_{\ell\gamma}$.  The contributions
expected are small compared to those discussed in
Sections~\ref{diboson}--\ref{pthru}.  All simulated candidates are
found to be two-body photon-lepton events, as would be expected for a
process with a two-body final state.  Contributions to multi-body
photon-lepton events are bounded from above by $0.01$ at the 68\%
confidence level, and are henceforth assumed to be negligible.

\linespace{1.25}   
\begin{table}[!ht]
\begin{center}
\begin{tabular}{lcr}\hline\hline
   & 
$N_{MC}$ (8.4 fb$^{-1}$) & 
\multicolumn{1}{c}{$N_{\ell\gamma}$} \\ 
\hline
\multicolumn{3}{c}{Two-Body Events}\\ 
\hline
$e\gamma X$         &    10 & $0.07\pm0.02$ \\
$\mu\gamma X$       &     3 & $0.03\pm0.01$ \\
\hline
\multicolumn{3}{c}{Multi-Body Events}\\ 
\hline
$e\gamma X$         & 0 & $\lt 0.01$ \\
$\mu\gamma X$       & 0 & $\lt 0.01$ \\
$e\gamma\mett X$    & 0 & $\lt 0.01$ \\
$\mu\gamma\mett X$  & 0 & $\lt 0.01$ \\
\hline\hline
\end{tabular}
\caption
{The contribution to photon-lepton candidates,
$N_{\ell\gamma}$, of heavy-flavored hadrons decaying to leptons, for the
various categories analyzed.
Included is the number of candidate events $N_{MC}$ produced by the 
simulation for each category.}
\label{hf table}
\end{center}
\end{table}
\linespace{2.0}

\section{Analysis of Photon-Lepton Candidates}
\label{analysis}

\linespace{1.0}
\begin{figure}[!ht]
\begin{center}
\includegraphics*[width=0.75\textwidth]{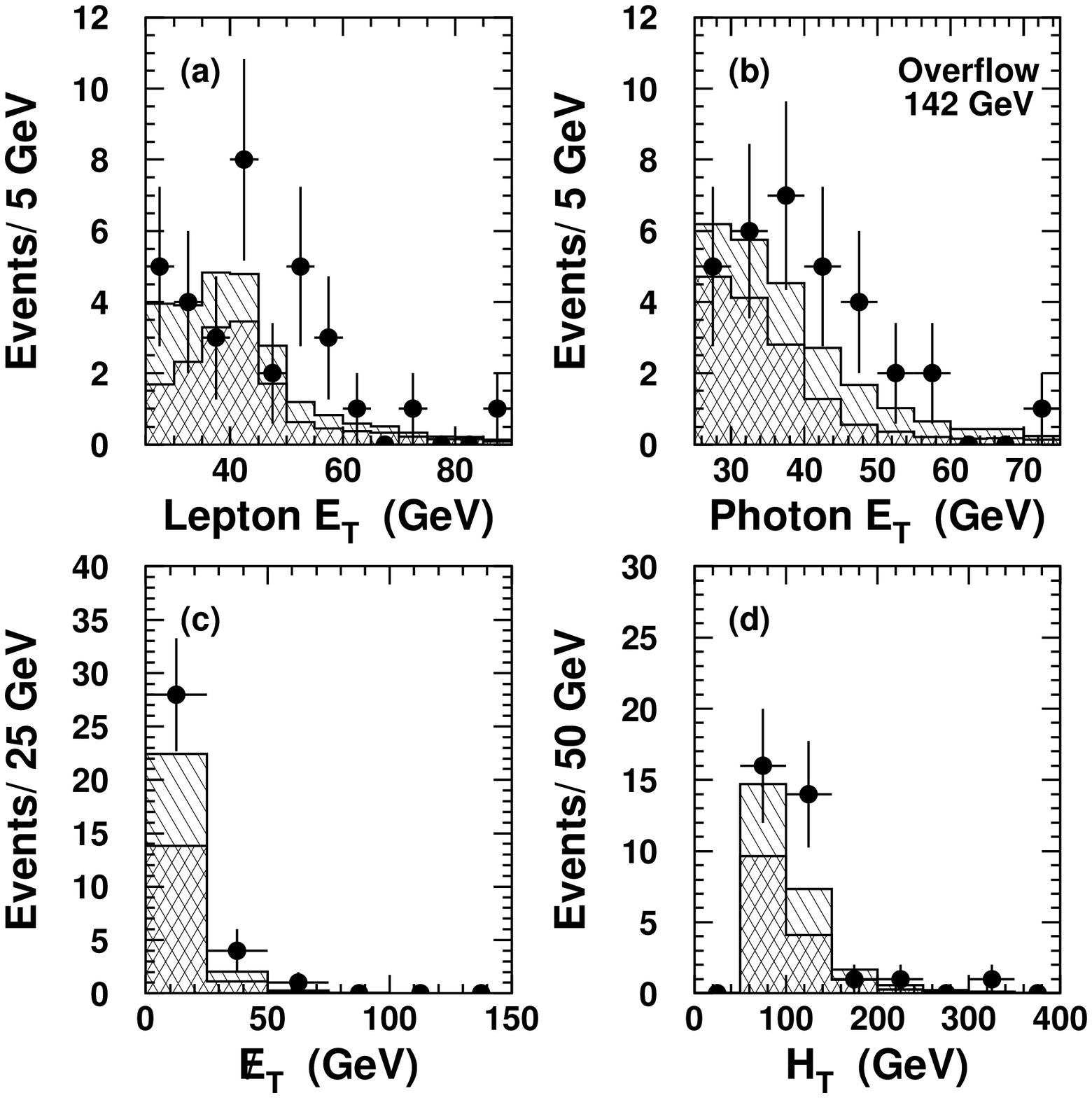}
\end{center}
\caption{The distributions for 
(a) lepton $\Et$, (b) photon $\Et$,
(c) $\mett$, and (d) $\Ht$
in two-body photon-lepton events.  
The points are CDF data,
the hatched histogram is the total predicted mean background, and 
the cross-hatched histogram is the predicted mean diboson background.}
\label{emu 2b obj}
\end{figure}
\linespace{2.0}

\linespace{1.0}
\begin{figure}[!ht]
\begin{center}
\includegraphics*[width=0.75\textwidth]{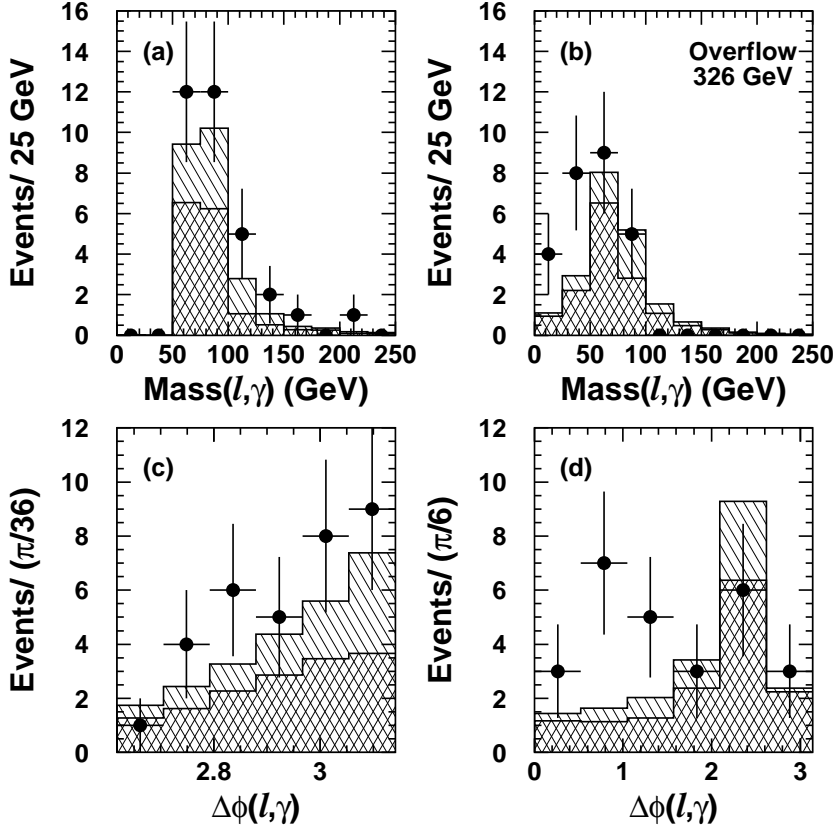}
\end{center}
\caption{The distributions for 
(a) $M_{\ell\gamma}$ in two-body photon-lepton events,
(b) $M_{\ell\gamma}$ in inclusive multi-body photon-lepton events,
(c) $\Delta\varphi_{\ell\gamma}$ 
in two-body photon-lepton events, and   
(d) $\Delta\varphi_{\ell\gamma}$ 
in inclusive multi-body photon-lepton events.   
The points are CDF data,
the hatched histogram is the total predicted mean background, and 
the cross-hatched histogram is the predicted mean diboson background.}
\label{emu mang}
\end{figure}
\linespace{2.0}

\linespace{1.0}
\begin{figure}[!ht]
\begin{center}
\includegraphics*[width=0.75\textwidth]{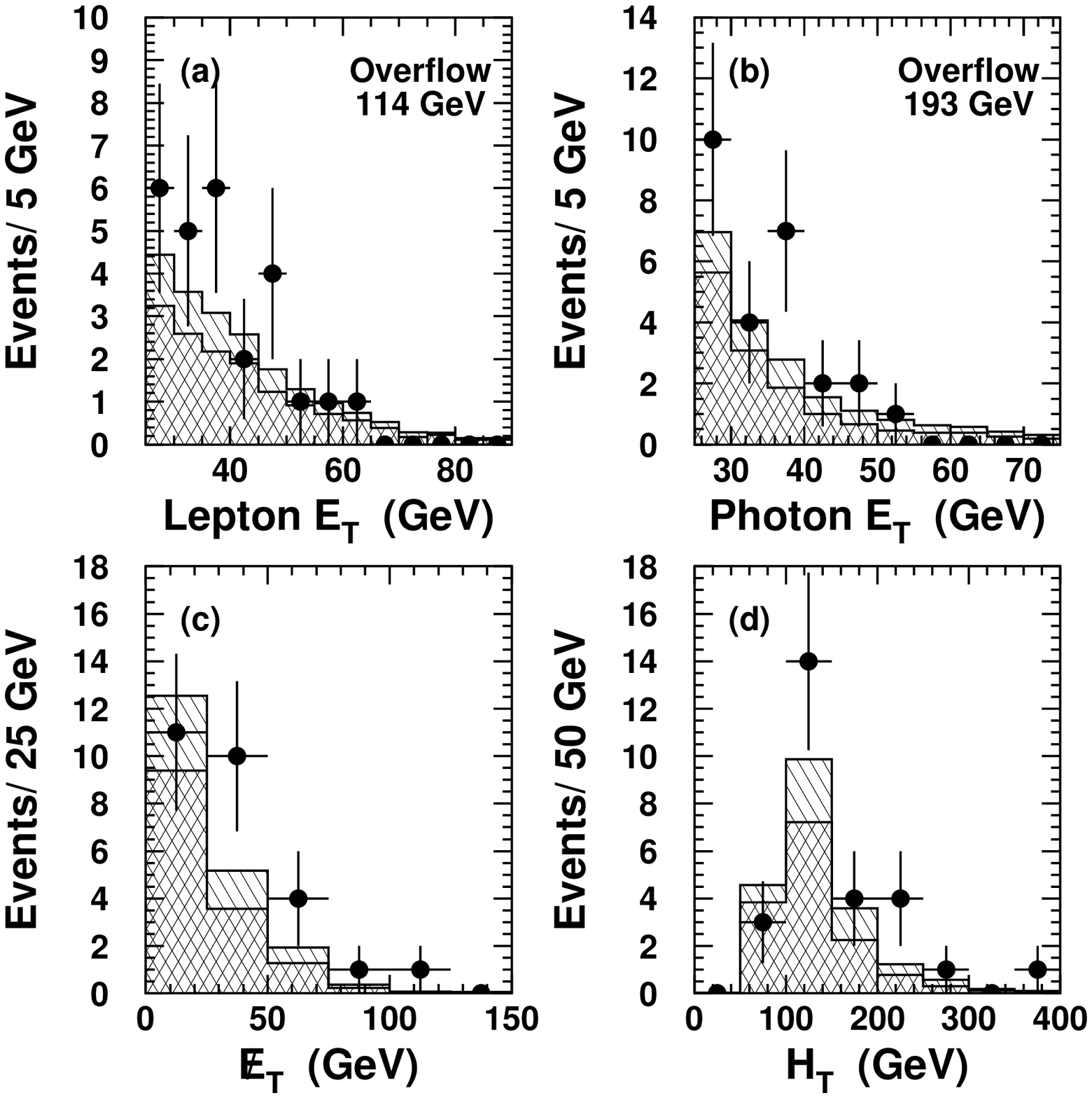}
\end{center}
\caption{The distributions for 
(a) lepton $\Et$, (b) photon $\Et$,
(c) $\mett$, and (d) $\Ht$
in inclusive multi-body photon-lepton events.  
The points are CDF data,
the hatched histogram is the total predicted mean background, and 
the cross-hatched histogram is the predicted mean diboson background.}
\label{emu mb obj}
\end{figure}
\linespace{2.0}

The objectives of this analysis are the comparison of the observed
event totals, in the various photon-lepton samples described in
Section~\ref{photon lepton id}, with the totals predicted by the
standard model, and the similar comparison of the distributions of
kinematic properties in those samples.  New physics in small samples
of events would most likely manifest itself as an excess of observed
events over expected events.  In the absence of a specific alternative
model, the significance of an observed excess is computed from the
likelihood of obtaining the observed number of events, assuming that the
null hypothesis (i.e., the standard model) is correct.  This
``observation likelihood'', denoted here by $P(N\geq N_{0}|\mu_{SM})$,
is defined as that fraction of the Poisson distribution of expected
events (with a mean $\mu_{SM}$ predicted by the standard model) which
yields outcomes $N$ greater than or equal to that observed in CDF
data, $N_{0}$.  A small observation likelihood is indicative of a
sample which is potentially better explained by physics beyond the
standard model.

For each photon-lepton sample, the mean event total predicted by the
standard model, $\mu_{SM}$, is the sum of each of the sources
discussed in Section~\ref{background}.  The uncertainty in $\mu_{SM}$
is the standard deviation of a large ensemble of calculations.  For
each calculation in the ensemble, each quantity used to compute
photon-lepton event sources (simulation systematics, integrated
luminosity, photon and lepton misidentification rates, etc.) varies
randomly as a Gaussian distribution, where the center of the
distribution is the mean value of the quantity and the width is the
uncertainty of the quantity.  This ensemble of calculations accounts
for correlated uncertainties between the various contributing sources,
such as the uncertainty in the integrated luminosity used to normalize
the various simulated event totals.  The observation likelihood
$P(N\geq N_{0}|\mu_{SM})$ is again computed from a large ensemble of
calculations.  For each calculation in the ensemble, each quantity
used to compute photon-lepton event sources again varies randomly as a
Gaussian distribution, and the resulting mean event total is used to
randomly generate a Poisson distributed outcome $N$.  The fraction of
calculations in the ensemble with outcomes $N \geq N_{0}$ gives
$P(N\geq N_{0}|\mu_{SM})$.

The total standard model predictions for the distributions of
kinematic properties are the sums of the distributions of the
corresponding properties of each of the sources discussed in
Section~\ref{background}.  For the contribution from jets
misidentified as photons, the appropriately weighted distributions of
jet properties in lepton-jet events are used in the predicted
distributions of photon properties.  Similarly, for the contribution from
electrons misidentified as photons the distributions of electron
properties in electron-electron events are used to predict
distributions of photon properties, and for the contribution from hadrons
misidentified as muons the distributions of track properties in
photon-track events are used to predict distributions of muon
properties.

\subsection{Two-Body and Inclusive Multi-Body Photon-Lepton Events}

\linespace{1.25}

\begin{table}[!ht]
\begin{center}
\begin{tabular}{llll}\hline\hline
Process & 
\multicolumn{1}{c}{$e\gamma X$} & 
\multicolumn{1}{c}{$\mu\gamma X$} & 
\multicolumn{1}{c}{$\ell\gamma X$} \\ \hline
W$+\gamma$& 
$\ \,1.2\pml0.2$& $\ \,1.5\pml0.2$ & $\ \,2.7\pml0.3$ \\
Z$+\gamma$&
$\ \,5.4\pml0.6$& $\ \,7.1\pml0.8$ & $12.5\pml1.2$ \\ 
$\ell$+jet, jet $\rightarrow \gamma$ &
$\ \,1.9\pml0.3$& $\ \,1.5\pml0.3$ & $\ \,3.3\pml0.7$ \\ 
$Z \rightarrow ee, e \rightarrow \gamma$  &
$\ \,4.1\pml1.1$&\multicolumn{1}{c}{---}& $\ \,4.1\pml1.1$ \\ 
Hadron$+\gamma$ &
\multicolumn{1}{c}{---}& $\ \,1.4\pml0.7$ & $\ \,1.4\pml0.7$ \\ 
$\pi/K$ Decay$+\gamma$ &
\multicolumn{1}{c}{---}& $\ \,0.8\pml0.9$ & $\ \,0.8\pml0.9$ \\ 
$b/c$ \  \ Decay$+\gamma$   & 
$\ \,0.07\pm0.02$ & $\ \,0.03\pm0.01$ & $\ \,0.10\pm0.03$ \\ 
\hline
Predicted $\mu_{SM}$ 
& $12.6\pml1.4$ & $12.3\pml1.8$ & $24.9\pml2.4$ \\  
Observed $N_{0}$& 
\multicolumn{1}{c}{20} & 
\multicolumn{1}{c}{13} & 
\multicolumn{1}{c}{33} \\ 
$P(N\geq N_{0}|\mu_{SM})$ & 
\multicolumn{1}{c}{0.043} & 
\multicolumn{1}{c}{0.46} & 
\multicolumn{1}{c}{0.093} \\ 
\hline\hline
\end{tabular}
\end{center}
\caption{The mean number $\mu_{SM}$ of two-body photon-lepton events
predicted by the standard model, the number $N_0$ observed in CDF data,
and the observation likelihood $P(N\geq N_{0}|\mu_{SM})$.  
There exist correlated uncertainties between the different photon-lepton
sources.} 
\label{2blg result}
\end{table}
\linespace{2.0}

The predicted and observed totals for two-body
photon-lepton events are compared in Table~\ref{2blg result}.  The mean
predicted contributions from each of the sources discussed in
Section~\ref{background} are also listed.  Half of the predicted total
originates from $Z^{0}\gamma$ production where one of the charged
leptons has evaded identification; the other half originates from
roughly equal contributions of $W\gamma$ production, misidentified
jets, misidentified electrons, and misidentified charged hadrons.  The
observed photon-electron total is somewhat higher than
predicted, with an observation likelihood of 4.3\%; the
observed photon-muon total is in excellent agreement with the predicted
total, however, so that the observation likelihood of the
two-body photon-lepton event total increases to 9.3\%.  
%The probability that
%the 33 total observed events would exhibit an asymmetry between
%electrons and muons of (20-13)/33 = 21\% or greater where none is
%expected, is 30\%.

The predicted and observed distributions of the
kinematic properties of two-body photon-lepton events are compared in
Figures~\ref{emu 2b obj} and~\ref{emu mang}.  Superimposed upon the
distributions of the total contribution predicted by the standard model
are the distributions of the contribution from standard model diboson
production.  

Figure~\ref{emu 2b obj} shows the distributions of 
photon $\Et$, lepton $\Et$, and $\mett$ for the events.  The observed
distributions of photon and lepton $\Et$ exhibit the range of values expected
from the standard model.  The number of two-body photon-lepton events
observed with $\mett \lt 25\GeV$ is in good agreement with the
predicted total.  There are 5 events observed with
$\mett \gt 25\GeV$, whereas 2.3 events are expected, a result 
which is potentially related to that observed in 
multi-body $\ell\gamma\mett$ events described below. 

The distribution of the total $\Et$ of all objects in the event,
$\Ht$, is also included in Figure~\ref{emu 2b obj}.  
It is defined as the sum of the magnitudes
of $\mett$ and the transverse energies of all electrons, muons,
photons, and jets in the event:
\begin{eqnarray}
\Ht & \equiv & \mett +  
\sum_{e} E^{e}_{T} + 
\sum_{\mu} cp^{\mu}_{T} + \nonumber \\*
& &
\sum_{\gamma} E^{\gamma}_{T} + 
\sum_{j} E^{j}_{T}(cor).  
\end{eqnarray}
The jets included in this sum are required to have 
$E^{j}_{T}(raw) \gt 8\GeV$
and $|\eta_{j}| \lt 2.4$, just as in Equation~\ref{eq:metj}.
Large $\Ht$ is correlated with the production of massive particles,
virtual or real.  The observed data exhibit the range of $\Ht$ values
expected.

\linespace{1.25}

\begin{table}[!ht]
\begin{center}
\begin{tabular}{lrrr}\hline\hline
Process & 
\multicolumn{1}{c}{$e\gamma X$} & 
\multicolumn{1}{c}{$\mu\gamma X$} & 
\multicolumn{1}{c}{$\ell\gamma X$} 
\\ \hline
W$+\gamma$& 
$2.4\pm0.3$ & $2.5\pm0.3$ & $5.0\pm0.6$ \\
Z$+\gamma$&
$5.0\pm0.5$& $4.6\pm0.5$ & $9.6\pm0.9$ \\ 
$\ell$+jet, jet $\rightarrow \gamma$ &
$1.7\pm0.3$& $1.5\pm0.3$ & $3.2\pm0.6$ \\ 
$Z \rightarrow ee, e \rightarrow \gamma$  &
$1.7\pm0.5$&\multicolumn{1}{c}{---}& $1.7\pm0.5$  \\ 
Hadron$+\gamma$ &
\multicolumn{1}{c}{---}&$0.5\pm0.3$&$0.5\pm0.3$ \\ 
$\pi/K$ Decay$+\gamma$ &
\multicolumn{1}{c}{---}& $0.3\pm0.3$&$0.3\pm0.3$ \\ 
$b/c$ \  \ Decay$+\gamma$   & 
$\lt 0.01$ & $\lt 0.01$ & $\lt 0.01$  \\ 
\hline
Predicted $\mu_{SM}$ 
& $10.9\pm1.0$ & $9.3\pm1.0$ & $20.2\pm1.7$ \\  
Observed $N_0$& 
\multicolumn{1}{c}{11} & 
\multicolumn{1}{c}{16} & 
\multicolumn{1}{c}{27}   \\ 
$P(N\geq N_{0}|\mu_{SM})$ & 
\multicolumn{1}{c}{0.52} & 
\multicolumn{1}{c}{0.037} & 
\multicolumn{1}{c}{0.10}  \\ 
\hline\hline
\end{tabular}
\end{center}
\caption{The mean number $\mu_{SM}$ of inclusive multi-body photon lepton
events predicted by the standard model, the number $N_0$ observed in CDF
data, and the observation likelihood $P(N\geq N_{0}|\mu_{SM})$.
There exist correlated uncertainties between the different photon-lepton
sources.} 
\label{mblg result}
\end{table}
\linespace{2.0}

The predicted and observed totals for inclusive multi-body
photon-lepton events are compared in Table~\ref{mblg result}.  The
magnitude of the predicted total is similar to that of two-body
photon-lepton events.  About half of the predicted total originates
from $Z^{0}\gamma$ production, a quarter from $W\gamma$ production,
and the remaining quarter from particles misidentified as photons or
leptons.  In this sample the observed photon-muon total is higher than
predicted, with an observation likelihood of 3.7\%; all of the
difference can be attributed to events with large $\mett$, as
discussed below.  The observed photon-electron total is in excellent
agreement with the predicted total, and the observation likelihood of
the inclusive multi-body photon-lepton total increases to 10\%.

The predicted and observed distributions of the kinematic properties
of inclusive multi-body photon-lepton events are compared in
Figures~\ref{emu mang} and~\ref{emu mb obj}.  The difference between
the observed and predicted totals can be entirely attributed to events
with $\mett \gt 25\GeV$; the observed events with lower $\mett$ agree
with predictions.  There is also a larger proportion of observed
events than expected with smaller photon-lepton azimuthal separation,
$\Delta \varphi_{\ell\gamma}$, for which the contributions from
misidentified photons or leptons are largely absent.

\subsection{Multi-Body $\ell\gamma\mett$ Events}

\linespace{1.0}
\begin{figure}[!ht]
\begin{center}
\includegraphics*[width=0.75\textwidth]{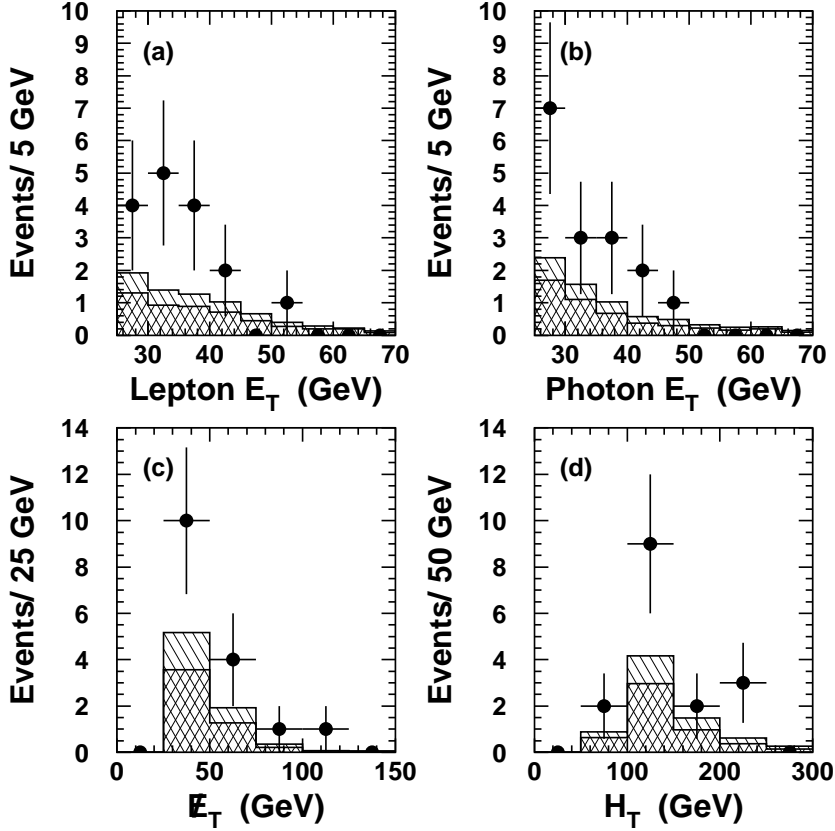}
\end{center}
\caption{The distributions for 
(a) lepton $\Et$, (b) photon $\Et$,
(c) $\mett$, and (d) $\Ht$
in multi-body $\ell\gamma\mett$ events.  
The points are CDF data,
the hatched histogram is the total predicted mean background, and 
the cross-hatched histogram is the predicted mean diboson background.}
\label{emu met obj}
\end{figure}
\linespace{2.0}

\linespace{1.0}
\begin{figure}[!ht]
\begin{center}
\includegraphics*[width=0.75\textwidth]{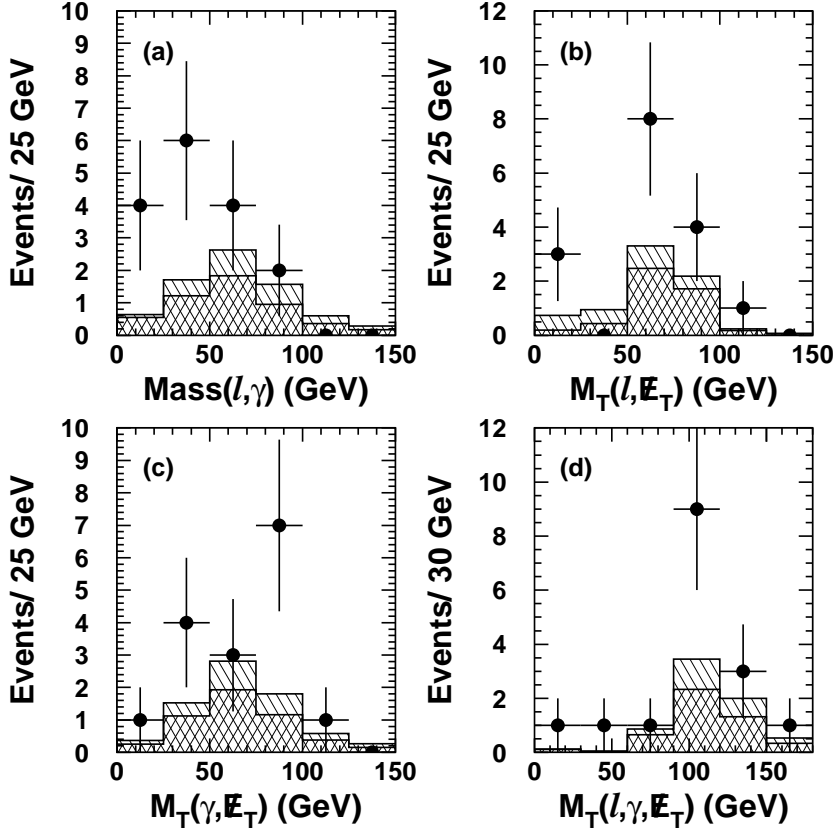}
\end{center}
\caption{The distributions for 
(a) photon-lepton mass, 
(b) lepton-$\mett$ transverse mass,
(c) photon-$\mett$ transverse mass, and 
(d) $\ell\gamma\mett$ transverse mass
in multi-body $\ell\gamma\mett$ events.  
The points are CDF data,
the hatched histogram is the total predicted mean background, and 
the cross-hatched histogram is the predicted mean diboson background.}
\label{emu met mass}
\end{figure}
\linespace{2.0}

\linespace{1.0}
\begin{figure}[!ht]
\begin{center}
\includegraphics*[width=0.75\textwidth]{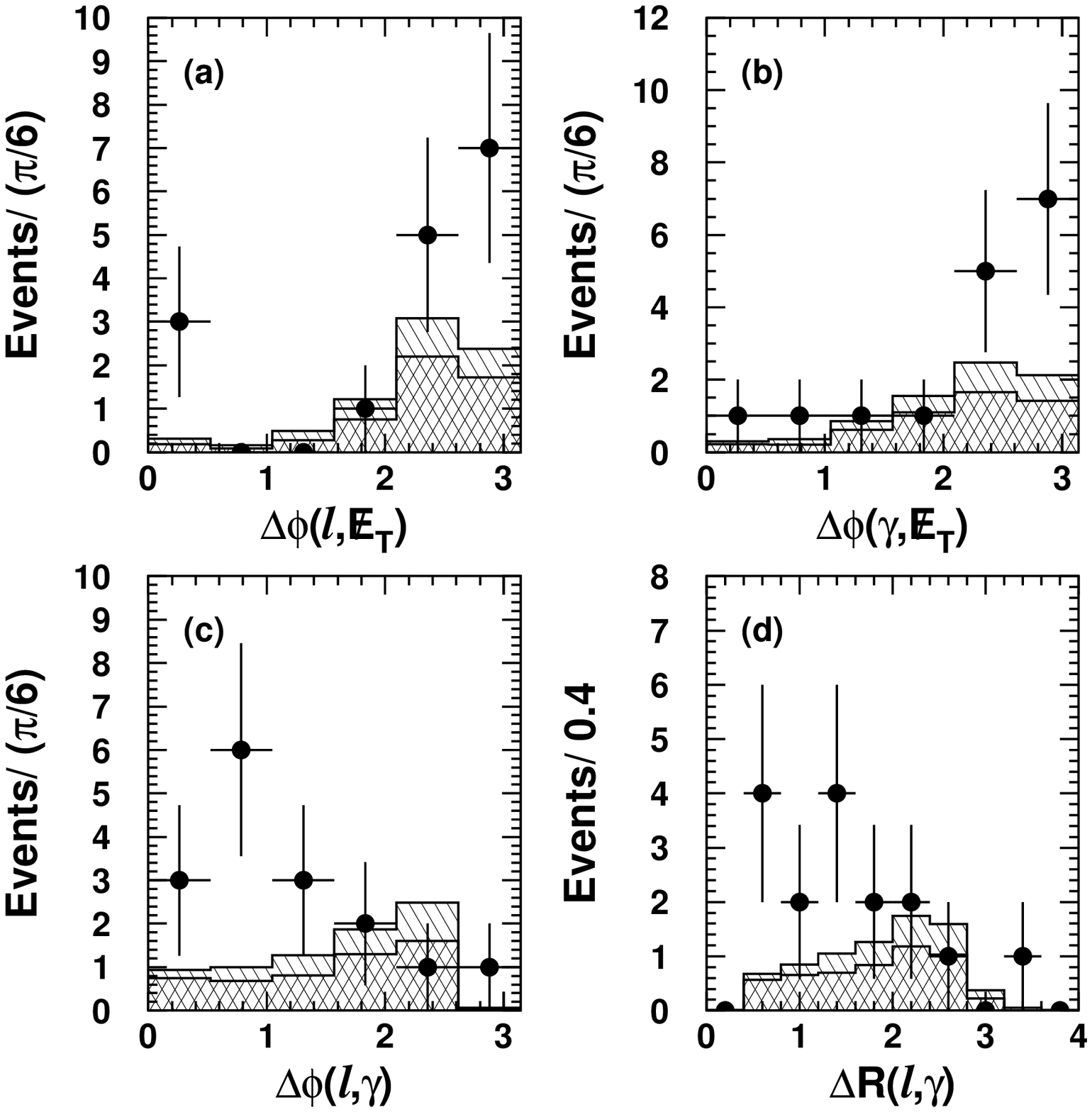}
\end{center}
\caption{The distributions for 
(a) $\Delta\varphi(\ell\mett)$,  
(b) $\Delta\varphi(\gamma\mett)$,  
(c) $\Delta\varphi_{\ell\gamma}$, and  
(d) $\Delta R_{\ell\gamma}$  
in multi-body $\ell\gamma\mett$ events.  
The points are CDF data,
the hatched histogram is the total predicted mean background, and 
the cross-hatched histogram is the predicted mean diboson background.}
\label{emu met ang}
\end{figure}
\linespace{2.0}

The predicted and observed totals for multi-body $\ell\gamma\mett$
events are compared in Table~\ref{mblg result2}.  For photon-electron
events, requiring $\mett \gt 25\GeV$ suppresses the contribution from
$Z^{0}\gamma$ production and from electrons misidentified as photons,
which have no intrinsic $\mett$, while preserving the contribution
from $W\gamma$ production.  As a result, 57\% of the predicted
$e\gamma\mett$ total arises from $W\gamma$ production, 31\% from jets
misidentified as photons, only 3\% from $Z^{0}\gamma$ production, and
the remaining 9\% from other particles misidentified as photons.  The
observed $e\gamma\mett$ total agrees with the predicted total, with a
25\% probability that the predicted mean total of $3.4$ events yields
5 observed events.  Included in the 5 events observed is the
$ee\gamma\gamma\mett$ event.

For photon-muon events, requiring $\mett \gt 25\GeV$ does not
completely eliminate the contribution from $Z^{0}\gamma$, for if the
second muon has $|\eta| \gt 1.2$ and $\Pt \gt 25\GeVc$ it evades all
forms of muon detection and induces the necessary amount of $\mett$.
The rate at which this occurs is estimated well by $Z^{0}\gamma$ event
simulation, however, since it is solely a function of the CDF detector
acceptance for such a second muon.  Of the 4.6 multi-body photon-muon
events predicted to originate from $Z^{0}\gamma$ production, 2.2
events are predicted to contain a second visible muon, 1.0 are
predicted to induce more than 25\GeV of $\mett$ as above, and 1.4 are
predicted to induce less than 25\GeV of $\mett$.  As shown in
Table~\ref{mlg result}, 1 event is observed with a second visible
muon, in agreement with $Z^{0}\gamma$ predictions. 
The predicted total for multi-body $\mu\gamma\mett$ events consists
of 47\% $W\gamma$ production, 24\% events with jets misidentified as
photons, 23\% $Z^{0}\gamma$ production, and the remaining 7\% from
other particles misidentified as muons.    
 
The observed $\mu\gamma\mett$ total is much higher than predicted (11
observed vs. 4 expected), with an observation likelihood of only
0.54\%; the observation likelihood of the $\ell\gamma\mett$ total is
only slightly higher at 0.72\%.

% The probability that the 16 total observed events would
%exhibit an asymmetry between muons and electrons of (11-5)/16 = 37.5\%
%or greater where none is expected, is 7.7\%, so it is reasonable to
%interpret the difference in expected and observed totals as arising
%from electron and muon events, and not from the muon events alone.
\linespace{1.25}

\begin{table}[!ht]
\begin{center}
\begin{tabular}{llll}\hline\hline
Process & 
\multicolumn{1}{c}{$e\gamma\mett X$} & 
\multicolumn{1}{c}{$\mu\gamma\mett X$} & 
\multicolumn{1}{c}{$\ell\gamma\mett X$}\\ \hline
W$+\gamma$& 
 $1.9\pml0.3$ & $2.0\pml0.3$ & $3.9\pml0.5$ \\
Z$+\gamma$&
 $0.3\pml0.1$ & $1.0\pml0.2$ & $1.3\pml0.2$\\ 
$\ell$+jet, jet $\rightarrow \gamma$ &
 $1.1\pml0.2$ & $1.0\pml0.2$ & $2.1\pml0.4$\\ 
$Z \rightarrow ee, e \rightarrow \gamma$  &
 $0.10\pm0.04$&\multicolumn{1}{c}{-}& $0.10\pm0.04$ \\ 
Hadron$+\gamma$ &
\multicolumn{1}{c}{-}&$0.2\pml0.1$&$0.2\pml0.1$\\ 
$\pi/K$ Decay$+\gamma$ &
\multicolumn{1}{c}{-}&$0.1\pml0.1$&$0.1\pml0.1$\\ 
$b/c$ \  \ Decay$+\gamma$   & 
$\ \ \ \ \ \ \lt 0.01$ &$\ \ \ \ \ \ \lt 0.01$ &$\ \ \ \ \ \ \lt 0.01$ \\ 
\hline
Predicted $\mu_{SM}$ 
& $3.4\pml0.3$ & $4.2\pml0.5$ & $7.6\pml0.7$ \\  
Observed $N_0$ & 
\multicolumn{1}{c}{5} & 
\multicolumn{1}{c}{11} & 
\multicolumn{1}{c}{16} \\ 
$P(N\geq N_{0}|\mu_{SM})$& 
\multicolumn{1}{c}{0.26} & 
\multicolumn{1}{c}{0.0054} & 
\multicolumn{1}{c}{0.0072} \\ 
\hline\hline
\end{tabular}
\end{center}
\caption{The mean number $\mu_{SM}$ of multi-body $\ell\gamma\mett$
events predicted by the standard model, the number $N_0$ observed in
CDF data, and the observation likelihood $P(N\geq N_{0}|\mu_{SM})$.
There exist correlated uncertainties between the different
photon-lepton sources.}
\label{mblg result2}
\end{table}
\linespace{2.0}

The predicted and observed distributions of the kinematic properties
of multi-body $\ell\gamma\mett$ events are compared in
Figures~\ref{emu met obj}--\ref{emu met ang}.  The photon $\Et$,
lepton $\Et$, $\mett$, and $\Ht$ observed are within the range
expected from the standard model.  The observed photon $\Et$ spectrum
has more events near the 25\GeV threshold than expected.  However,
nearly all photon candidates are one standard deviation or more above
threshold in terms of the 3\% CEM energy resolution~\cite{jwb thesis}.
The masses of combinations of objects in observed $\ell\gamma\mett$
events are characterized by photon-lepton mass less than 100\GeVcsq,
lepton-$\mett$ transverse mass greater than 50\GeVcsq, photon-$\mett$
transverse mass between 80 and 100\GeVcsq, and $\ell\gamma\mett$
transverse mass between 90 and 120\GeVcsq.  The observed angular
distributions favor smaller azimuthal photon-lepton separation and
larger lepton-$\mett$ and photon-$\mett$ azimuthal separations than
expected from the standard model.  The difference in observed and
predicted totals is therefore difficult to attribute to misidentified
photons or leptons, which as shown in Figure~\ref{emu met ang} tend to
have the larger photon-lepton azimuthal separation that is
characteristic of a two-body final state.
 
\subsection{Events with Additional Leptons or Photons}

\linespace{1.25}

\begin{table*}[!ht]
\begin{center}
\begin{tabular}{lllll}\hline\hline
Process & 
\multicolumn{1}{c}{$ee\gamma X$} & 
\multicolumn{1}{c}{$\mu\mu\gamma X$} & 
\multicolumn{1}{c}{$\ell\ell\gamma X$} &
\multicolumn{1}{c}{$e\mu\gamma X$} \\ \hline
%\multicolumn{1}{c}{$e\gamma\gamma X$} & 
%\multicolumn{1}{c}{$\mu\gamma\gamma X$} & 
%\multicolumn{1}{c}{$\ell\gamma\gamma X$}\\ \hline
Z$+\gamma$ &
$3.3\pml0.4$ & $2.2\pml0.3$ & $5.5\pml0.6$ & $0.05\pm0.01$ \\
% $0.012\pm0.012$ &$0.004\pm0.004$ & $0.016\pm0.016$ \\ 
$\ell$+jet, jet $\rightarrow \gamma$ &
$0.19\pm0.04$ & $0.13\pm0.03$ & $0.32\pm0.07$ & \multicolumn{1}{c}{---} \\
% \multicolumn{1}{c}{---} & \multicolumn{1}{c}{---} & \multicolumn{1}{c}{---} \\ 
\hline
Predicted $\mu_{SM}$ 
& $3.5\pml0.4$ & $2.3\pml0.3$ & $5.8\pml0.6$ & $0.05\pm0.01$ \\
%  $0.012\pm0.012$ & $0.004\pm0.004$ & $0.016\pm0.016$ \\  
Observed $N_0$ & 
\multicolumn{1}{c}{4} & 
\multicolumn{1}{c}{1} & 
\multicolumn{1}{c}{5} &
\multicolumn{1}{c}{0} \\
%\multicolumn{1}{c}{1} & 
%\multicolumn{1}{c}{0} & 
%\multicolumn{1}{c}{1} \\ 
$P(N\geq N_0|\mu_{SM})$ & 
\multicolumn{1}{c}{0.45} & 
\multicolumn{1}{c}{0.90} & 
\multicolumn{1}{c}{0.68} & 
\multicolumn{1}{c}{0.95} \\
%\multicolumn{1}{c}{0.013} & 
%\multicolumn{1}{c}{1.0} & 
%\multicolumn{1}{c}{0.015}\\
\hline\hline
\end{tabular}
\end{center}
\caption
{The mean number $\mu_{SM}$ of multi-body events with additional
leptons or photons predicted by the standard model, the number $N_0$
observed in CDF data, and the observation likelihood
$P(N\geq N_{0}|\mu_{SM})$.  There exist correlated uncertainties
between the different photon-lepton sources.}
\label{mlg result}
\end{table*}
\linespace{2.0}

\linespace{1.25}
\begin{table}[!ht]
\begin{center}
\begin{tabular}{lrrr}\hline\hline
Process &  
\multicolumn{1}{c}{$e\gamma\gamma$} & 
\multicolumn{1}{c}{$\mu\gamma\gamma$} & 
\multicolumn{1}{c}{$\ell\gamma\gamma$}\\ \hline
Z$+\gamma$ &
  $0.012\pm0.012$ &$0.004\pm0.004$ & $0.016\pm0.016$ \\ 
$\ell$+jet, jet $\rightarrow \gamma$ &
 --- & --- & --- \\ 
\hline
Predicted $\mu_{SM}$
 & $0.012\pm0.012$ & $0.004\pm0.004$ & $0.016\pm0.016$ \\  
Observed $N_0$ &   
\multicolumn{1}{c}{1} & 
\multicolumn{1}{c}{0} & 
\multicolumn{1}{c}{1} \\ 
$P(N\geq N_0|\mu_{SM})$ &  
\multicolumn{1}{c}{0.013} & 
\multicolumn{1}{c}{1.0} & 
\multicolumn{1}{c}{0.015} \\ 
\hline\hline
\end{tabular}
\end{center}
\caption{The mean number $\mu_{SM}$ of multi-body events with
additional photons predicted by the standard model, the number $N_0$
observed in CDF data, and the observation likelihood
$P(N\geq N_{0}|\mu_{SM})$.  Expected contributions from jets
misidentified as photons are negligible.}
\label{mlg result2}
\end{table}
\linespace{2.0}

The predicted and observed totals for multi-body
multi-lepton events are compared in Table~\ref{mlg result}.  Nearly all 
of the predicted total is expected from $Z^{0}\gamma$ production.
Approximately 6 events are expected; 5 events are observed,
including the $ee\gamma\gamma\mett$ event.  Both the electron and muon
channels are in good agreement with the standard model.
No $e\mu\gamma$ events were expected, and none were observed.

The predicted and observed totals for
multi-body multi-photon events are compared in Table~\ref{mlg result2}.
Only a small (0.01 event) contribution is expected from $Z\gamma$ production;
the single event observed is the $ee\gamma\gamma\mett$ event.  Judged
solely as an event with one lepton with $\Et \gt 25\GeV$ and two
photons with $\Et \gt 25\GeV$, the observation likelihood of
this event is 1.5\%.  Judged as an event with an additional lepton and
large $\mett$, the observation likelihood is much smaller, as
described in a previous analysis~\cite{eegg}.

\section{Conclusion}

We have performed an inclusive study of events containing at least one
photon and one lepton ($e$ or $\mu$) in proton-antiproton collisions,
motivated by the possibility of uncovering heretofore unobserved
physical processes at the highest collision energies.  In particular,
the unexplained $ee\gamma\gamma\mett$ event, uncovered early on in the
CDF analysis of the 1994-5 run of the Fermilab Tevatron, indicated
that the samples of previously unexamined particle combinations
involving leptons and photons could contain potentially related, and
therefore possibly novel, processes.  The definition of the
photon-lepton samples studied was chosen \textit{a priori}, including
the kinematic range of particles analyzed and the particle
identification techniques employed.  Wherever possible, the methods 
of previously published studies of leptons or photons at large transverse
momentum were adopted.  The questions of interest were also defined 
\textit{a priori}, namely whether the event totals of the 
photon-lepton subsamples enumerated in Figure~\ref{lgamma path} agree
with standard model predictions. As a supplemental result, the
distributions of the kinematic properties of the various photon-lepton
subsamples are presented in Section~\ref{analysis}.

The answers to those questions are as follows.  A two-body
photon-lepton sample, meant to encompass physical processes with two
energetic particles in the final state, was observed to have a total
(33 events) consistent with that of standard model predictions (25
events).  Specifically, the observed total was greater then the
predicted mean total, but the observation likelihood within the
standard model of a total greater than or equal to that observed was
more than 9\%.  A multi-body photon-lepton sample, meant to encompass
physical processes with three or more energetic particles in the final
state, was also observed to have an inclusive total (27 events)
consistent with standard model predictions (20 events).  The observed
total was again higher than the predicted mean total, but the
likelihood of a total greater than or equal to that observed was 10\%.

Several subsets of the multi-body photon-lepton sample were studied
for the presence of additional particles.  A subset of multi-body
photon-lepton events with additional leptons (5 $ee\gamma$ or
$\mu\mu\gamma$ events and 0 $e\mu\gamma$ events) was observed to have
good agreement with standard model predictions (6 events and 0 events,
respectively).  A subset of multi-body photon-lepton events with
additional photons was studied, yielding only the unexplained
$ee\gamma\gamma\mett$ event, whereas the predicted mean total of
inclusive $\ell\gamma\gamma$ events (requiring the presence of neither
$\mett$ nor a second lepton) is 0.01, an observation likelihood of
1\%.  This event and estimations of its likelihood have been analyzed
elsewhere~\cite{eegg}.

Finally, a subset of the multi-body photon-lepton sample, consisting
of those events with $\mett \gt 25\GeV$, was observed to have a total
(16 events) that is substantially greater than that predicted by the standard
model ($7.6\pm0.7$ events).  The likelihood of a total greater than or
equal to that observed was 0.7\%.  Moreover, the excess events in the
observed inclusive multi-body photon-lepton sample can be completely
accounted for by the excess in the multi-body $\ell\gamma\mett$
sample; observed multi-body photon-lepton events with $\mett \lt
25\GeV$ agree well with the standard model.

\linespace{1.25}
\begin{table}[!ht]
\begin{center}
\begin{tabular}{lrrc}\hline\hline
Category & 
\multicolumn{1}{c}{$\mu_{SM}$} & 
\multicolumn{1}{c}{$N_0$} & 
\multicolumn{1}{c}{$P(N\geq N_0|\mu_{SM})$ \%} \\ \hline
All $\ell\gamma X$& \multicolumn{1}{c}{---}  & 77 & --- \\
\hline
Z-like $e\gamma$ & \multicolumn{1}{c}{---}  & 17 & --- \\
Two-Body $\ell\gamma X$        & $24.9\,\ \pm2.4\,\ $ & 33 &  $\,\  9.3$ \\ 
Multi-Body $\ell\gamma X$      & $20.2\,\ \pm1.7\,\ $ & 27 & 10.0 \\ 
\hline
Multi-Body $\ell\ell\gamma X$  & $5.8\,\ \pm0.6\,\ $   &  5 & 68.0 \\ 
Multi-Body $\ell\gamma\gamma X$& $0.02\pm0.02$ &  1 & $\,\  1.5$ \\ 
Multi-Body $\ell\gamma\mett X$ & $7.6\,\ \pm0.7\,\ $   & 16 & $\,\  0.7$ \\ 
\hline\hline
\end{tabular}
\end{center}
\caption{The results for all photon-lepton categories analyzed, including 
the mean number of events $\mu_{SM}$ predicted by the standard model,
the number $N_0$ observed in CDF data, 
and the observation likelihood $P(N\geq N_{0}|\mu_{SM})$.} 
\label{lgamma result}
\end{table}
\linespace{2.0}

That the standard model prediction yields the observed total of a
particular sample of events with 0.7\% likelihood (equivalent to 2.7
standard deviations for a Gaussian distribution) is an interesting
result, but it is not a compelling observation of new
physics.  Multi-purpose particle physics experiments analyze dozens of
independent samples of events, making a variety of comparisons with
the standard model for each sample.  In the context of this analysis
alone, five mostly independent subsamples of photon-lepton events were
analyzed.  This large number of independent comparisons with the
standard model for the same collection of data increases the
chance that outcomes with $\sim 1\%$ likelihood occur.  However,
once a particular comparison has been identified as anomalous, the
same comparison performed with subsequent experiments is no longer
subject to the dilution of its significance by the number of other
independent comparisons performed concurrently.  Hence an observation
of increased significance in the forthcoming run of the
Fermilab Tevatron would confirm decisively the failure of the standard
model to describe $\ell\gamma\mett$ production; an observation of no
significant excess would confirm the present result as a statistical 
fluctuation.

\begin{acknowledgments}
We thank the Fermilab staff and the technical staffs of the
participating institutions for their vital contributions.  We thank
U.~Baur and S.~Mrenna for their prompt response to our need for
programs to calculate the standard model $W\gamma$ and $Z\gamma$
backgrounds used in this analysis.  
This work was supported by the U.S. Department of Energy and National
Science Foundation; the Italian Istituto Nazionale di Fisica Nucleare;
the Ministry of Education, Culture, Sports, Science, and Technology of
Japan; the Natural Sciences and Engineering Research Council of
Canada; the National Science Council of the Republic of China; the
Swiss National Science Foundation; the A. P. Sloan Foundation; the
Bundesministerium fuer Bildung und Forschung, Germany; the Korea
Science and Engineering Foundation (KoSEF); the Korea Research
Foundation; and the Comision Interministerial de Ciencia y Tecnologia,
Spain.
                                                                  
\end{acknowledgments}

%\input{acknowledgments}            

%\linespace{1.0}

%\linespace{2.0}

%\input{bibliography_pp}

\begin{thebibliography}{99}
%\raggedright

% introduction.tex
\bibitem{SM} S.~L. ~Glashow,
Nucl. Phys. {\bf 22} 588, (1961); S. Weinberg,
Phys. Rev. Lett. {\bf 19} 1264, (1967);
A. Salam, Proc. 8th Nobel Symposium, Stockholm, (1979).
      
\bibitem{cdfblurb}
F. Abe \textit{et al.}, Nucl. Instrum. Methods Phys. Res.,
Sect.~A \textbf{271}, 387 (1988).  The CDF coordinate system defines 
$r$, $\varphi$, and $z$ as cylindrical coordinates, with the
$z$-axis along the direction of the proton beam.  The angle $\theta$ is
the polar angle relative to the $z$-axis, the pseudorapidity
is defined as $\eta = -\ln(\tan(\theta/2))$, 
and the transverse energy is defined as $\Et = E \sin\theta$.  
Missing transverse energy ($\mett$) is the vector opposite to
the vector sum of the transverse energies of all objects in an event,
$\mett = -\Sigma \Et$.

\bibitem{eegg}
% CDF diphoton plus met PRD
F.~Abe \textit{et al.}, Phys. Rev. D \textbf{59}, 092002 (1999).
%D.~Toback, Ph.~D. thesis, University of Chicago, 1997.
The PEM electron candidate, while satisfying all standard selection criteria, 
is found to have tracking data inconsistent with that of control samples, as
described in these references.  

\bibitem{susy} 
S.~Ambrosanio, G.~L.~Kane, G.~D.~Kribs, S.~P.~Martin, and S.~Mrenna,
Phys. Rev. Lett. \textbf{76}, 3498 (1996);
G.~L.~Kane and S.~Mrenna, Phys. Rev. Lett. \textbf{77} 3502 (1996);
S.~Ambrosanio, G.~L.~Kane, G.~D.~Kribs, S.~P.~Martin, and S.~Mrenna, 
Phys. Rev. D \textbf{55}, 1372 (1997).

      

\bibitem{gg}
%SEARCH FOR GAUGE MEDIATED SUSY BREAKING
%TOPOLOGIES AT S**(1/2) SIMILAR TO 189-GEV.
R. Barate \textit{et al.}, Eur. Phys. J. C \textbf{16} 71, (2000);
% D0 higgs search
B.~Abbott \textit{et al.}, Phys. Rev. Lett. \textbf{82}, 2244 (1999);
% D0 gmsb limits
B.~Abbott \textit{et al.}, Phys. Rev. Lett. \textbf{80}, 442 (1998);
% D0 monopole search
B.~Abbott \textit{et al.}, Phys. Rev. Lett. \textbf{81}, 524 (1998);
% CDF diphoton plus met PRL
F.~Abe \textit{et al.}, Phys. Rev. Lett \textbf{81}, 1791 (1998);
% CDF PJW diphoton analysis
P.~J.~Wilson, in \textit{ICHEP 98}, Proceedings of the 
29th International Conference in High-energy Physics, 
Vancouver, Canada, 1998, edited by 
A.~Astbury, D.~Axen and J.~Robinson (World Scientific, Singapore, 1998); 
% D0 diphoton plus met
S.~Abachi \textit{et al.}, Phys. Rev. Lett. \textbf{78}, 2070 (1997);
%SEARCH FOR ANOMALOUS PRODUCTION OF HIGH MASS PHOTON PAIRS IN E+ E-
%COLLISIONS AT LEP.
P.D. Acton \textit{et al.}, Phys. Lett. \textbf{B311} 391, (1993).
\bibitem{gj} 
% D0 photon plus jets plus met
B.~Abbott \textit{et al.}, Phys. Rev. Lett. \textbf{82}, 29 (1999).
\bibitem{gb}
% CDF photon bb technicolor search
F.~Abe \textit{et al.}, Phys. Rev. Lett \textbf{83}, 3124 (1999).

%detector.tex
\bibitem{VTX} 
F.~Snider \textit{et al.}, Nucl. Instrum. Methods \textbf{A268}, 75 (1988).
This is the reference for the previous generation of the device.  
The replacement for the 1994-5 data sample has more modules, each with
a shorter drift length but is otherwise similar.

\bibitem{Drell-Yan} 
F.~Abe \textit{et al.}, Phys. Rev. D \textbf{59}, 052002 (1999).

\bibitem{CFT} 
G.~W.~Foster \textit{et al.}, 
Nucl. Instrum. Methods Phys. Res. A \textbf{268}, 33 (1988).

%selection.tex

\bibitem{cdf lepton id}
%top mass preprint
%T.~Affolder \textit{et al.}, Report No. FERMILAB-PUB-00-127-E (2000);
%top mass prd
T.~Affolder \textit{et al.}, Phys. Rev. D \textbf{63}, 032003 (2001);
%top evidence
F.~Abe \textit{et al.}, Phys. Rev. D \textbf{50}, 2966 (1994).

%\bibitem{cdf photon id}
% CDF diphoton plus met PRD
%F.~Abe \textit{et al.}, Phys. Rev. D \textbf{59}, 092002 (1999);
% CDF photon bb technicolor search
%Phys. Rev. Lett \textbf{83}, 3124 (1999).

\bibitem{W mass}
% W mass preprint
%T.~Affolder \textit{et al.}, Report No. FERMILAB-PUB-00-158-E (2000);
% W mass PRD
T.~Affolder \textit{et al.}, Phys. Rev. D \textbf{64}, 052001 (2001);
% W mass 1a
F.~Abe \textit{et al.}, Phys. Rev. D \textbf{52}, 4784 (1995).
 
\bibitem{trigger tower} Trigger towers subtend 0.2 in $\eta$ by 
$15^{\circ}$ in $\varphi$.

\bibitem{jwb thesis}
J.~Berryhill, Ph.D. thesis, University of Chicago, 2000.

\bibitem{mk thesis}
M.~Kruse, Ph.D. thesis, Purdue University, 1996.

\bibitem{r papers}
J.~Wahl, Ph.~D. thesis, University of Chicago, 1999;
F.~Abe \textit{et al.}, Phys. Rev. D \textbf{52}, 2624 (1995).
%S.~Kopp, Ph.~D. thesis, University of Chicago, 1994.

\bibitem{jet id}
% 3-jet 1a result
F.~Abe \textit{et al.}, Phys. Rev. D \textbf{45}, 1448 (1992).

%\bibitem{mett 25}
%F.~Abe \textit{et al.}, Phys. Rev. D \textbf{59}, 092002 (1999);
%\textbf{52}, 2624 (1995);
%J.~Wahl, Ph.~D. thesis, University of Chicago, 1999;
%D.~Toback, Ph.~D. thesis, University of Chicago, 1997;
%S.~Kopp, Ph.~D. thesis, University of Chicago, 1994.
 
%diboson_mc.tex
\bibitem{diboson mc} U.~Baur and S.~Mrenna, private communication.
The source code is available publicly at 
\url{http://moose.ucdavis.edu/mrenna/code/}.

\bibitem{diboson matrix}
U.~Baur and E.~L.~Berger, Phys. Rev. D \textbf{47}, 4889 (1993);
\textbf{41}, 1476 (1990).

\bibitem{PYTHIA}
T.~Sjostrand, Computer Physics Commun. \textbf{82} (1994) 74;
S.~Mrenna, Computer Physics Commun. \textbf{101} (1997) 232.
An archive of program versions and documentation is available publicly at
\url{http://www.thep.lu.se/tf2/staff/torbjorn/Pythia.html}.
          
\bibitem{CTEQ5}
H.~L.~Lai \textit{et al.}, Eur. Phys. J. C \textbf{12}, 375 (2000).
The source code is available publicly at 
\url{http://www.phys.psu.edu/~cteq/CTEQ5Table/}.
                   
\bibitem{TAUOLA}
\textsc{tauola} version 2.5 (June 1994), S.~Jadach \textit{et al.}, 
Computer Phys. Commun. \textbf{76} 361 (1993).

\bibitem{MRST} A.D. Martin, R.G. Roberts, W.J. Stirling, and R.S. Thorne,
Eur. Phys. J. C \textbf{4}, 463 (1998).  A 5\% uncertainty is recommended for
$W$ and $Z^{0}$ boson production at the Tevatron.  
The parton species and momenta contributing to $W\gamma$ and 
$Z^{0}\gamma$ production are very similar.
                          
\bibitem{K factor} 
U. Baur, T. Han, and J. Ohnemus,
Phys. Rev. D \textbf{57} (1998) 2823; 
U. Baur, T. Han, and J. Ohnemus,
Phys. Rev. D \textbf{48} (1993) 5140; 
J. Ohnemus,
Phys. Rev. D \textbf{47} (1993) 940.

%\bibitem{EOT}
%F.~Abe \textit{et al.}, Phys. Rev. D \textbf{59}, 092002 (1999).  

\bibitem{lumint}
D.~Cronin-Hennessy \textit{et al.}, Nucl. Instrum. Methods Phys. Res.,
Sect. A \textbf{443/1}, 37-50 (2000).
    
% fake photon
%\bibitem{Wahl trigger} 
%J.~Wahl, Ph.D. thesis, University of Chicago, 1999.

% pthru.tex 
\bibitem{kpi int length} 
D.E. Groom \textit{et al.}, Eur. Phys. J. C
\textbf{15}, 1 (2000).  Computer-readable data files may be found at
\url{http://pdg.lbl.gov/xsect/contents.html}.

\bibitem{ktopi} 
D.~Buskulic \textit{et al.}, Z. Phys. \textbf{C66}, 355 (1995);
P.~Abreu \textit{et al.}, Nucl. Phys. \textbf{B444}, 3 (1995);
R.~Akers \textit{et al.}, Z. Phys. \textbf{C63}, 181 (1994);
H.~Aihara \textit{et al.}, Phys. Rev. Lett. \textbf{61}, 1263 (1988);
D.~Antreasyan \textit{et al.}, Phys. Rev. Lett. \textbf{38}, 115 (1977); 
J.W.~Cronin \textit{et al.}, Phys. Rev. \textbf{D11}, 3105 (1975).

% hf.tex
\bibitem{QQ}
\textsc{QQ} version 8.08 (June 1991), 
P.~Avery, K.~Read, G.~Trahern, Cornell Inernal Report No. CSN-212 (1985),
unpublished. 

\bibitem{cdf photonc}
S.~Kuhlmann, Report No. FERMILAB-CONF-99-165-E (1998);
F. Abe \textit{et al.}, Phys. Rev. D \textbf{60}, 092003 (1999);
Phys. Rev. Lett. \textbf{77}, 5005 (1996).
%R.~Hamilton, Ph.D. thesis, Harvard University, 1996.

%appendixa
%\bibitem{Toback trigger}
%D.~Toback, Ph.D. thesis, University of Chicago, 1997, p.29.

%appendixb 
%\bibitem{lep id}
%M.~Kruse, Ph.D. thesis, Purdue University, 1996.

\end{thebibliography}
\end{document}